\RequirePackage{fix-cm}
\documentclass[12pt]{iopart}
\RequirePackage{graphicx}
\RequirePackage{mathptmx}      
\RequirePackage[numbers,sort&compress]{natbib}
\RequirePackage[colorlinks,citecolor=blue,urlcolor=blue,linkcolor=blue]{hyperref}

\def\bibhang{24pt}%
\usepackage[T1]{fontenc} 

\usepackage{pbox}

\usepackage[margin=10pt]{subcaption}
\usepackage[percent]{overpic}
\usepackage{dblfloatfix}

\usepackage{amssymb}
\usepackage{enumerate}

\newcommand*\cleartoleftpage{%
  \clearpage
  \ifodd\value{page}\hbox{}\newpage\fi
}

\begin{document}

\title[Observation of an excess at 30~GeV in the opposite sign di-muon spectra of ${\rm Z}^{0} \to b\overline{ b} + {\rm X}$]{Observation of an excess at 30~GeV in the opposite sign di-muon spectra of ${\rm Z}^{0} \to b\overline{ b} + {\rm X}$ events recorded by the ALEPH experiment at LEP}

\author{Arno Heister}
\ead{Arno@slhc.info}

\begin{abstract}
The re-analysis of the archived data recorded at the ${\rm Z}^{0}$ resonance by the ALEPH experiment at LEP during the years 1992-1995 shows an excess in the opposite sign di-muon mass spectra at 30.40~$\pm$~0.46~GeV in events containing b quarks. The excess has a natural width of 1.78 $\pm$~1.14~GeV.

The di-muon excess has a local significance around $5\,\sigma$ ($Z_{\rm asym}$), depending on the background model used. The significances for background models based on a kernel density approximation stay close to $3\,\sigma$ ($Z_{\rm freq,\,lee}$), when including a look elsewhere effect. Another method to obtain a significance value results in at least $2.6\,\sigma$ ($Z_{\rm Bi}$). A compatible, but smaller excess is visible in the opposite di-electron mass spectrum as well.

This paper uses the data collected by the ALEPH experiment in the years 1992-1995, which have been archived to allow their use for physics analyses after the closure of the collaboration.
\end{abstract}


\submitto{\JPG}

\maketitle

\section{Introduction}
\label{sec:introduction}
Models of new phenomena in high energy physics, which predict event signatures unique and easy to distinguish from Standard Model event topologies, were always of great interest by experimental particle physicists. In particular, models which predict event signatures containing muons in the final state are popular, because muons are easy to detect in high energy collisions. To make those signatures stick out further, additional features of special event classes of the proposed models are highlighted. Often those event classes contain extra particles with a non-negligible lifetime making them look different from e.g.~Standard Model QCD events.  

In 2006, after the shutdown of the Large Electron-Positron Collider (LEP), "hidden valley" models were proposed \cite{Strassler:ys,Han:2007yu,Strassler:yq}. Examples of such models predict new particles, which have (electrically-neutral) bound states, low masses and long lifetimes. Eventually, such events can be found in the archived data of past experiments at LEP and the Tevatron. Certainly, the running LHC experiments are looking for such and other signatures of new physics phenomena.

The ALEPH experiment was one of the four main experiments at LEP.  A re-analysis of the archived ALEPH data recorded during the years 1992-1995 at the ${\rm Z}^{0}$ resonance is presented here~\cite{ALEPH-archived-data}. The outcome is an excess in the invariant mass spectra  of the opposite sign di-muon pairs at 30.40~GeV in events where at least one b-quark jet is additionally present. The natural width of the excess is 1.78~GeV. The background is modeled using events where a muon and an electron with opposite signs are present. These events had to pass the same selection criteria as used for the signal. A compatible but smaller excess is visible in the opposite di-electron mass spectrum as well. The present analysis suggests the following event topology: ${\rm Z}^{0} \to {\rm b} \, \overline{\rm b} \, \mu^{+}\mu^{-}$ or ${\rm Z}^{0} \to {\rm b} \, \overline{\rm b} \, {\rm e}^{+}{\rm e}^{-}$.

This paper is organized as follows: In Sec.~\ref{sec:Experimental_Procedure} an overview of the ALEPH experiment, as well as details of the archived data used in the present analysis are given. Details of the analysis are presented in Sec.~\ref{sec:analysis}. Crosschecks and further studies of the excess are described in Sec.~\ref{sec:crosschecks}. A summary is given in Sec.~\ref{sec:summary}.

\section{Experimental procedure}\label{sec:Experimental_Procedure}

\subsection{The ALEPH detector}
\label{sec:alephdetector}
The ALEPH detector (Apparatus for LEp PHysics) was one of the four multi-purpose detectors at LEP and was located at the experimental area of Point 4 near Echenevex (France) \cite{Decamp:1990jra}. The detector had a cylindrical shape with approximately $12 {\rm\;m}$ diameter by $12 {\rm\;m}$ length and consisted of independent and modular sub-detectors arranged in layers around the beam-pipe (radius $5.3 {\rm\;cm}$), each specializing in a different task~\cite{Buskulic:1994wz}.

Charged particles were tracked with three devices inside a super-conducting solenoid having an axial field of $1.5 {\rm\;T}$. The innermost tracking detector was the mini vertex detector (VDET), which was installed in 1991. It consisted of 2 layers of silicon wafers with strip readout in 2 dimensions (radii: $\approx 6.3 {\rm\;cm}$ and $10.8 {\rm\;cm}$). It was followed by the inner tracking chamber (ITC), a cylindrical multi-wire drift chamber able to provide up to eight precise $r$-$\phi$ points per track. Finally, the time projection chamber TPC, a cylindrical imaging drift chamber, provided up to 21 three dimensional coordinates of the particle trajectories. In hadronic ${\rm Z}^{0}$ decays, tracks crossing at least four pad rows in the TPC are reconstructed with an efficiency of 98.6\%. For tracks with two VDET coordinates a transverse momentum resolution $\Delta p_{t}/p_t = 6\cdot 10^{-4}\,p_t \oplus 0.005\,(p_t{\rm\:in\:GeV})$ could be achieved.

The energy of neutral and charged particles was measured by an electromagnetic calorimeter (ECAL) inside and a hadron calorimeter (HCAL) outside the magnetic coil. Both were sampling calorimeters with different longitudinal segmentation. The ECAL read out towers pointed to the nominal interaction point and had an average granularity of $0.9^{\circ}\,\times\,0.9^{\circ}$. The resulting energy resolution was $\sigma(E)/E = 0.0009 + 0.18/\sqrt E\,(E{\rm\:in\:GeV})$. The typical granularity of the projective HCAL towers was $3.7^{\circ}\,\times\,3.7^{\circ}$ corresponding to $4 \times 4$ ECAL towers. The obtained energy resolution was $\sigma(E)/E = 0.85/\sqrt E\,(E{\rm\:in\:GeV})$.

Muons were identified using the tracking capabilities of the HCAL together with the muon chambers (MUON) outside the HCAL. The average muon identification efficiency was 86\%. For muons traversing both double layers of streamer tubes an accuracy in the track direction measurement of $\approx$ 10-15~mrad could be achieved. The momentum components $p_{T}$, $p_{Z}$ of the muons were measured by means of the tracking system.

The ALEPH trigger was designed to detect events stemming from ${\rm e}^+{\rm e}^-$ collisions with very high efficiency. For each physics channel the trigger logic and redundancy allowed trigger efficiencies near 100\% to be obtained. The trigger decision was based on:
\begin{enumerate}
\item Total-energy trigger: energy deposits in the electromagnetic calorimeter
\item Electron-track trigger: track segments in the drift chambers with corresponding energy deposits in the electromagnetic calorimeter
\item Muon-track trigger: track segments in the drift chambers with corresponding energy deposits in the hadronic calorimeter
\item Back-to-back trigger: two back-to-back track segments in the drift chambers
\end{enumerate}
The use of the Total-energy and Muon-track trigger to collect hadronic ${\rm Z}^0$ events had an efficiency of ($99.99 \pm 0.01$)\%.

The luminosity was measured using Bhabba events. They were triggered with a rate of 2-3 Hz. The overall trigger rate was 4-5 Hz with ${\rm Z}^{0}$ events (at the peak) and 2-photon events contributing about 0.5 Hz each. The remaining contribution stemmed from cosmic rays, noise and beam related background. Because of the low trigger rate no reduction of the event rate was needed to satisfy the bandwidth limitations of the ALEPH readout.

\subsection{Data sample and hadronic event selection}
\label{sec:data-sample}
From 1989 to 1995 CERN's electron-positron collider (LEP) was operated at the centre-of-mass energy of the ${\rm Z}^0$ resonance, corresponding to about 91.2~GeV (LEP1 phase). LEP delivered about 16M ${\rm Z}^0$ bosons to the four experiments ALEPH, DELPHI, L3 and OPAL during these years~\cite{ALEPH:2005ab}. The present analysis uses the archived ALEPH experiment data recorded during 1992-1995 ($\approx$ 3.7 M events) as well as  about 8 Million simulated events from the archived ALEPH Monte Carlo (MC) data, where the hadronic decays ${\rm Z}^{0} \to {\rm q}\overline{\rm q}$ were fully simulated.

Hadronic ${\rm Z}^{0}$ decay modes are  identified by means of the standard ALEPH hadronic selection criteria, based on the observation of at least five good charged particle tracks~\cite{Decamp:1990ky,Decamp:1991aj,Buskulic:1993gu}. After this selection about 1.9 M hadronic events of the archived ALEPH data remain for further analysis\footnote{The standard run quality requirements of the ALEPH Heavy Flavor group are used. One expects about 2.2M hadronic events passing this selection according to e.g.~Ref.~\cite{Barate:1997kr}. This discrepancy has to be double checked with ALEPH experts. However, this does not affect the present result.}. 

\section{Di-Muon mass analysis}
\label{sec:analysis}
\subsection{Overview}
\label{sec:analysis_overview}
This analysis searches for at least two opposite sign (OS) leptons (muons or electrons) in hadronic ${\rm Z}^{0}$ decays (see Sec. \ref{sec:data-sample}) positively identified as ${\rm Z}^{0} \to {\rm b}\overline{\rm b} + {\rm X}$ final state as described below. A di-lepton pair is classified as {\it signal} in case it consists of two same flavor opposite sign leptons, i.e. an opposite sign di-muon or opposite sign di-electron, or as {\it background} in case of different flavor opposite signs, i.e. an electron/muon pair with opposite signs.

This analysis is based on the standard  ALEPH analysis software ALPHA~\cite{ALEPH-alpha}. It utilizes charged tracks reconstructed by means of ALPHA. The same track selection is employed as used for tagging the presence of long-lived particles by means of the routine QIPBTAG optimized for primary vertex reconstruction and b-jet identification\footnote{For identifying b-quark signatures a vertex detector is important. Because the ALEPH vertex detector VDET was installed in 1991 the LEP1 data starting from 1992 are used for the present study (see also Sec.~\ref{sec:alephdetector})}. In order to search inclusively in hadronic ${\rm Z}^{0}$ decays, the requirement that charged tracks are close to the thrust axis of the event is removed. This allows di-lepton configurations which have a large opening angle and hence possibly large invariant mass. 

Electron and muon tracks are identified by means of the ALEPH Lepton ID for the reprocessed data~\cite{ALEPH-lepton-id}. Important selection criteria for electrons and muons are summarized in Tab.~\ref{table:aleph_lepton_id}. If not mentioned otherwise, always the ALEPH standard reconstruction is used, e.g.~for the primary vertex determination, etc.

\begin{table*}[t]
   \begin{subtable}[h]{0.5\linewidth}
   \centering
    \begin{tabular}{rll}
        \hline\noalign{\smallskip}
        {\bf Electrons} \\
        \noalign{\smallskip}\hline\noalign{\smallskip}
 	$p_{\rm T,\,\rm track }$ & $> 2\,{\rm\; GeV}$  \\ 
        \# TPC hits  &$\ge 5$  \\
        $|cos\,\theta_{\rm track}|$ &$< 0.95$  \\
        $|{\rm d}_{0,\, {\rm track}}|$ &$< 0.5$  \\
        $|{\rm Z}_{0, \,{\rm track}}|$ &$< 5$ \\  
        \noalign{\smallskip}\hline
  \end{tabular}  
  \subcaption{\label{table:aleph_electron_id}}
\end{subtable}
   \begin{subtable}[h]{0.5\linewidth}
   \centering
     \begin{tabular}{rll}
        \hline\noalign{\smallskip}
        {\bf Muons} \\
        \noalign{\smallskip}\hline\noalign{\smallskip}
        $p_{\rm T,\,\rm track }$ & $> 2.5\,{\rm\; GeV}$  \\ 
        \# TPC hits  & $\ge 5$  \\
        $|cos\,\theta_{\rm track}|$ & $< 0.95$  \\
        $|{\rm d}_{0, \,{\rm track}}|$ & $< 0.5$  \\
        $|{\rm Z}_{0, \,{\rm track}}|$ & $< 5$  \\
        \noalign{\smallskip}\hline
   \end{tabular}  
   \subcaption{\label{table:aleph_muon_id}}
\end{subtable}
 \caption{Important parameters of the used ALEPH lepton identification for electrons (left) and muons (right) as described in Ref.~\cite{ALEPH-lepton-id}. The observable ${\rm d}_{0,\, {\rm track}}$ is the distance of closest approach of the track to the beam axis. ${\rm Z}_{0, \,{\rm track}}$ is the z coordinate of the track point where ${\rm d}_{0,\, {\rm track}}$ was measured.\label{table:aleph_lepton_id}}
\end{table*}

\begin{figure*}[b]
      \centering
      \subcaptionbox{$P_{\rm H,\,mass}$ obtained from ALEPH data and its description by the ALEPH MC data. The plot is consistent with Fig.~2 in Ref.~\cite{Barate:1997kr}.\label{fig:btag_data_mc}}
        {\includegraphics[width=0.49\textwidth]{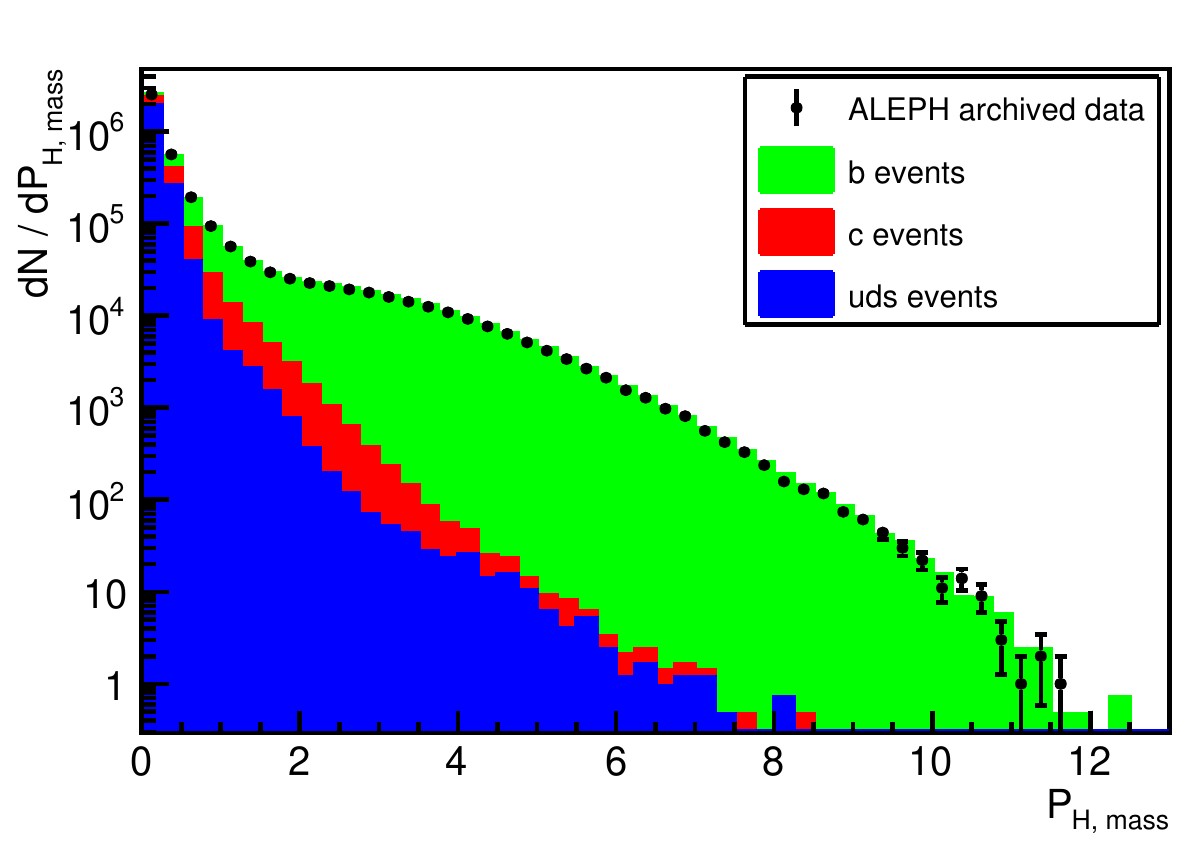}}
      \subcaptionbox{The b-purity as a function of the b-efficiency for the b-tag observables $P_{\rm H,\,mass}$ and $P_{\rm H,\,mass,\,max}$. The used b-tag working points are marked in the plot.\label{fig:btag_efficiency}}
        {\includegraphics[width=0.49\textwidth]{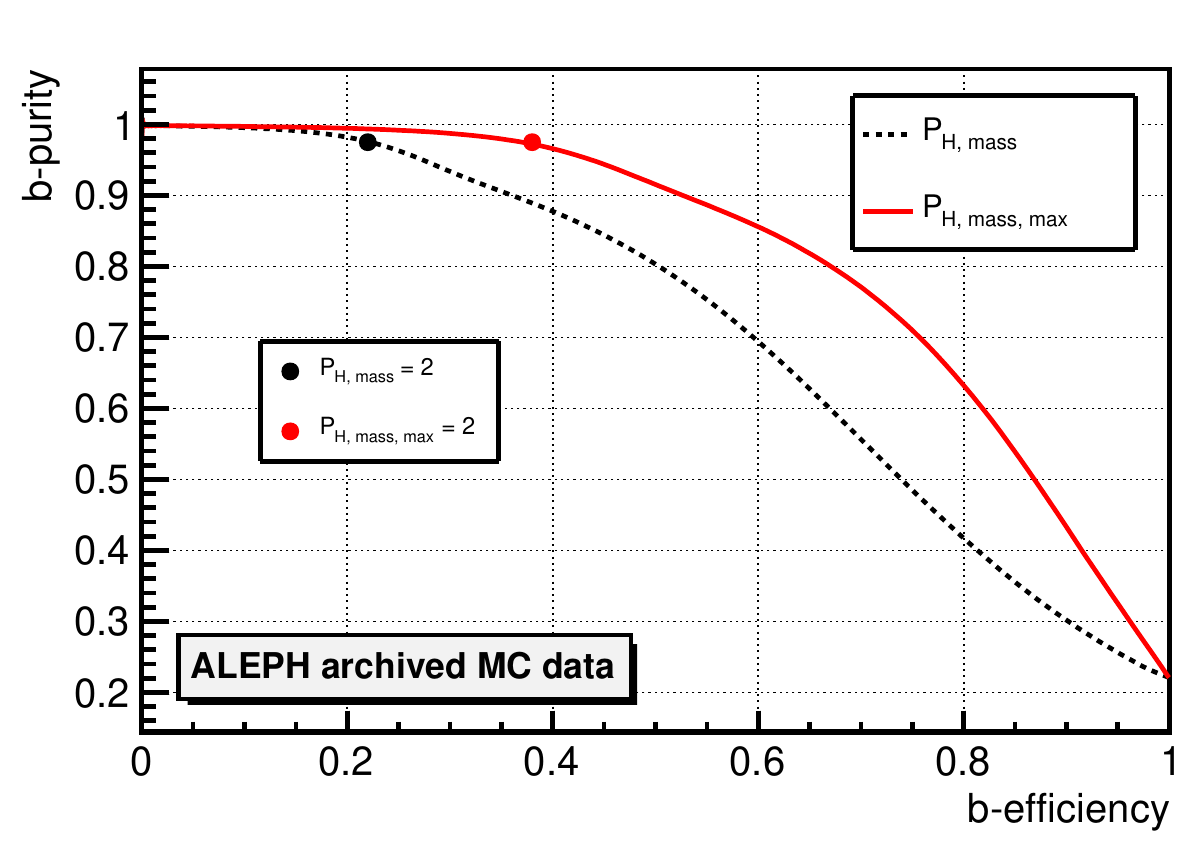}}
      \caption{The b-tag observables $P_{\rm H,\,mass}$ and $P_{\rm H,\,mass,\,max}$.}
\end{figure*}

Three criteria are used for the final selection (see Sec.~\ref{sec:analysis_btag} and \ref{sec:analysis_semileptonic}): a) A {\it lifetime mass tag} to identify ${\rm Z}^{0} \to {\rm b}\overline{\rm b} + {\rm X}$ decays, b) events should not have much missing momentum and c) a requirement that the two leptons should be close to the reconstructed primary vertex . The latter two requirements are used to suppress semi-leptonic decays, especially from ${\rm b}\overline{\rm b}$ final states.

\subsection[Identification of ${\rm Z}^{0} \to {\rm b}\overline{\rm b} + {\rm X}$ decays]{\boldmath Identification of ${\rm Z}^{0} \to {\rm b}\overline{\rm b} + {\rm X}$ decays}
\label{sec:analysis_btag}

The ALEPH lifetime mass tag is used to identify the ${\rm b}\overline{\rm b}$ final state~\cite{Barate:1997kr}. Hadronic events are divided into two hemispheres by means of a plane perpendicular to the thrust axis of the event. Events are classified as ${\rm Z}^{0} \to {\rm b}\overline{\rm b} + {\rm X}$  events by means of the observable $P_{\rm H,\,mass}$, which is defined for each of the two hemispheres per event:
\begin{eqnarray*}
  P_{\rm H,\,mass} = 0.7\,\log_{10} \mu_{H} + 0.3\,\log_{10} P_{\rm H}
\end{eqnarray*}
The observable  $\mu_{H}$ in the formula above is constructed by adding up the tracks inside a hemisphere in order of decreasing inconsistency with the primary vertex until their invariant mass exceeds $1.8 {\rm\;GeV}$ (the mass of a ${\rm c}$-hadron). The probability $P_{\rm track}$ of the last track added before exceeding the mass limit is defined as $\mu_{H} := P_{\rm track}$. The observable $P_{\rm H}$ (hemisphere lifetime probability) is calculated from tracks associated to one hemisphere. Details about the definition of $P_{\rm track}$ and  $P_{\rm H}$ can be found in~\cite{Jacobsen:1991ex,Brown:1992,Barate:1997kr}. Fig.~\ref{fig:btag_data_mc} shows the distribution of the observable $P_{\rm H,\,mass}$ obtained from ALEPH data and its reasonably good description by the ALEPH MC data.

For the present analysis, events with ${\rm b}\overline{\rm b}$ in the final state are selected, if for one of the two hemispheres $P_{\rm H,\,mass, max} > 2$. The b-efficiency for this b-tag working point is about 38\% (Fig.~\ref{fig:btag_efficiency}). Using as b-tag $P_{\rm H,\,mass} > 2$ for one arbitrarily selected hemisphere results in a b-efficiency of about 22\%, which is consistent with Ref.~\cite{Barate:1997kr}. The c- and light-quark contamination in the selected events for both working points is only a few percent, i.e. about 2-3 \% of the events are selected wrongly as b's.

\begin{figure*}[b]
      \centering
      \subcaptionbox{\label{fig:P_di-muon}}
        {\includegraphics[width=0.49\textwidth]{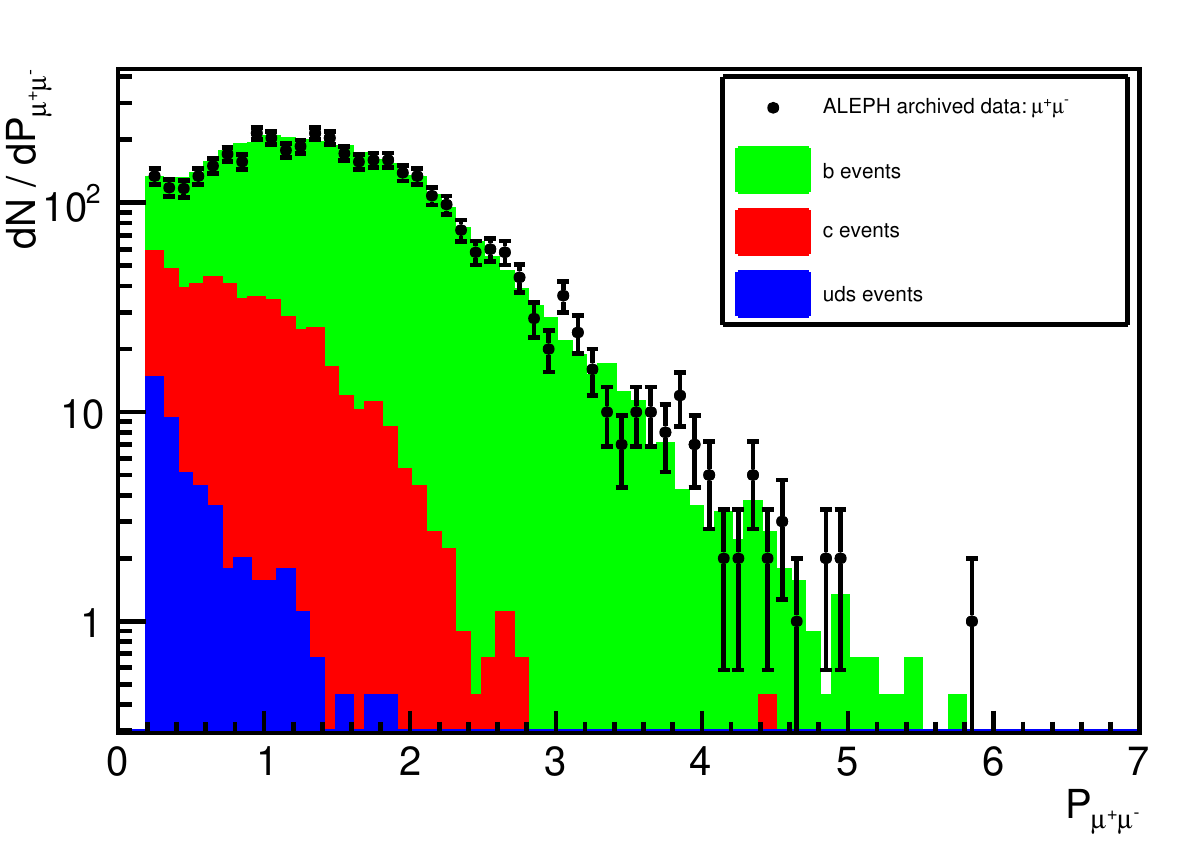}}
      \subcaptionbox{\label{fig:P_miss}}
        {\includegraphics[width=0.49\textwidth]{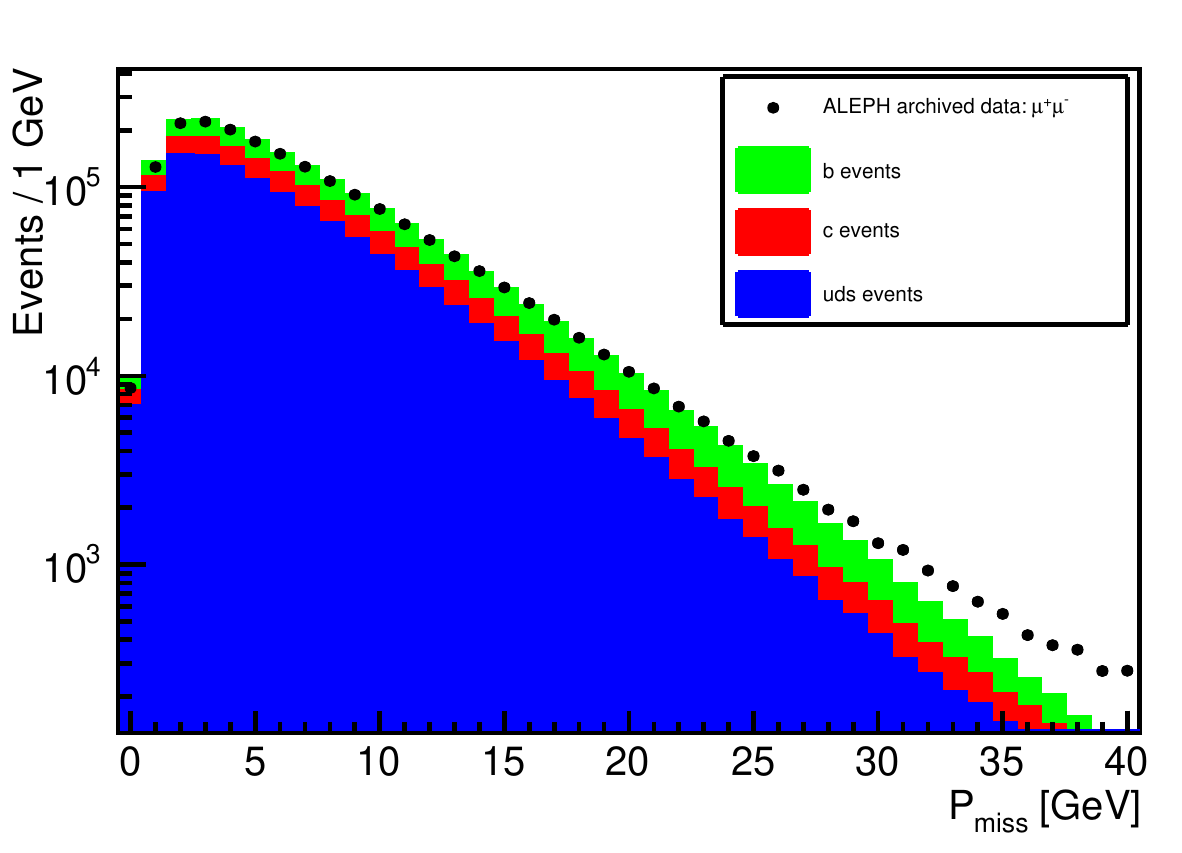}}
      \caption{The observables $P_{\rm \mu^{+}\mu^{-}}$ and $P_{\rm miss}$.}
\end{figure*}

\subsection{Suppression of semi-leptonic decays}
\label{sec:analysis_semileptonic}
In case of semi-leptonic decays of b- and c-quarks a neutrino is produced, which cannot be directly detected. To suppress this kind of decays in the final event selection the missing 3-momentum $P_{\rm miss}$ is required to be low: $P_{\rm miss} < 18\,{\rm\; GeV}$.

In addition, reconstructed lepton tracks from such semi-leptonic decays will stem from an additional secondary vertex where the quark decayed. Therefore, in order to suppress them, the probability of the two lepton tracks to stem from the same primary vertex is required to be high. The significance of the signed 3D impact parameter for the two reconstructed lepton tracks is used to calculate their lifetime probability, i.e. the confidence level $P_{\rm \mu^{+}\mu^{-}}$.The calculation is done in the same way as for a hemisphere or jet lifetime probability~\cite{Jacobsen:1991ex,Brown:1992}. It is required that $P_{\rm \mu^{+}\mu^{-}} < 2.5$.

Fig.~\ref{fig:P_di-muon} and \ref{fig:P_miss} show the distribution of the observables used to discriminate against b semi-leptonic decays, as well as their good description by the simulation in the region of interest. Both working points ensure the desired suppression of semi-leptonic b-decays.

If not mentioned otherwise, all selection criteria described here are applied to events entering the distribution shown in the following.

\subsection{Choice of the background model: semi-leptonic decays}
\label{sec:background model}
The background (model) is obtained from ALEPH data only using hadronic ${\rm b}\overline{\rm b}$ final states, where in addition an electron and muon track with opposite signs are found. All opposite sign electron-muon pairs in the event are used for the shown distributions. It turns out that such hadronic ${\rm Z}^{0}$ events with additional opposite sign electron-muon pairs originate mostly from semi-leptonic b-quark decays (see also Sec.~\ref{sec:crosschecks}). One of the two b-quarks from the ${\rm Z}^{0}$ decays into electron + quark + neutrino final state and the other b-quark into muon + quark + neutrino.

\begin{figure*}[!htbp]
      \centering
      {\includegraphics[width=0.65\textwidth]{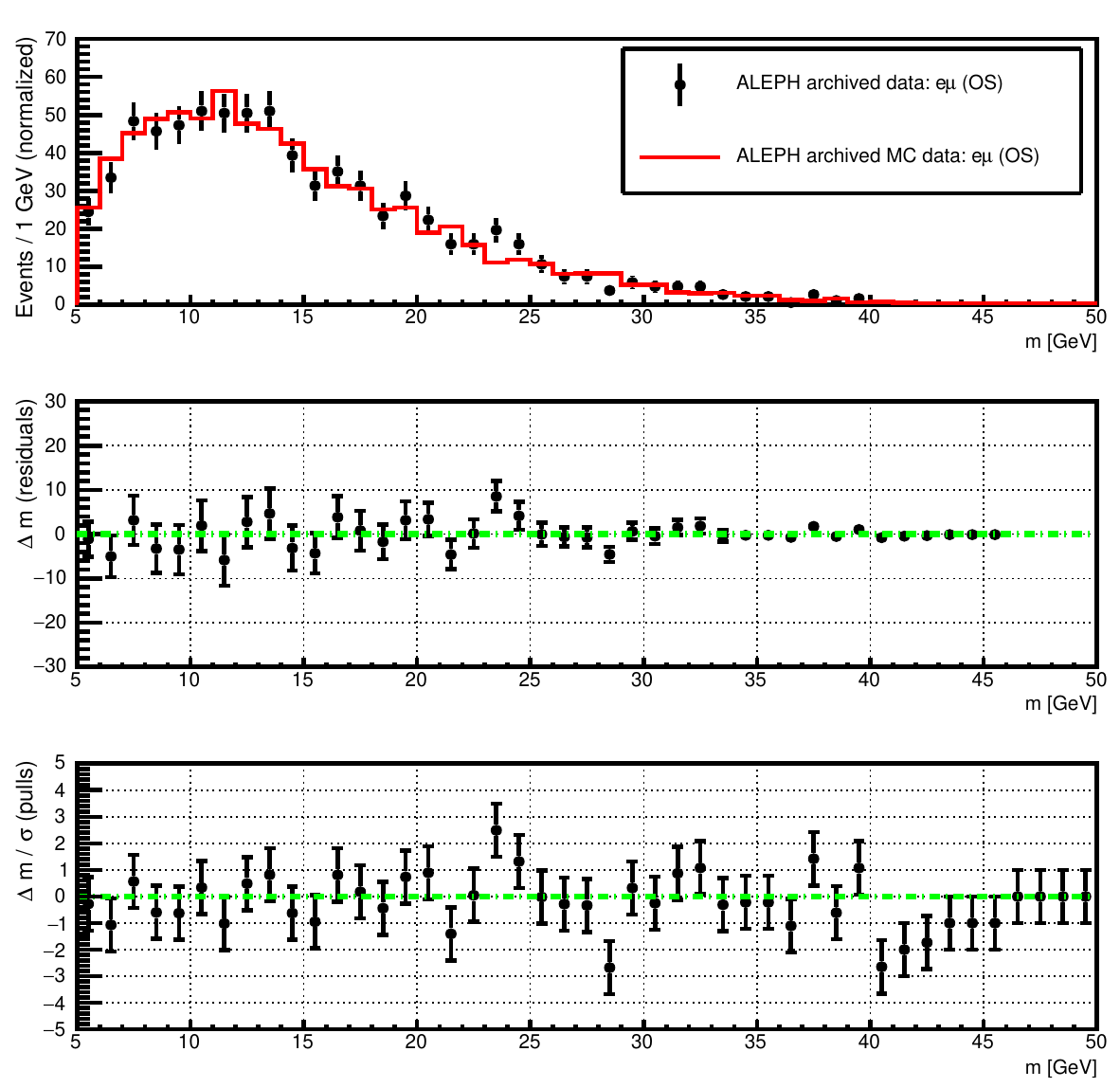}}
      \caption{Comparison of reconstructed electron-muon masses from ALEPH data with MC data. The overall event numbers are normalized to the yield of the opposite sign di-muon mass spectrum obtained from ALEPH data.\label{fig:background_data_mc}}
      {\includegraphics[width=0.65\textwidth]{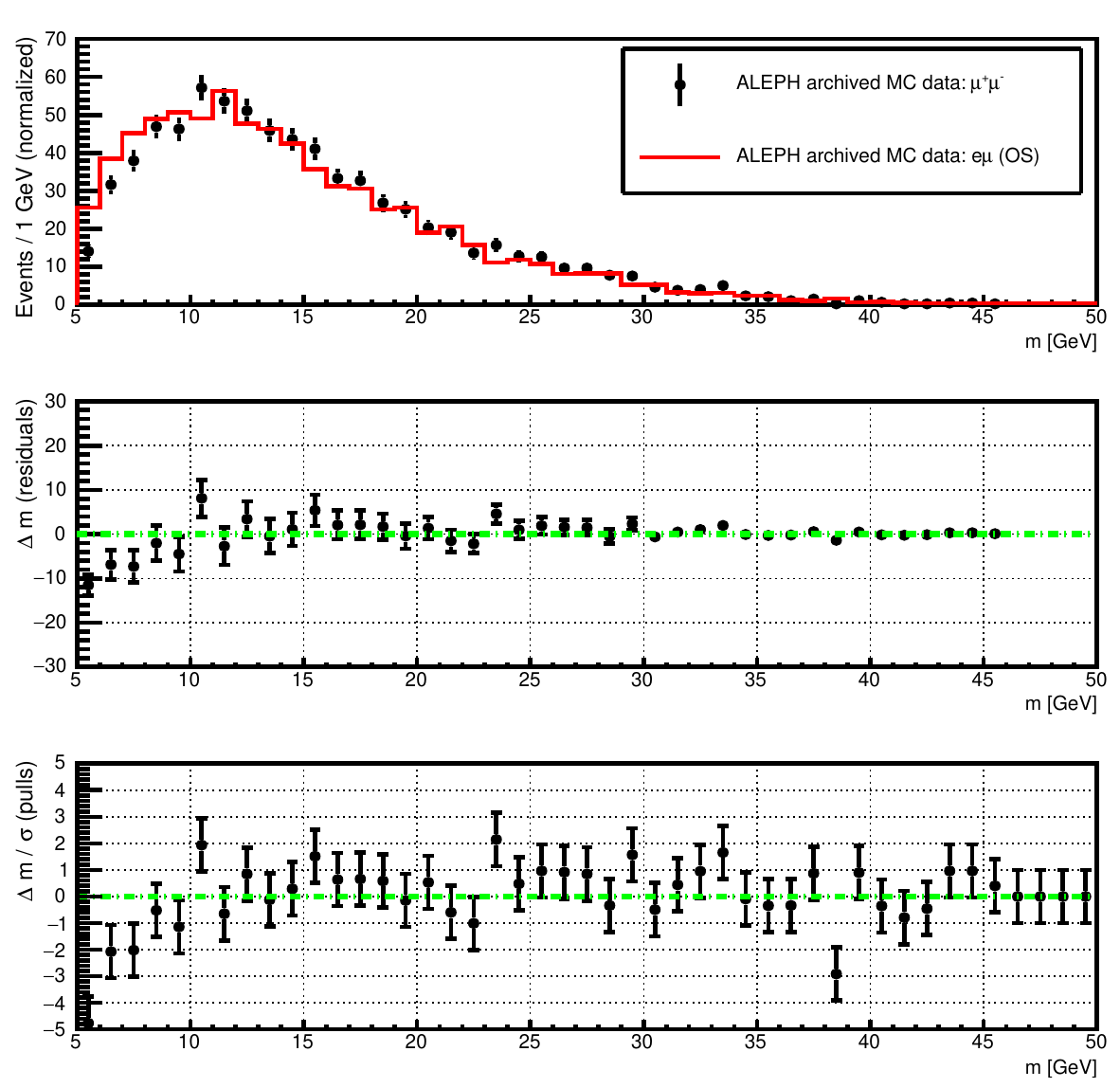}}
      \caption{Comparison of the opposite sign di-muon mass spectrum to the electron-muon mass spectrum for ALEPH MC data only.The overall event numbers are normalized to the yield of the opposite sign di-muon mass spectrum obtained from ALEPH data.\label{fig:signal_background_mc}}
\end{figure*}

\begin{figure*}[!htbp]
      \centering
       {\includegraphics[width=0.65\textwidth]{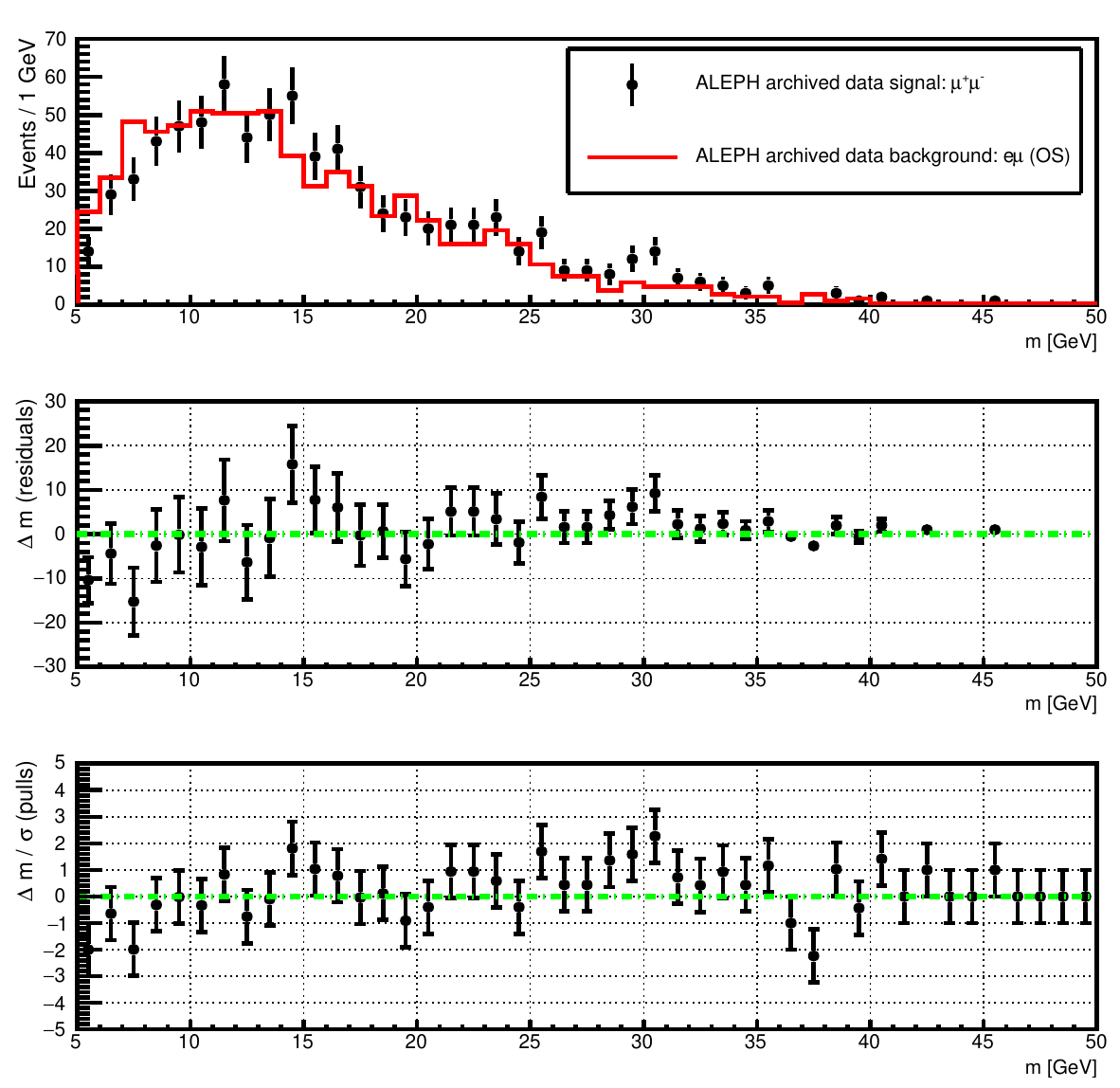}}
      \caption{Comparison of the reconstructed di-muon mass spectrum with the electron-muon mass spectrum using ALEPH data only. The background distributions are normalized to the opposite sign di-muon mass spectrum as described in Sec.~\ref{sec:background_norm}.\label{fig:signal_background_data}}      
      {\includegraphics[width=0.65\textwidth]{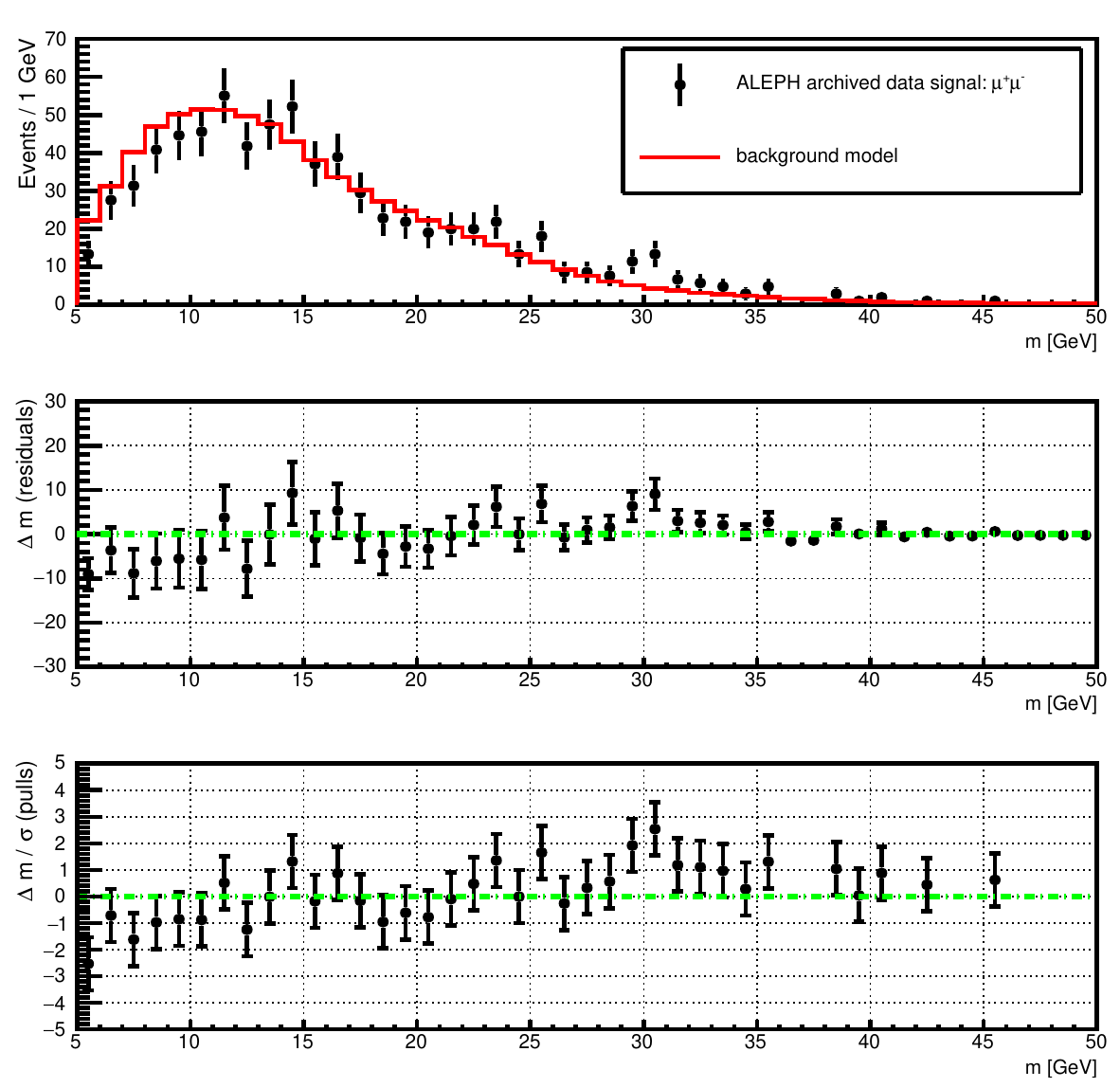}}
      \caption{Comparison of the reconstructed di-muon mass spectrum to the background model computed from the electron-muon mass spectrum using ALEPH data. The background distributions are normalized to the opposite sign di-muon mass spectrum as described in Sec.~\ref{sec:background_norm}.\label{fig:signal_background_model_data}}
\end{figure*}

\subsubsection{Background model: Normalization}
\label{sec:background_norm}
The probability of two semi-leptonic b-quark decays, one into a final state with an electron and the other into a muon is twice as high as compared to the case when both semi-leptonic decays have a muon in the final state. Additionally one has to take into account the slightly less efficient ALEPH reconstruction of electrons in comparison to muons, which is about 6\% less using the standard ALEPH lepton identification~\cite{ALEPH-lepton-id}. The lepton identification efficiency is independent from the reconstructed di-lepton mass in the range from 5 to 50~GeV, both for muons and electrons. In the following this normalization is used when comparing same lepton flavor di-lepton mass distributions (signal) with the mass spectrum obtained from the opposite sign electron-muon pairs (combinatorial background).

\subsubsection{Background model: Construction}
The opposite sign electron-muon mass distribution obtained from ALEPH data is used as the background model in this analysis. The checks below serve to demonstrate that the above choice of background is valid.

Fig.~\ref{fig:background_data_mc} shows the electron-muon invariant mass distribution obtained from ALEPH data compared to the same distribution using ALEPH MC data. The distributions agree well within their errors. A deviation is visible around 24~GeV, where the description by the simulated data is poor.\footnote{For all tests presented in this section, only the error of the data points is used in the residual and pull distributions. It follows that the corresponding errors in the distributions are slightly too small (Fig.~\ref{fig:background_data_mc} to ~\ref{fig:signal_background_model_data}).} A small downwards fluctuation around 28~GeV is visible as well.

The opposite sign di-muon mass spectrum compared to the opposite sign electron-muon mass spectrum using ALEPH MC data only is shown in Fig.~\ref{fig:signal_background_mc}. Also here the distributions agree well within their errors. The deviation around 24~GeV is less visible. Around 10~GeV the di-muon distribution shows some discrepancy from the electron-muon one. Here the well known Y resonances~\cite{Agashe:2014kda} present in the ALEPH simulation for opposite sign di-muons are causing this effect.

Fig.~\ref{fig:signal_background_data} shows the comparison of the di-muon spectrum with the electron-muon mass spectrum, both obtained from ALEPH data only. The normalization of the electron-muon events is computed as described in Sec.~\ref{sec:background_norm}. With the exception of one small deviation around 26~GeV and above 36~GeV for 2 bins, the distributions agree pretty well from 15 to 50~GeV. There is a one bin deviation around 26~GeV (to be expected with the available statistics) and a larger excess around 30~GeV.

To smoothen the statistical fluctuations visible in the electron-muon mass spectrum shown in Fig.~\ref{fig:signal_background_data} and thus reduce this influence on the final result a background model is constructed. The model is obtained by means of a one-dimensional kernel estimation probability density function (p.d.f.), which models the electron-muon mass spectrum as a superposition of Gaussian kernels, one for each data point, each contributing 1/N to the total integral of the p.d.f.~\cite{Cranmer:2000du}. For the computation of the model the electron-muon mass data points are used unbinned. The comparison of this background model with the di-muon mass spectrum is shown in Fig.~\ref{fig:signal_background_model_data}. As expected the background model does not show anymore the fluctuations seen in the bare electron-muon mass spectrum shown in Fig.~\ref{fig:signal_background_data}. The agreement of the di-muon mass spectrum with the background model is still reasonable as shown in Fig.~\ref{fig:signal_background_model_data}. The excess around 30~GeV is clearly visible as before.

\subsubsection{Background model: Summary}
In the region 15 to 50~GeV, where no effects from J/$\psi$ and/or Y resonances are expected, the background model template obtained from ALEPH data only using the p.d.f. from the one-dimensional kernel density approximation provides within errors a reasonably good description of (a) the opposite sign electron-muon mass spectrum and (b) the opposite sign di-muon mass spectrum (excepting the region around 30~GeV). With this p.d.f. we are able to overcome uncertainties due to the statistical fluctuations visible in the electron-muon mass spectrum because of the limited statistics of the ALEPH data.
 
\subsection{Results}
\label{sec:results}
The ROOT framework including the RooFit and RooStats packages~\cite{ROOT,Antcheva:2009zz,Moneta:2010pm} was utilized to construct and fit a signal + background model to the di-muon mass spectrum. The background model and its features were presented in Sec.~\ref{sec:background model}. The signal model is constructed by a convolution of a Gaussian distribution with a Breit-Wigner distribution. The width of the Gaussian distribution is designed to model the ALEPH detector resolution, whereas the width of the Breit-Wigner distribution models the natural width of the resonance. Since both widths are highly correlated a {\it penalty function} is added, which constrains and models the ALEPH detector resolution for different di-muon masses (see ALEPH track transverse momentum resolution described in Sec.~\ref{sec:alephdetector}):
\begin{eqnarray}
\sigma_{\rm res, constraint} = \sqrt{6 \cdot 10^{-4}} \cdot m_{\mu^{+}\mu^{-}}
\end{eqnarray}
An extended maximum likelihood (MLE) fit of the signal + background model to the opposite di-muon mass values in the range 15 to 50~GeV is used. The mass values are used unbinned in the fit.

\begin{figure*}[!h]
      \centering
       \includegraphics[width=10.1cm]{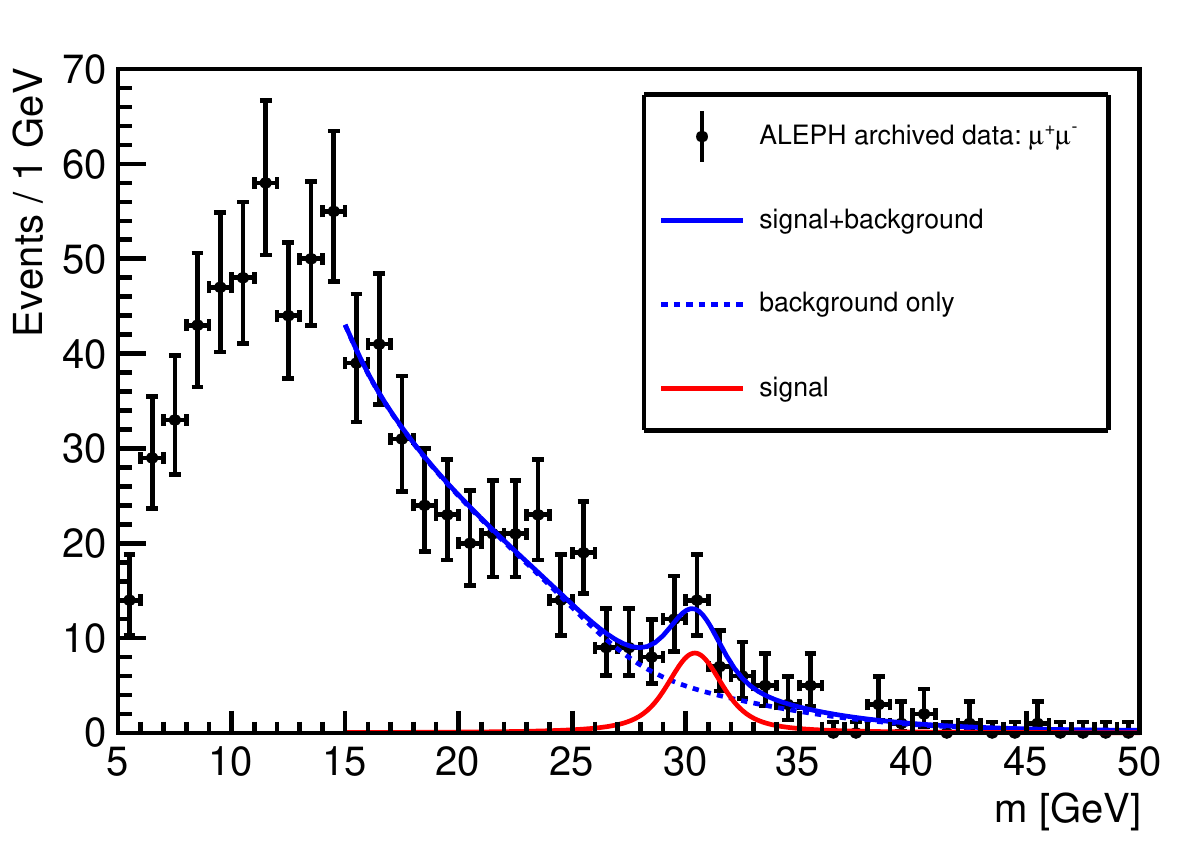}
      \caption{\label{fig:di-muon_signal}The result of the extended maximum likelihood fit of the signal + background model to the unbinned opposite sign di-muon mass spectrum.}
      \centering
      \subcaptionbox{\label{fig:di-muon_signal_binned}}
        {\includegraphics[width=0.46\textwidth]{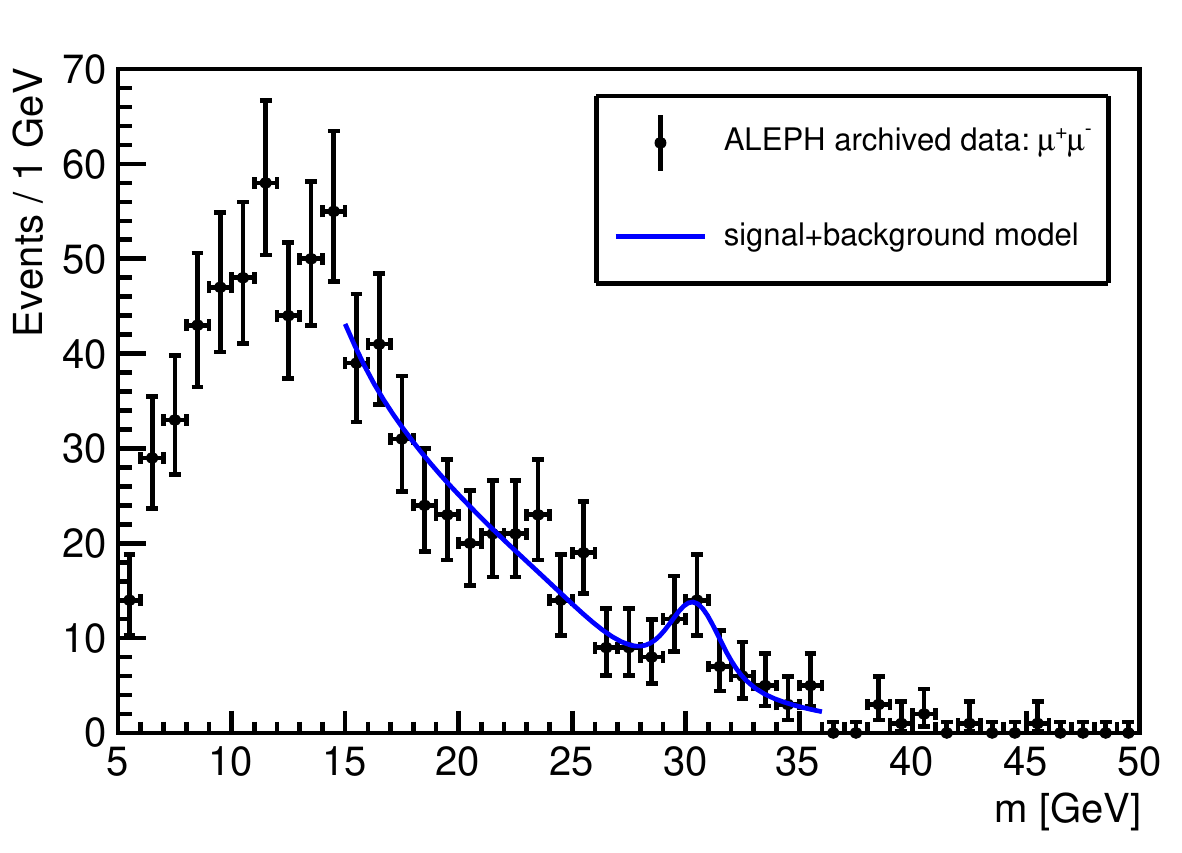}}
      \subcaptionbox{\label{fig:di-muon_signal_pull}}
        {\includegraphics[width=0.46\textwidth]{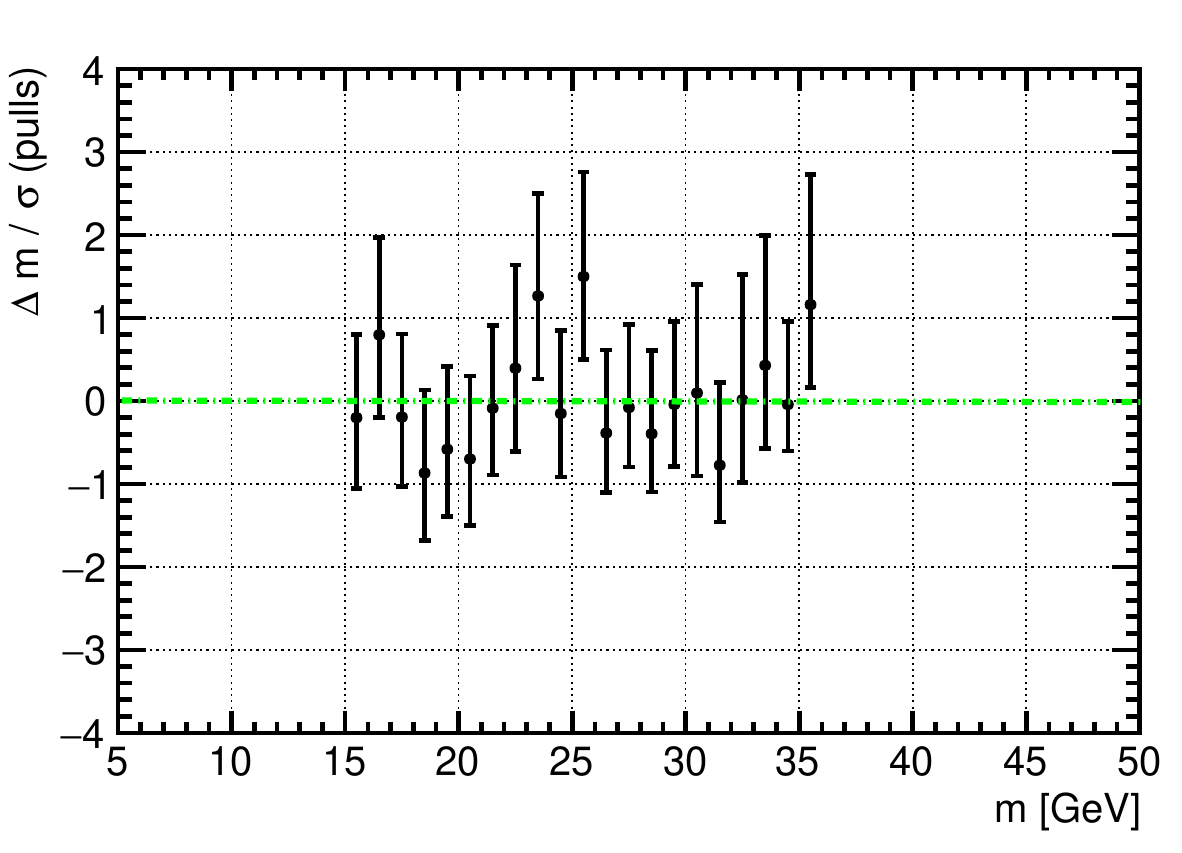}}
      \caption{The result of the $\chi^{2}$-fit of the signal + background model to the binned version of the opposite sign di-muon spectrum (left) and the corresponding distribution of pulls (right). $\chi^{2}/{\rm ndof} = 0.56$. To avoid bins with zero entries the fit range for this test is restricted to [15, 36]~GeV.}
 \begin{subtable}[c]{0.49\textwidth}
 \centering
 \vspace{1 mm}
   \scalebox{0.75}{ 
  \begin{tabular}{lrl}
        \hline\noalign{\smallskip}
        {\bf Parameter}& {\bf Value} & {\bf Error} \\
         \noalign{\smallskip}\hline\noalign{\smallskip}
        \# signal events & 32.31  & $\pm$ 10.87  \\ 
	\# background events (overall) & 1457.06 & $\pm$ 89.71 \\
	\noalign{\smallskip}\hline
	mass [GeV] & 30.40 & $\pm$ 0.46 \\
	\noalign{\smallskip}\hline
	width (Breit-Wigner) [GeV] & 1.78 & $\pm$ 1.14 \\
	\noalign{\smallskip}\hline
	width (Gaussian) [GeV] &	0.74 & $\pm$ 0.10 \\ 
        \noalign{\smallskip}\hline
  \end{tabular}}
  
  \subcaption{\label{table:dimuon_fit}}
\end{subtable}
   \begin{subtable}[h]{0.49\textwidth}
   \centering
 \centering
   \scalebox{0.75}{ 
  \begin{tabular}{ll}
        \hline\noalign{\smallskip}     
        {\bf Observable}& {\bf Value} \\
        \noalign{\smallskip}\hline\noalign{\smallskip}
 	$Z_{\rm Bi}$ & $2.63\,\sigma$\\
        \noalign{\smallskip}\hline
        $Z_{\rm asym}$ & $5.35\,\sigma$  \\
        p-value & $4.37725 \cdot 10^{-8}$  \\
        \noalign{\smallskip}\hline
  \end{tabular} 
  } 
   \subcaption{\label{table:dimuon_sig}}
\end{subtable}
 \caption{Parameter values of the extended maximum likelihood fit to the opposite sign di-muon mass spectrum obtained from ALEPH data (left) and significances of the excess (right).\label{table:dimuon}}
 \end{figure*}

Fig.~\ref{fig:di-muon_signal} shows the opposite sign di-muon mass spectrum together with the fitted signal + background model. The fit is performed in the range from 15 to 50~GeV. To compare the compatibility of the unbinned data with the fitted model the Kolmogorov-Smirnov (KS) test~\cite{A:1933uq,N:1948kx} is applied to mass values in the range 20 to 40~GeV around the visible excess. The obtained probability of equality for the KS test is 78\%, which is reasonably good.

To further scrutinize the applicability of the signal + background model to the opposite sign di-muon mass spectrum a $\chi^{2}$-fit to a binned version of the mass values is carried out, which provides a goodness-of-fit by means of the computed $\chi^{2}$ during the computation and the number of degrees of freedom (ndof) of the fit. Fig.~\ref{fig:di-muon_signal_binned} shows the result of the $\chi^{2}$-fit and Fig.~\ref{fig:di-muon_signal_pull} shows the corresponding pull distribution. To avoid bins with zero entries the fit range for this test is restricted to [15, 36]~GeV. Both distributions look reasonable. The $\chi^{2}/{\rm ndof} = 0.56$, which is also reasonably good.

The resulting parameters of the unbinned MLE fit are summarized in Table~\ref{table:dimuon_fit}. The excess at 30.40~GeV has a natural width of 1.78~GeV and its resolution of 0.74~GeV is compatible with the expected ALEPH detector performance in this mass range (see also Sec.~\ref{sec:alephdetector}).

\begin{figure*}[b]
      \centering
      \subcaptionbox{\label{fig:p0signal}}
        {\includegraphics[width=0.49\textwidth]{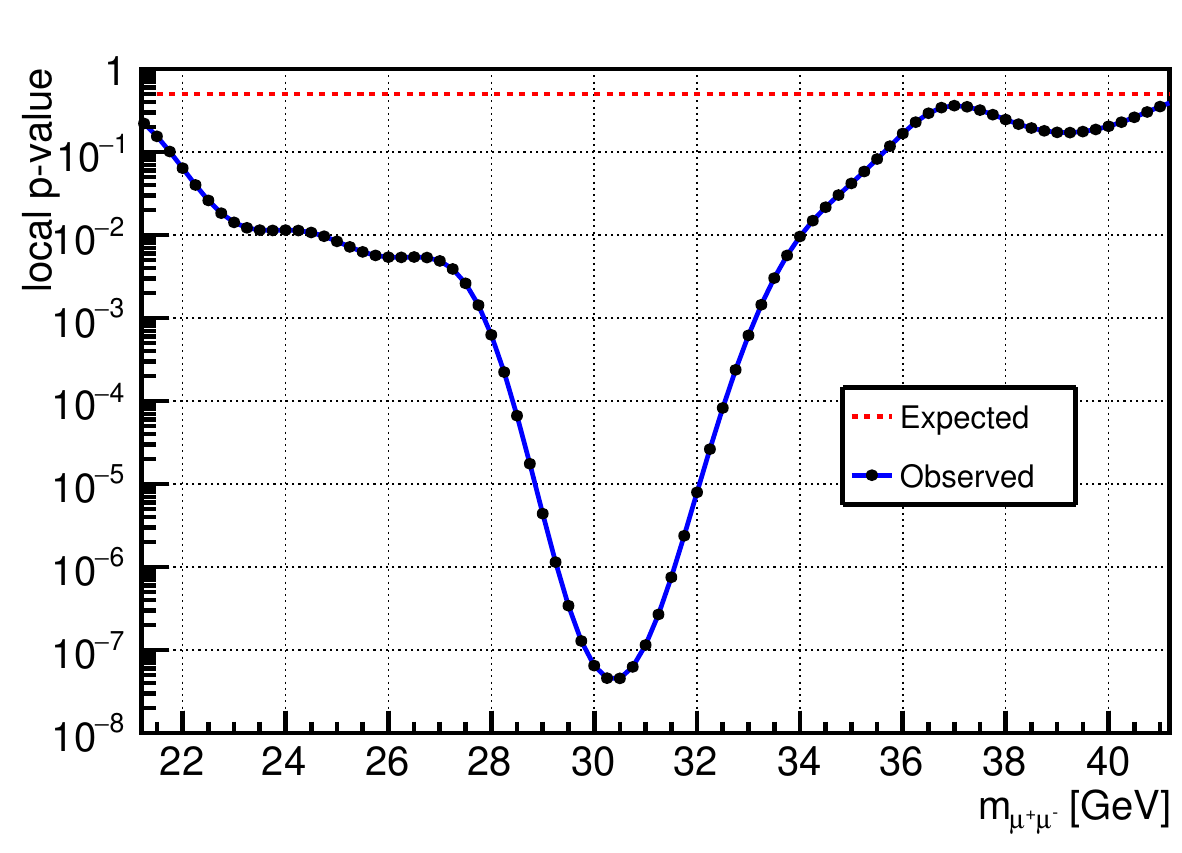}}
      \subcaptionbox{\label{fig:statsignal}}
        {\includegraphics[width=0.49\textwidth]{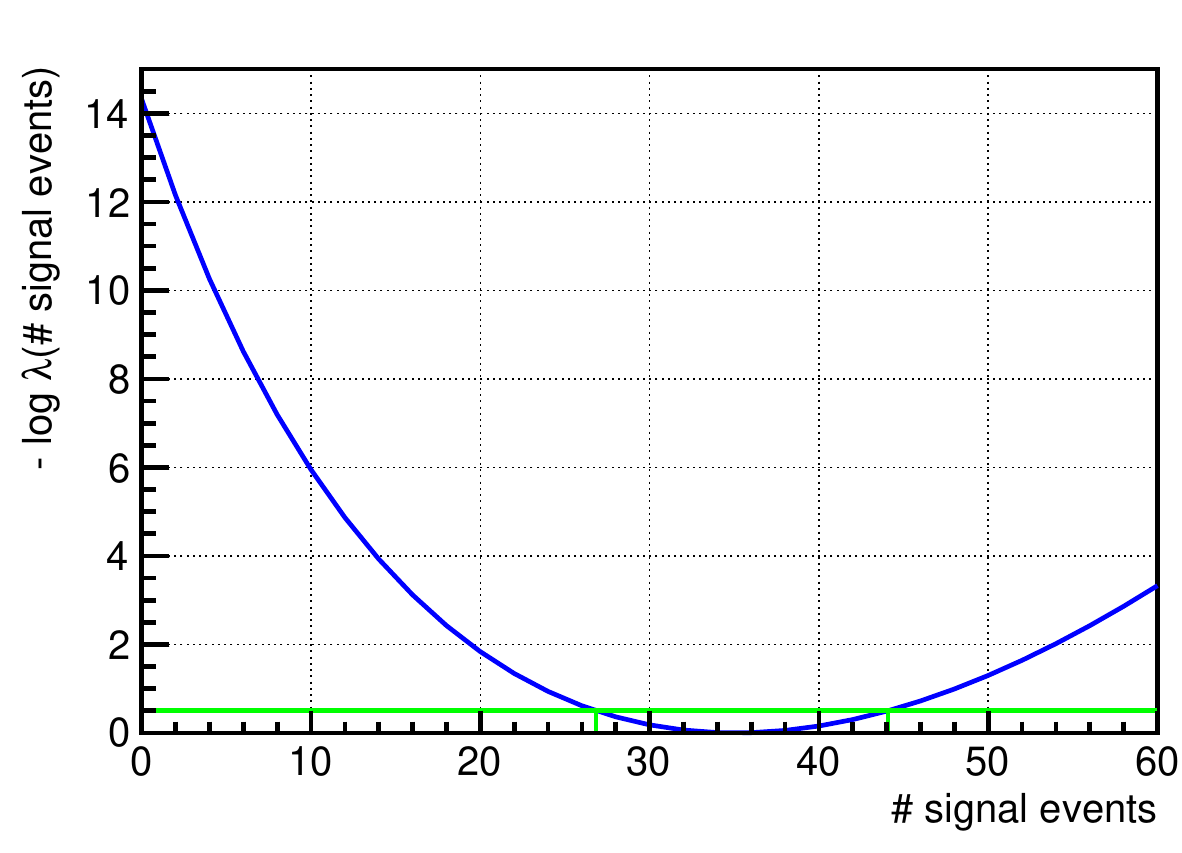}}
      \caption{The expected and observed local p-values as a function of the opposite sign di-muon mass ${\rm m}_{\mu^{+}\mu^{-}}$ (left) and the log profile likelihood curve as function of  the number of signal events (right). The $1\,\sigma$ interval (68\% Confidence Level) is obtained from the intersect of the $-\log \lambda$ curve with the horizontal dashed line $-\log \lambda = 0.5$. }
\end{figure*}

To rate the significance of the excess visible in Fig.~\ref{fig:di-muon_signal} two benchmarks are presented. $Z_{\rm Bi}$~\cite{Cousins:2008zz} is a robust observable in case no theoretical signal model can be used and hence the mean and width cannot be set constant in the fit. The numbers $n_{\rm on}$ and $n_{\rm off}$ to compute $Z_{\rm Bi}$ are obtained around the fitted mass using a $2\,\sigma$ wide area ($\tau = 1$). For details see~\cite{Cousins:2008zz}. A likelihood-based one-sided significance $Z_{\rm asym}$ using an asymptotic formula is computed utilizing the RooStats package~\cite{Moneta:2010pm,Cowan:2010js}. The corresponding local p-value is given as well. For the second benchmark a single parameter of interest, i.e. the number of signal events, is assumed to be free in the fit. The number of background events is treated as a nuisance parameter. The calculation is based on a result from Wilks~\cite{Wilks:1938dza} and Wald~\cite{10.2307/1990256}. Also $Z_{\rm asym}$ does not include any look elsewhere effect~\cite{Lyons:2008kq,Gross:2010qma,Cousins:2013kq}. Table~\ref{table:dimuon_sig} summarizes the values discussed.

The expected and observed local p-values as a function of the opposite sign di-muon mass ${\rm m}_{\mu^{+}\mu^{-}}$ utilizing the AsymptoticCalculator class of RooStats~\cite{Moneta:2010pm} are shown in Fig.~\ref{fig:p0signal}. Fig.~\ref{fig:statsignal} shows a plot of the log profile likelihood curve $-{\rm log} \lambda$ as function of the parameter of interest, i.e. the number of signal events. The $1\,\sigma$ interval (68\% Confidence Level) is obtained from the intersect of the $-\log \lambda$ curve with the horizontal dashed line $-\log \lambda = 0.5$.

To estimate the influence of systematic effects due to the shape of the used background model,  a second option of a kernel density estimation class in ROOT is used, namely the TKDE class~\cite{Jann:2008yu,Cranmer:db,kde1,kde3}. This class is used to construct an additional background model. The significance of the excess in the opposite sign di-muon mass turns out to be lower for this model, i.e. $Z_{\rm asym, TKDE} = 4.86$. The KS-test probability is 76\%. The goodness-of-fit of the binned $\chi^{2}$-fit is $\chi^{2}/{\rm ndof} = 0.61$.

\begin{figure*}[t]
\begin{subfigure}[b]{.5\linewidth}
     \centering
     \includegraphics[width=1\textwidth]{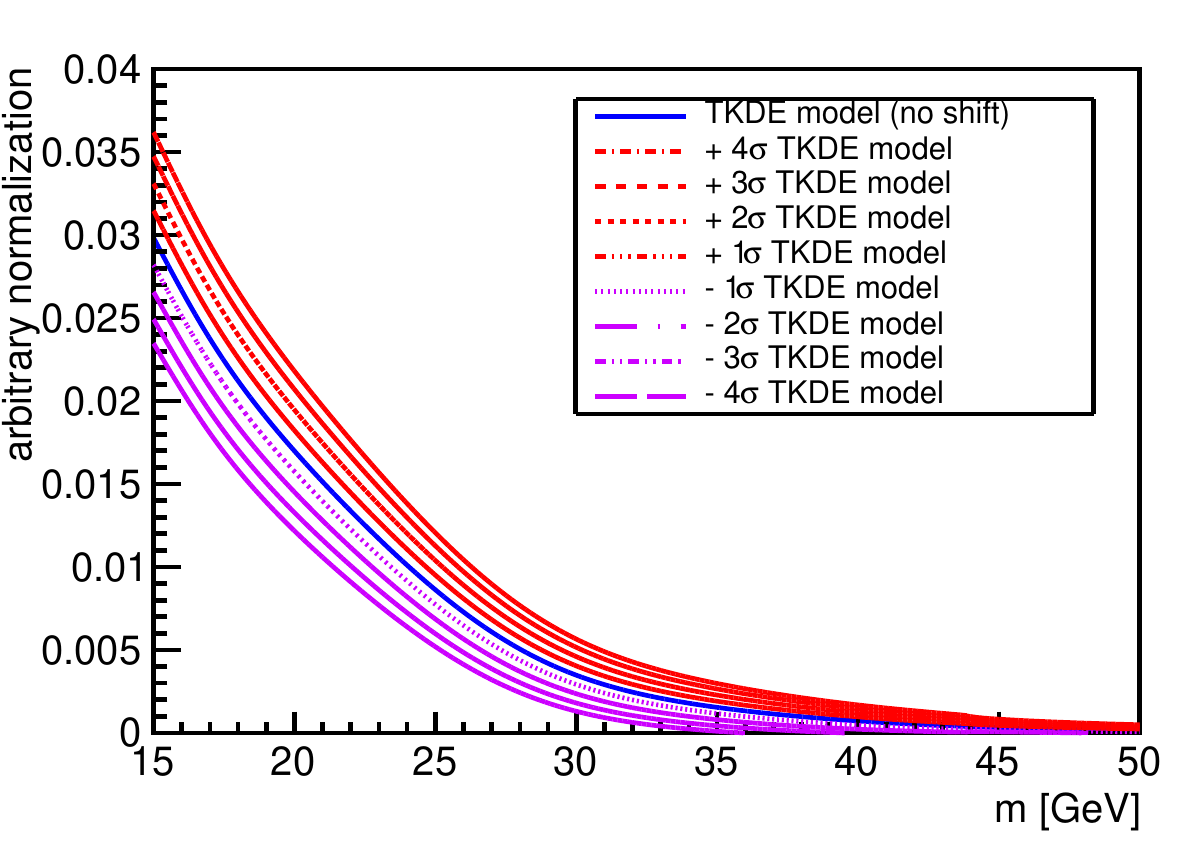}
     \caption{\label{fig:di-muon_TKDE}}
\end{subfigure}
\begin{subtable}[b]{.5\linewidth}
   \centering
   \scalebox{0.65}{ 
   \begin{tabular}{ccc}    
        \hline\noalign{\smallskip}     
        {\bf TKDE model shift} & {\boldmath \bf $Z_{\rm Bi, TKDE}$  $[\sigma]$} & {\boldmath \bf $Z_{\rm asym, TKDE}$  $[\sigma]$} \\
         \noalign{\smallskip}\hline\noalign{\smallskip}
 	$+4\,\sigma$ & 1.82 & 3.06  \\
 	$+3\,\sigma$ & 1.98 & 3.34  \\
 	$+2\,\sigma$ & 2.19 & 3.72  \\
 	$+1\,\sigma$ & 2.47 & 4.22  \\
 	no shift & 2.80 & 4.86 \\
        $-1\,\sigma$ & 3.27 & 5.79  \\
        $-2\,\sigma$ & 3.94 & 7.20  \\
        $-3\,\sigma$ & 4.95 & --  \\
        $-4\,\sigma$ & 5.71 & --  \\
        \noalign{\smallskip}\hline
    \end{tabular}}
    \captionsetup{skip=38pt}
    \caption{\label{table:di-muon_TKDE}}
\end{subtable}
\caption{The result of the extended maximum likelihood fit of the signal + background model to the unbinned opposite sign di-muon mass spectrum for an alternative KDE approach~\cite{Jann:2008yu,Cranmer:db,kde1,kde3}. The obtained significance for the cases, where this background model is artificially shifted in steps up to $\pm$ $4\,\sigma$, are given as well.}
\end{figure*}

Fig.~\ref{fig:di-muon_TKDE} shows the used TKDE background model plus the cases where this model was shifted by several $\sigma$'s. Tab. \ref{table:di-muon_TKDE} summarizes the obtained significance for the various background shifts, e.g.~an artificial shift upwards of the background model by $+1 \sigma$ gives $Z_{\rm asym, TKDE, +1 \sigma} = 4.22$. In the other direction $-1 \sigma$, we obtain a significance of $Z_{\rm asym, TKDE, -1 \sigma} = 5.79$. It follows that systematic effects are indeed present. Nevertheless, the significance of the observed excess remains always quite high.

\begin{figure*}[t]
\begin{subfigure}[b]{.5\linewidth}
     \centering
     \includegraphics[width=1\textwidth]{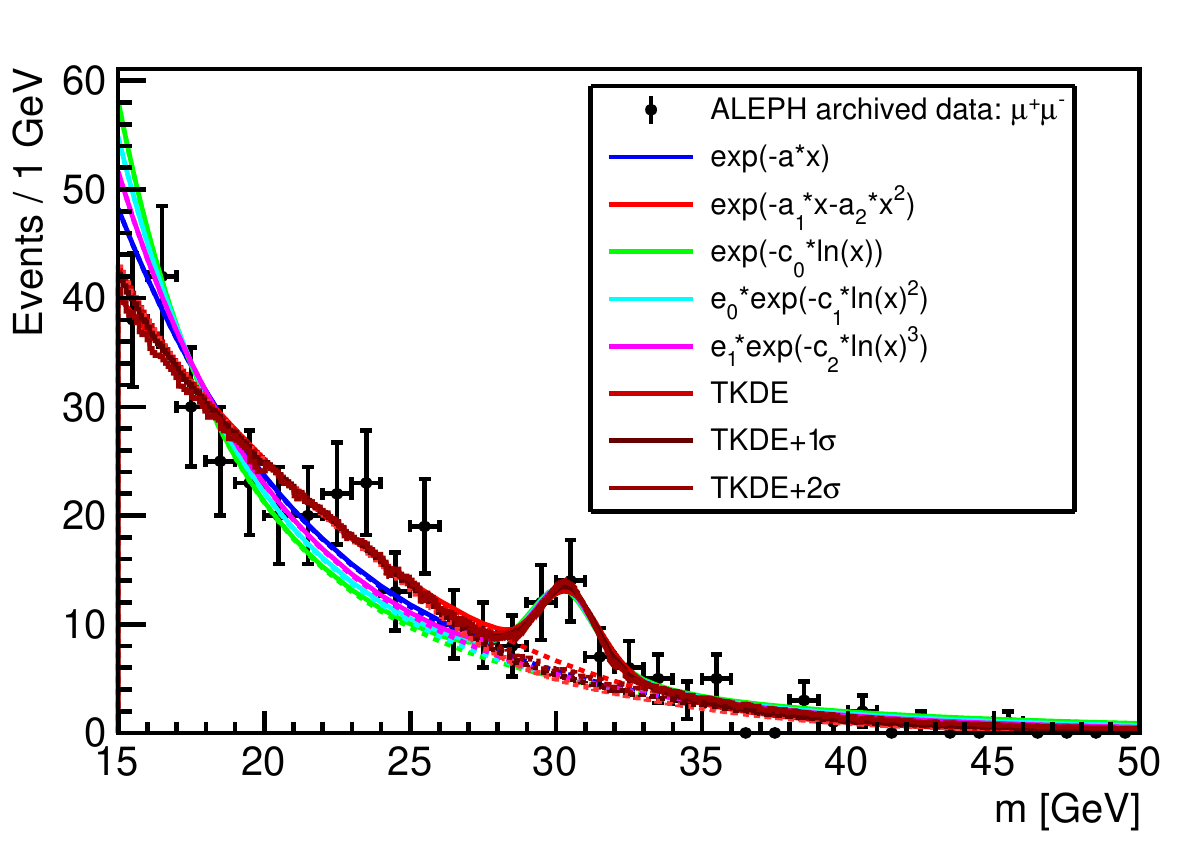}
     \caption{\label{fig:di-muon_extra_poly}}
\end{subfigure}
\begin{subtable}[b]{.5\linewidth}
    \centering
    \scalebox{0.65}{
    \begin{tabular}{lcc}    
        \hline\noalign{\smallskip}     
        {\bf background model} & {\boldmath \bf $Z_{\rm asym}$ $[\sigma]$} & {\boldmath \bf $\chi^{2}/{\rm ndof}$} \\ 
         \noalign{\smallskip}\hline\noalign{\smallskip}
 	$exp(-a \cdot x)$ & 3.5 & 0.91  \\
 	$exp(-a_{1} \cdot x-a_{2} \cdot x^{2})$ & 2.9 & 0.79  \\
 	$exp(-c_{0} \cdot ln(x))$ & 3.9 & 1.54  \\
 	$e_{0} \cdot exp(-c_{1} \cdot ln(x)^{2})$ & 3.8 & 1.38  \\
 	$e_{1} \cdot exp(-c_{2} \cdot ln(x)^{3})$ & 3.7 & 1.17 \\
        TKDE & 4.9 & 0.76  \\
        TKDE $+\,1\,\sigma$ & 4.2 & 0.67  \\
        TKDE $+\,2\,\sigma$ & 3.7 & 0.58  \\
        \noalign{\smallskip}\hline
    \end{tabular}}
    \captionsetup{skip=45pt}
    \caption{\label{table:di-muon_extra_poly}}
\end{subtable}
\caption{Significance $Z_{\rm asym}$ and goodness-of-fit evaluations $\chi^{2}/{\rm ndof}$ using different polynomial parametrizations and the TKDE method to model the background shape. The signal is modeled with a Breit-Wigner distribution convoluted with a Gaussian.\label{figmain:di-muon_extra_poly}
}
\end{figure*}

To further study the contribution to the systematic error of our result introduced by the choice of the background model, we evaluated and benchmarked the excess using additional polynomial parametrizations to model the background shape. The result, using the data of the opposite sign di-muon mass spectrum only, is given in Fig.~\ref{fig:di-muon_extra_poly} and Tab.~\ref{table:di-muon_extra_poly}. As usual the signal is modeled with a Breit-Wigner distribution convoluted with a Gaussian. The goodness-of-fit evaluation (Tab.~\ref{table:di-muon_extra_poly}) shows that the signal + background models are able to describe the mass spectrum more or less reasonably good. Except for one model the significance of the excess remains above $3\,\sigma$.

\begin{figure*}[b]
      \centering
      \subcaptionbox{\label{fig:simfitdimuon}}
        {\includegraphics[width=0.49\textwidth]{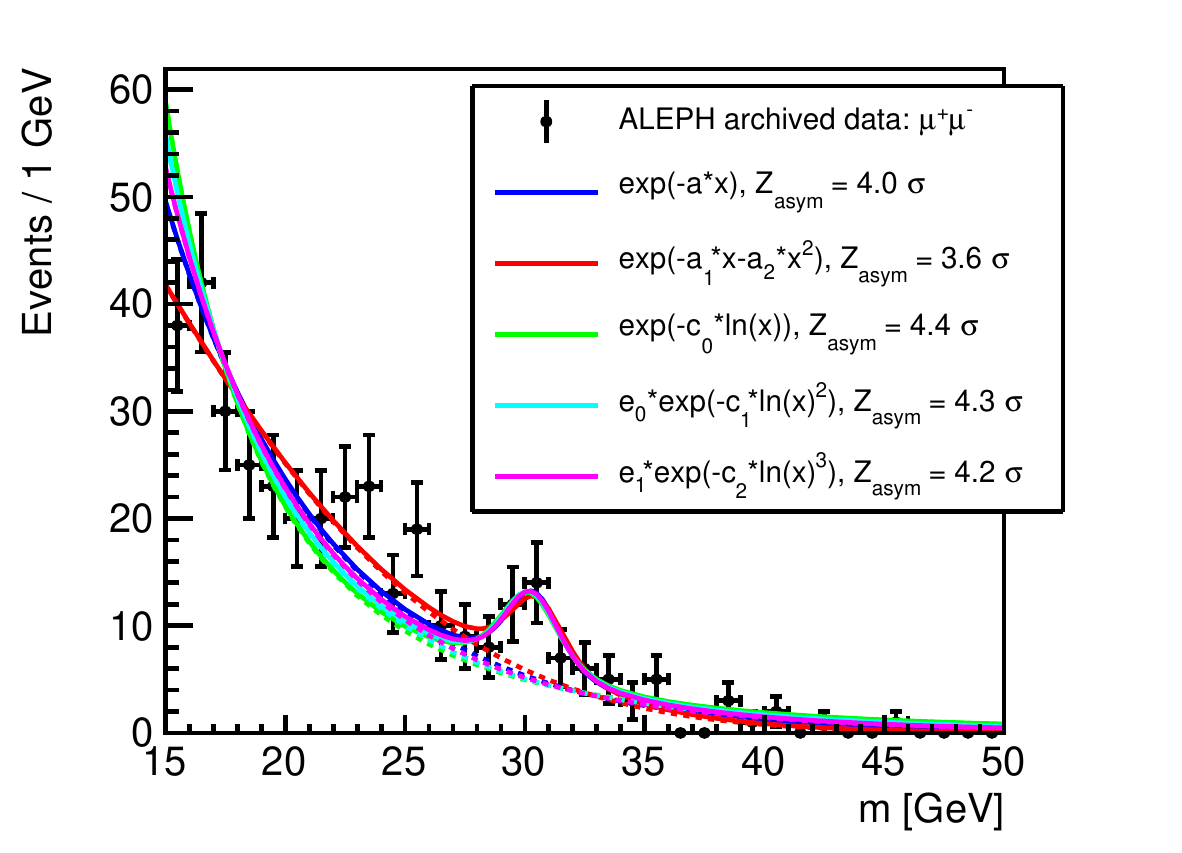}}
      \subcaptionbox{\label{fig:simfitbackground}}
        {\includegraphics[width=0.49\textwidth]{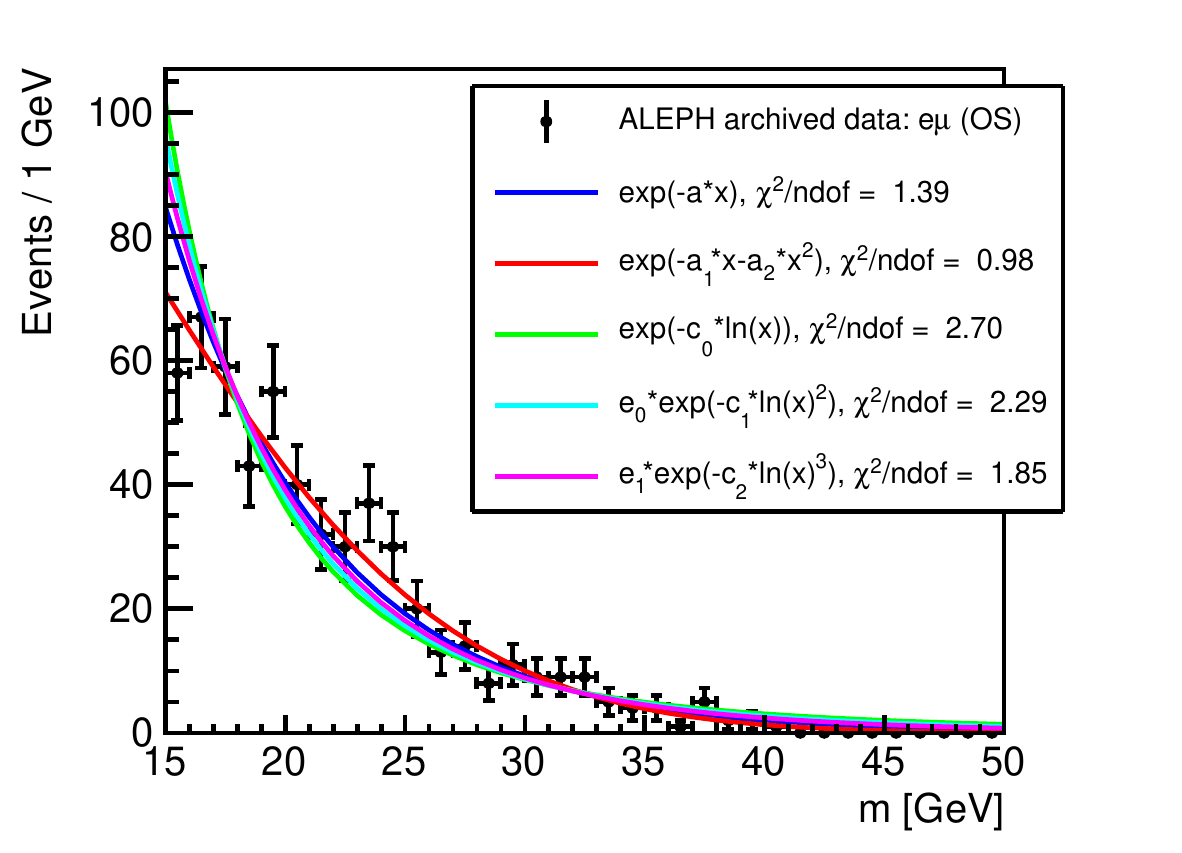}}
      \caption{The result using a simultaneous fit to the opposite sign di-muon (left) and opposite sign electron-muon (right) mass spectrum for different polynomial parametrizations of the background shape. The signal is modeled with a Breit-Wigner distribution convoluted with a Gaussian. The significances $Z_{\rm asym}$ and goodness-of-fit evaluations $\chi^{2}/{\rm ndof}$ are given in the figures.\label{figmain:di-muon_extra_poly_sig_back}}
\end{figure*}

Due to the less good description of the background shape for some polynomial pa\-ra\-me\-trizations and also because of the available statistics, the signal + background fits of the opposite sign di-muon mass spectrum result in less significant results when compared to background models using a one-dimensional kernel estimation probability density function. By means of a simultaneous fit to the opposite sign di-muon mass spectrum, i.e.~the signal distribution, and the opposite sign electron-muon mass spectrum, i.e.~the background, we can recover and improve the situation. The improved result using the combined information of the signal and background mass spectra, is shown in Fig.~\ref{figmain:di-muon_extra_poly_sig_back}. At least for the second parametrization, i.e.~the double exponential one, the goodness-of-fit evaluation gives a reasonably good result. The other models now show a larger discrepancy from the data, because of the improvement in the available statistics (di-muon plus electron-muon spectra).

\begin{figure}[t]
\centering
       \includegraphics[width=0.53\textwidth]{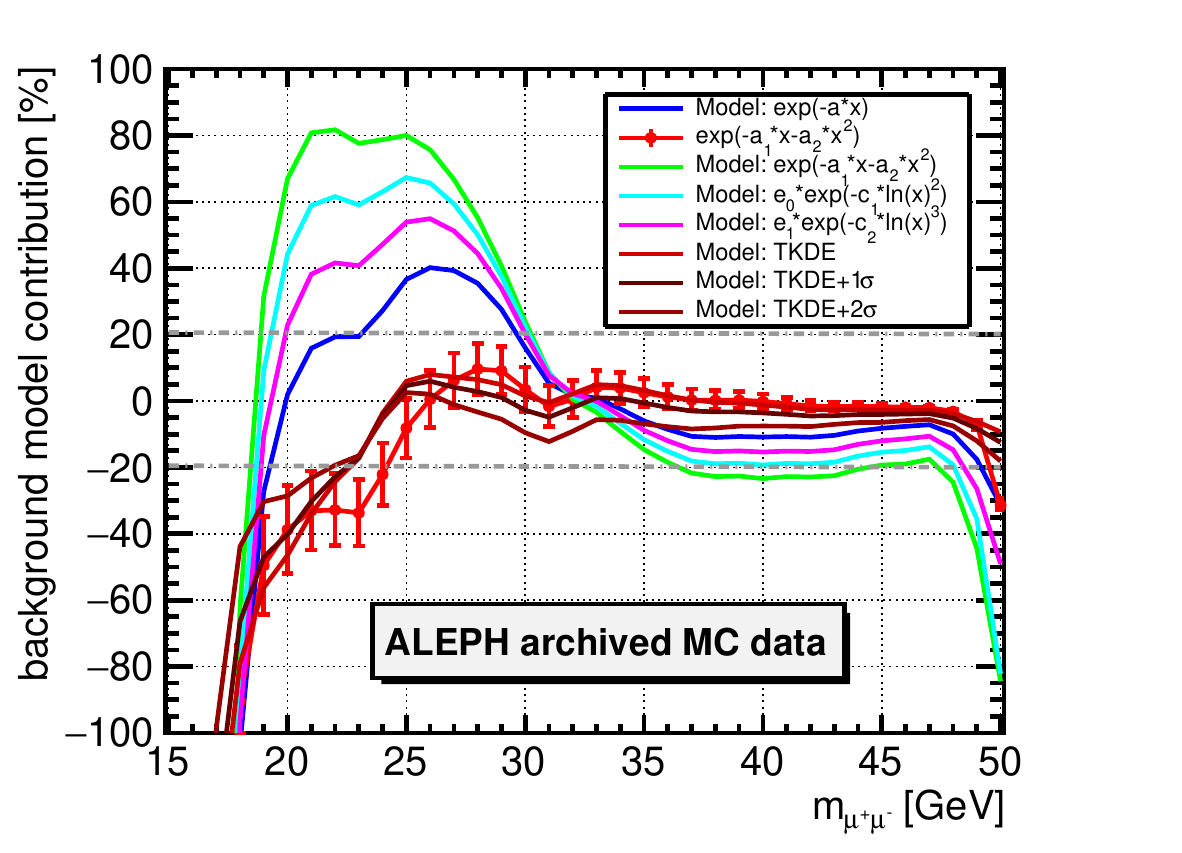}
      \caption{\label{fig:di-muon_spurious_signal} Evaluation of the contribution to the amount of measured signal events due to a specific choice of a background model using the ATLAS "spurious signal" method~\cite{ATLAS:2012ys,Aad:2015owa,ATLAS-CONF-2015-081}. For one graph the errors obtained by a MINOS error analysis are given\cite{1971smep.book.....E}. The other models have similar errors.}
\end{figure}

The ATLAS collaboration used a novel method to scrutinize choices of specific background models (e.g.~for their Higgs discovery in the di-photon final state) based on simulated data only~\cite{ATLAS:2012ys,Aad:2015owa,ATLAS-CONF-2015-081}. The evaluation using their method of the bias on the fitted signal yield introduced by a given background functional form ("spurious signal") is shown in Fig.~\ref{fig:di-muon_spurious_signal}. The signal + background models described earlier are used to scan the simulated ALEPH archived data for an excess in the mass region from 15 to 50~GeV in 1~GeV steps. To improve the available statistics for this test, the opposite sign di-muon, di-electron and electron-muon mass spectra were combined, since their shape is very similar as described in section~\ref{sec:background model}. The contributions to the amount of measured signal events for a specific choice of a background model are normalized to the signal event yield obtained in ALEPH archived data. For the double exponential parametrization of the background the errors obtained by a MINOS error analysis of the fitted signal yield are given\cite{1971smep.book.....E}. The other models have similar errors. In a wider mass region around 30~GeV the double exponential parametrization of the background shows a contribution to the excess event yield of less than 10\%. The TKDE models also have a quite low contribution in a wider range around the excess in ALEPH data.

\subsection{Evaluation of the look elsewhere effect}
It is mandatory to give an evaluation of a look elsewhere effect for every observed excess~\cite{Lyons:2008kq,Gross:2010qma,Cousins:2013kq}. Since there is no signal model available for the presented excess, one can at least obtain a look elsewhere effect by floating the fit parameters, namely the mean value of the excess and its width. We float the mean value in the range from 15 to 50~GeV.

\begin{figure*}[b]
      \centering
      \subcaptionbox{\label{fig:jspimumu}}
        {\includegraphics[width=0.49\textwidth]{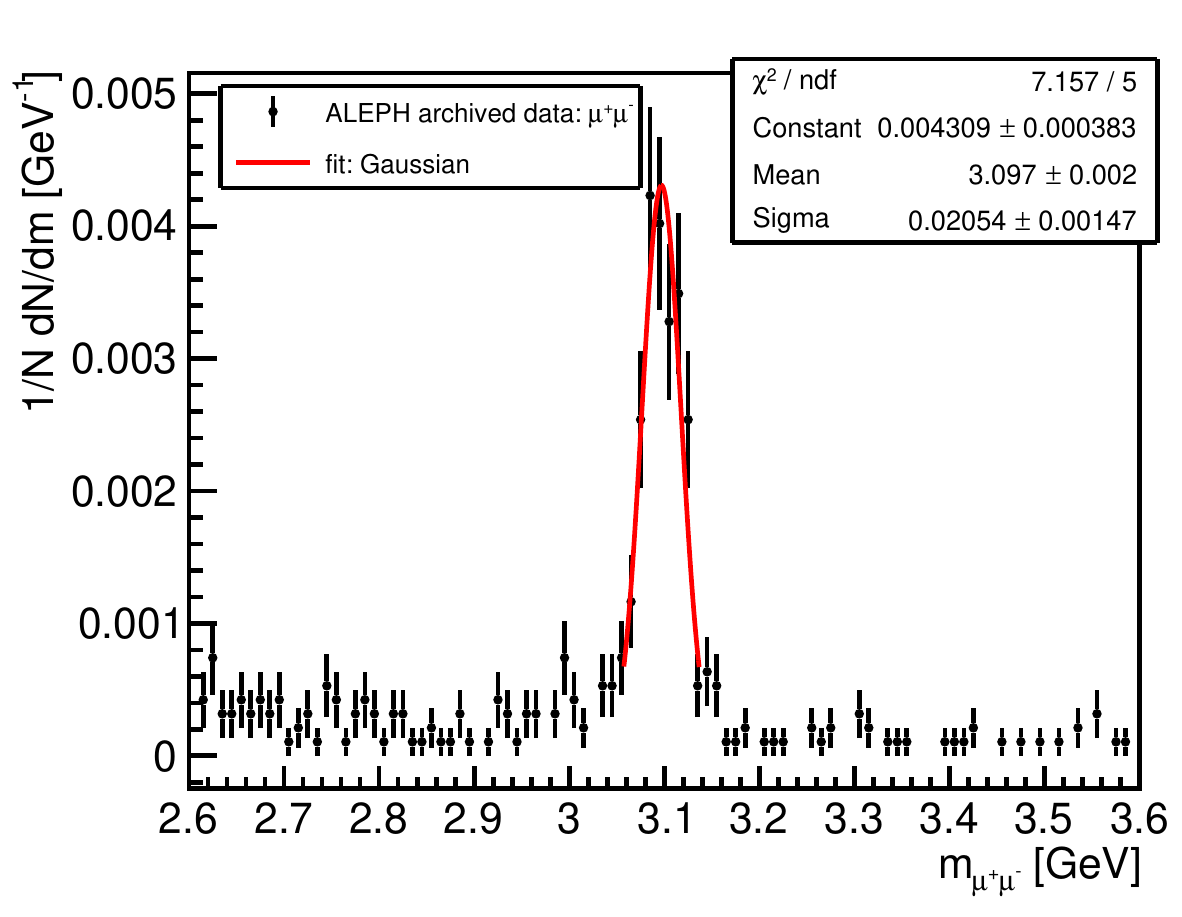}}
      \subcaptionbox{\label{fig:jpsiee}}
        {\includegraphics[width=0.49\textwidth]{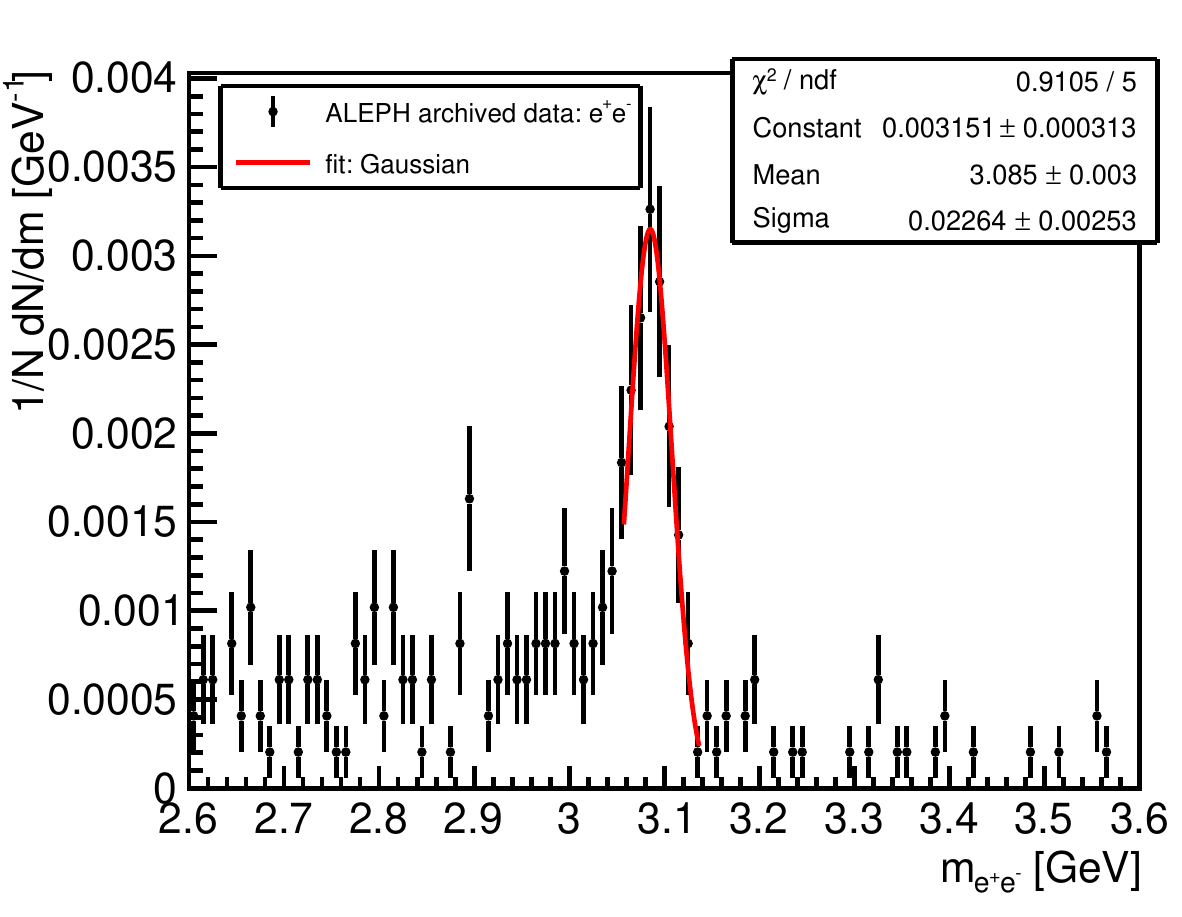}}
      \caption{Reconstructed J/$\psi$ decays from di-muons (left) and di-electrons (right).}
\end{figure*}

For all models adding the look elsewhere effect in the way described above reduces the obtained significance by 1.4 to $1.6\,\sigma$. As an example adding the look elsewhere effect to the significance obtained using as background model the single exponential parametrization shown in blue in Fig.~\ref{figmain:di-muon_extra_poly_sig_back} reduces the calculated significance from $Z_{\rm asym} = 4.0\,\sigma$ to $Z_{{\rm freq,\,lee}} = 2.6\,\sigma$. The significance $Z_{\rm freq,\,lee}$ is obtained using a frequentist-based calculation based on Toy Monte Carlo simulation, namely the FrequentistCalculator class of the RooStats package~\cite{ROOT,Antcheva:2009zz,Moneta:2010pm}. The double exponential pa\-ra\-me\-tri\-za\-tion shown in red in Fig.~\ref{figmain:di-muon_extra_poly_sig_back}, suffers slightly less by adding the look elsewhere effect. The significance drops from $Z_{\rm asym} = 3.6\,\sigma$ to $Z_{{\rm freq,\,lee}} = 2.4\,\sigma$. Models, which by construction represent the background much more accurate in a large mass range around the excess by means of a kernel density approximation, e.g.~the TKDE parametrization used in Fig.~\ref{figmain:di-muon_extra_poly}, have a change in the obtained significance when including the look elsewhere effect from  $Z_{\rm asym} = 4.9\,\sigma$ to $Z_{{\rm freq,\,lee}} = 2.9\,\sigma$.

\section{Crosschecks and further studies of the excess}
\label{sec:crosschecks}
We have checked that events in the signal region are spread over the whole data taking period from 1992 to 1995. The events are not from specific time slots in a run nor are they connected to a special detector region. We have looked at many displays of events from the signal region by means of the ALEPH offline event display DALI~\cite{ALEPH-DALI}. We saw reasonably good reconstructed jets from hadronic ${\rm Z}^{0}$ decays together with identified muons. Example event displays can be found in Appendix~\ref{App:event_displays}.

\begin{figure*}[b]
      \centering
      \subcaptionbox{\label{fig:minangle}}
        {\includegraphics[width=0.49\textwidth]{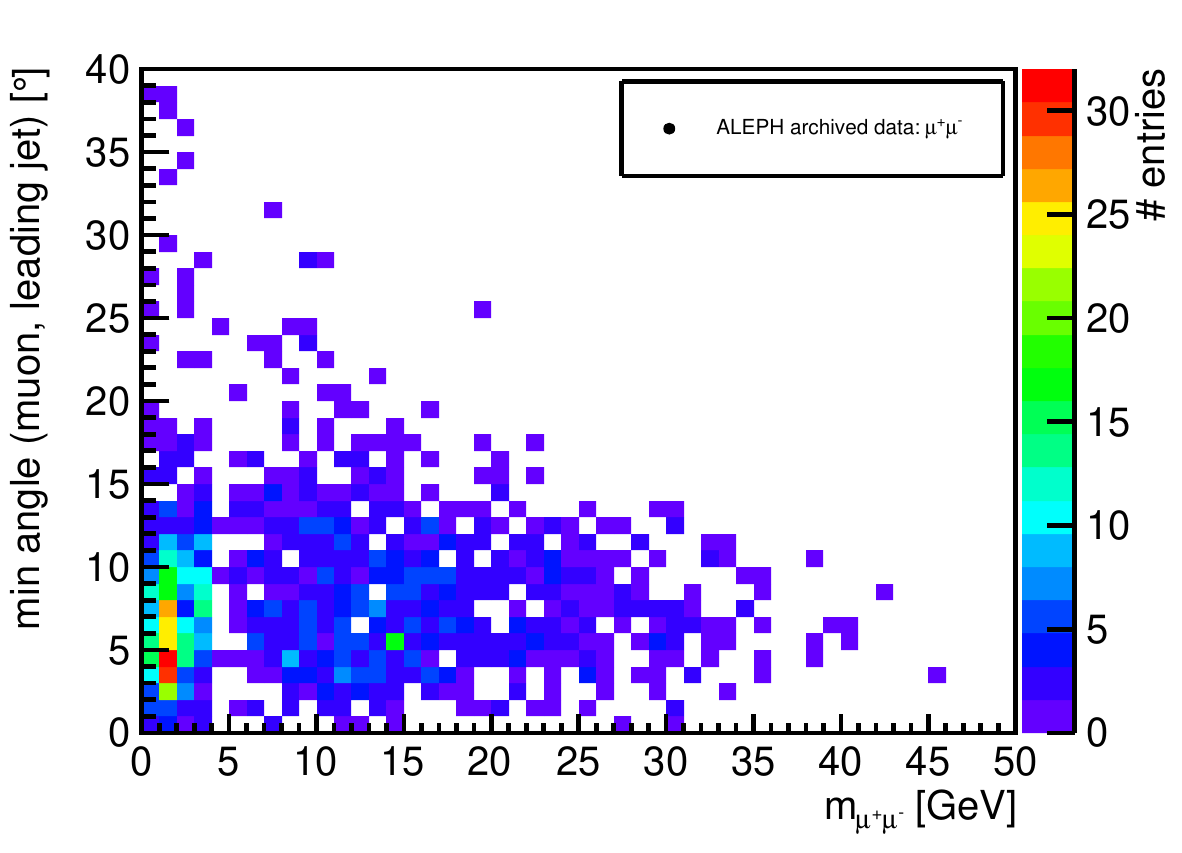}}
      \subcaptionbox{\label{fig:otherangle}}
        {\includegraphics[width=0.49\textwidth]{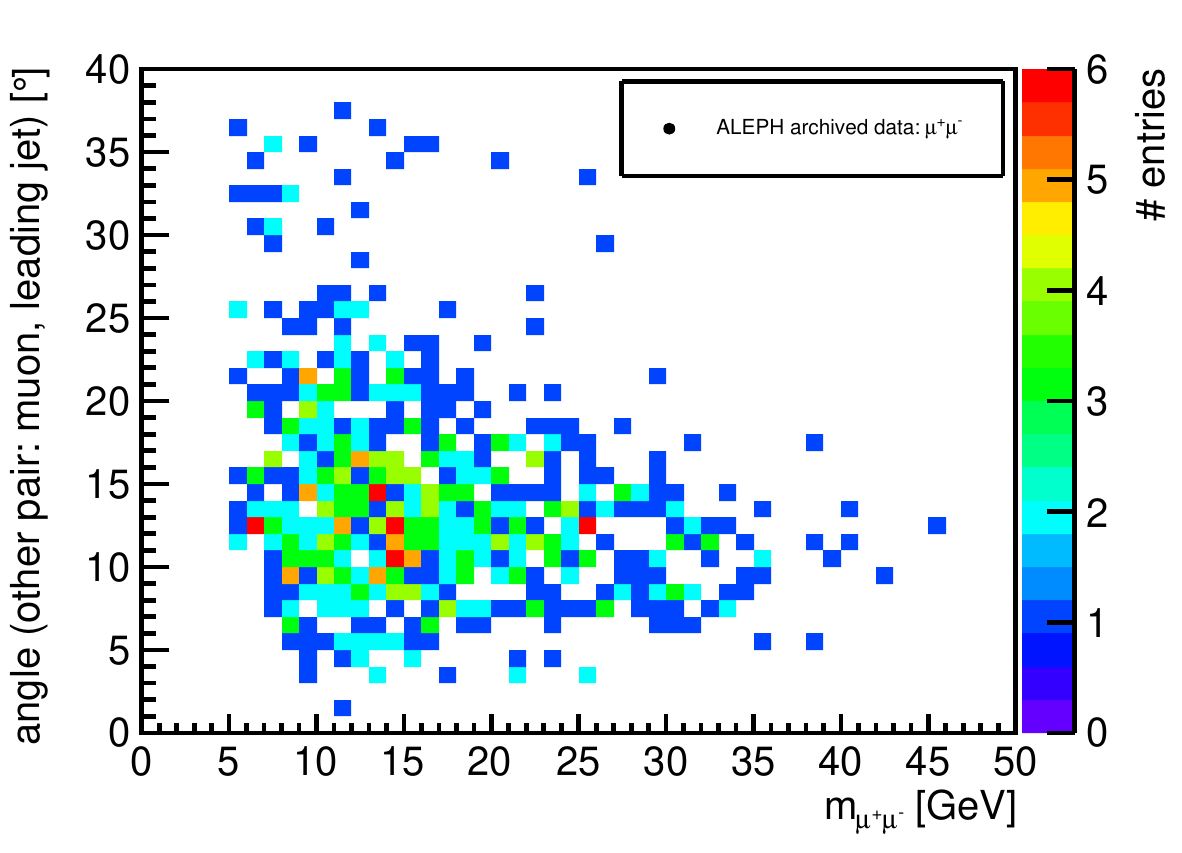}}
      \caption{The minimum angle between a muon and the leading jet versus opposite sign di-muon mass (left) and the angle of the other muon-jet combination versus opposite sign di-muon mass (right) obtained from ALEPH data.}
\end{figure*}

\subsection[J/$\psi$ reconstruction]{\boldmath J/$\psi$ reconstruction}
\label{sec:jpsi}
The opposite di-muon and di-electron mass spectra show the typical signatures of J/$\psi$ decays to these final states (Fig.~\ref{fig:jspimumu} and \ref{fig:jpsiee}). Both mass peaks are at the expected mean values of about 3.097~GeV~\cite{Agashe:2014kda} and the width of the fitted Gaussian shows the expected width of about 20 MeV compatible with the ALEPH mass resolution in this region (see Sec.~\ref{sec:alephdetector}). The reconstructed J/$\psi$ mass from di-electron tracks (Fig.~\ref{fig:jpsiee}) shows the typical tail on the left side of the peak due to Bremsstrahlung, so that the reconstructed mean is slightly shifted to a lower value.

\subsection{Lepton isolation}
\label{sec:isolation}
We used modern jet-algorithms, notably the generalized anti-$k_{t}$ algorithm for ${\rm e}^{+}{\rm e}^{-}$ collisions~\cite{Cacciari:2011ma} to reconstruct jets from ALEPH energy flow objects~\cite{ALEPH-alpha} excluding all leptons from the input objects. We found that in all cases one of the two muons from di-muons  is closer than $15^{\circ}$ to one of the leading jets in the appropriate event hemisphere (Fig.~\ref{fig:minangle}).

Generally, the higher the reconstructed opposite sign di-muon mass is, the more difficult it is to reconstruct and define the leading jets for both hemispheres. It turned out that, in case of a reconstructed high di-muon mass, at least one of the leading jets in one hemisphere tends to be broadened. This can also be seen in the example event displays shown in Appendix~\ref{App:event_displays}. For events where a leading jet could be defined in both hemispheres Fig.~\ref{fig:minangle} shows the minimum angle for a muon-leading jet combination as a function of the opposite sign di-muon invariant mass. In Fig.~\ref{fig:otherangle} the angle of the other muon-leading jet combination is plotted. It is seen that this angle is typically in the range of 5 to $20^{\circ}$.

In Fig.~\ref{fig:otherangle} it is also visible that for masses smaller than 30~GeV there is a marginal tendency that the muon-jet pairs are closer. Although the available event statistics is low, the event displays of di-muon events from the signal mass region around 30~GeV show the closeness of both muons to hadronic jet structures (see Appendix~\ref{App:event_displays}).

\begin{figure*}[!ht]
      \centering
      \subcaptionbox{\label{fig:ptrelbest}}
        {\includegraphics[width=0.48\textwidth]{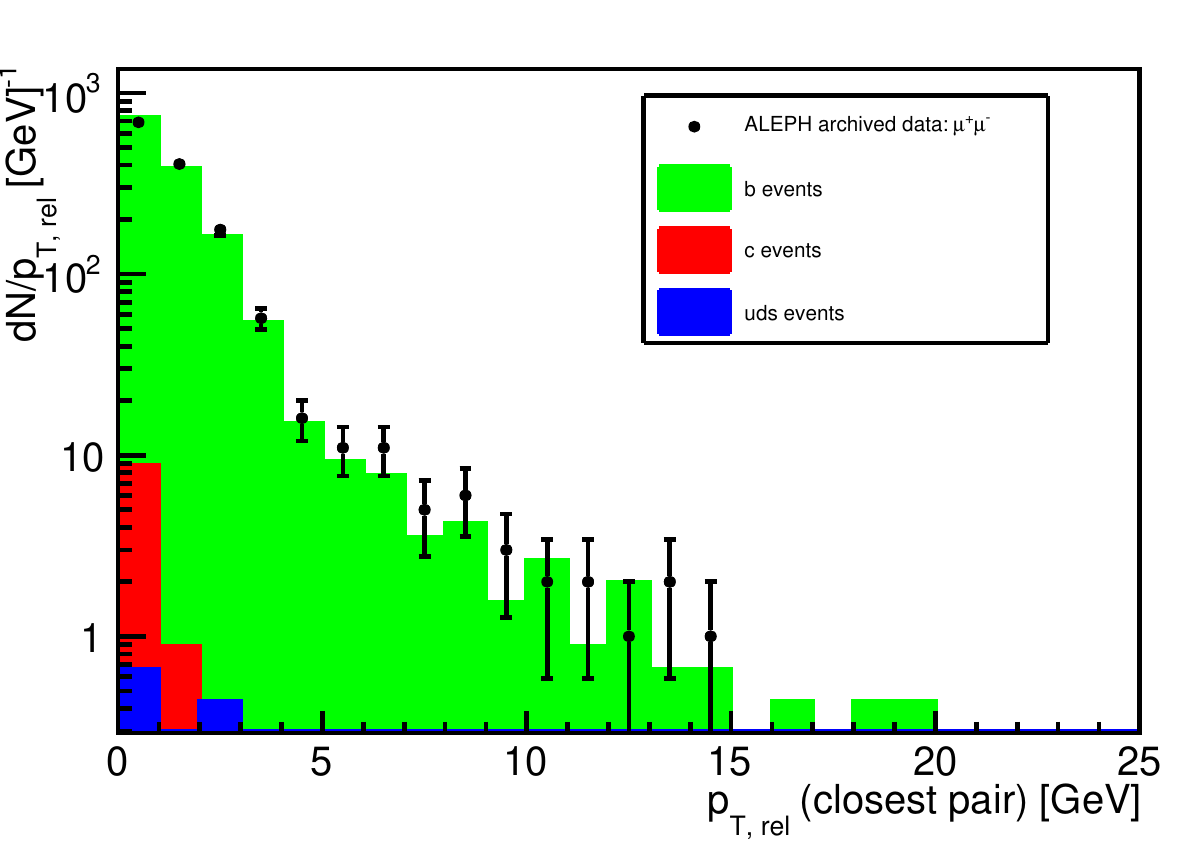}}
      \subcaptionbox{\label{fig:ptrelother}}
        {\includegraphics[width=0.48\textwidth]{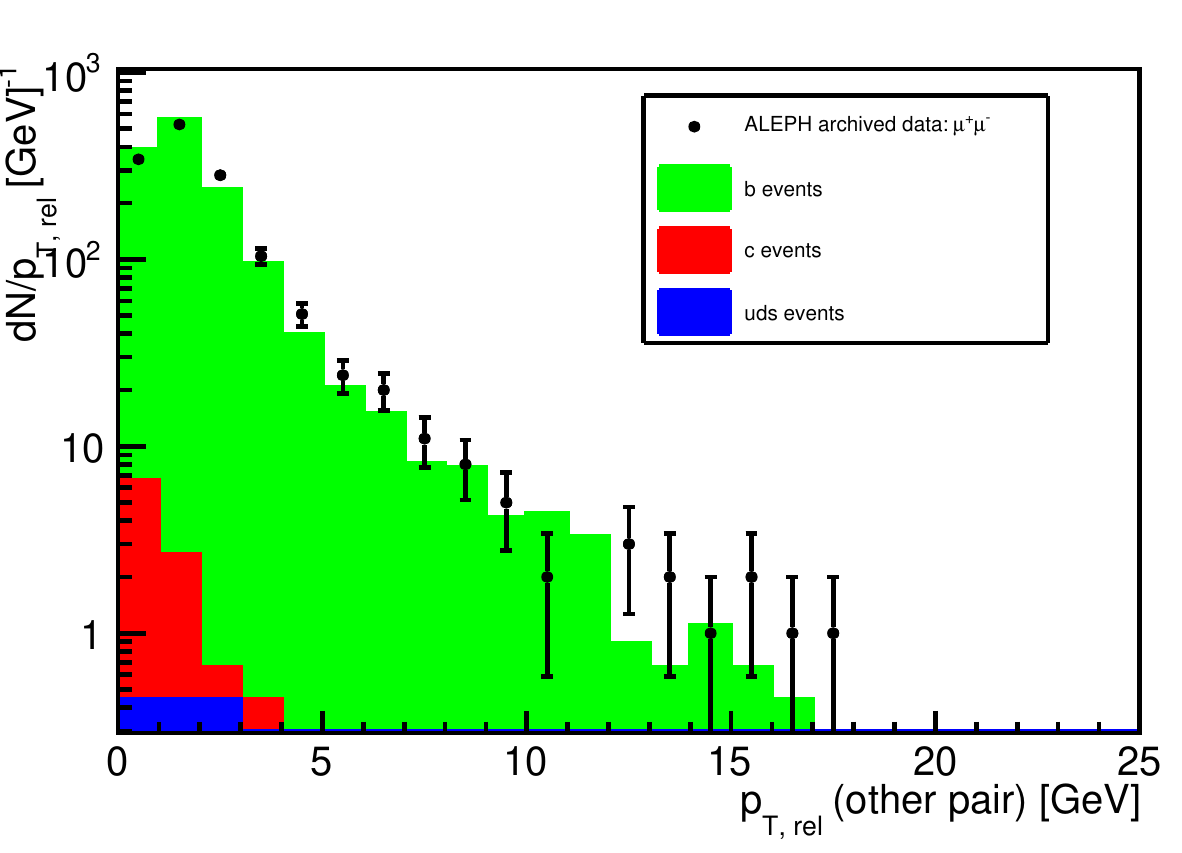}}

      \subcaptionbox{\label{fig:ptrelbestdiv}}
        {\includegraphics[width=0.48\textwidth]{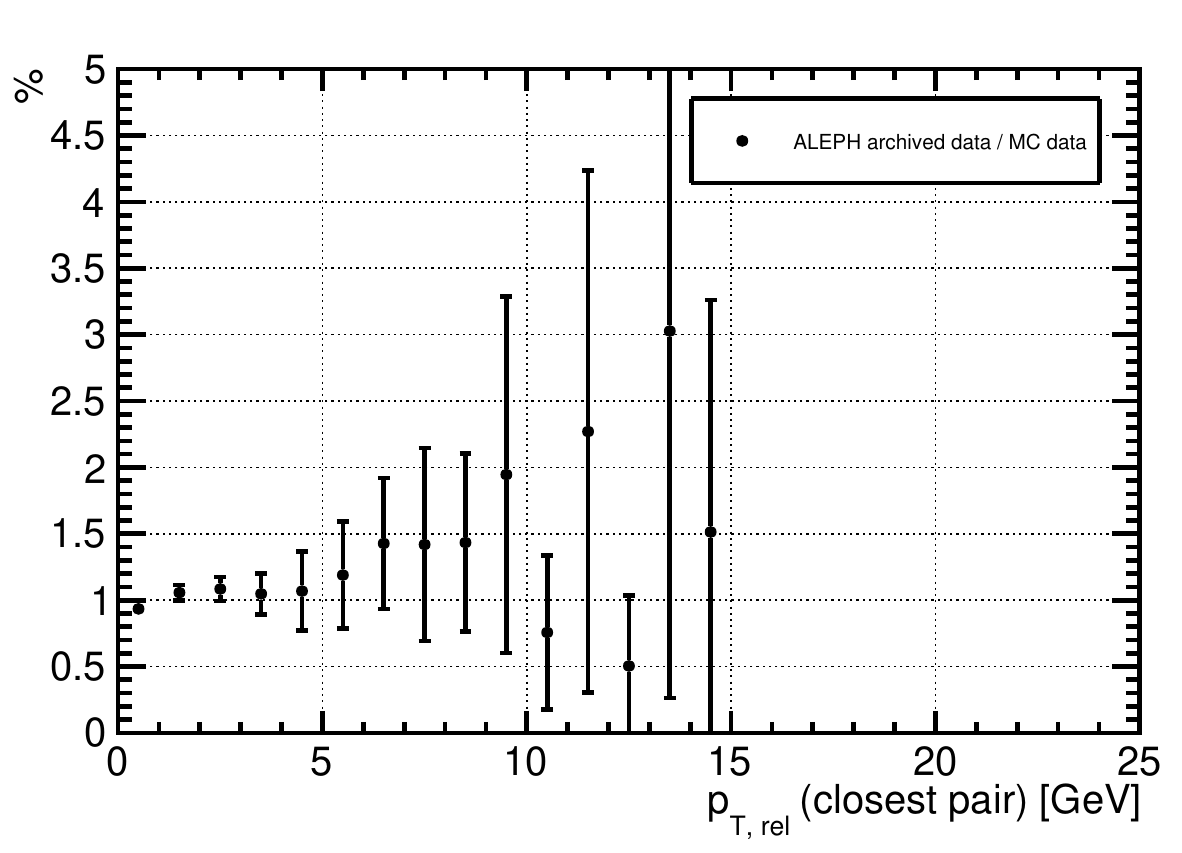}}
              \subcaptionbox{\label{fig:ptrelotherdiv}}
        {\includegraphics[width=0.48\textwidth]{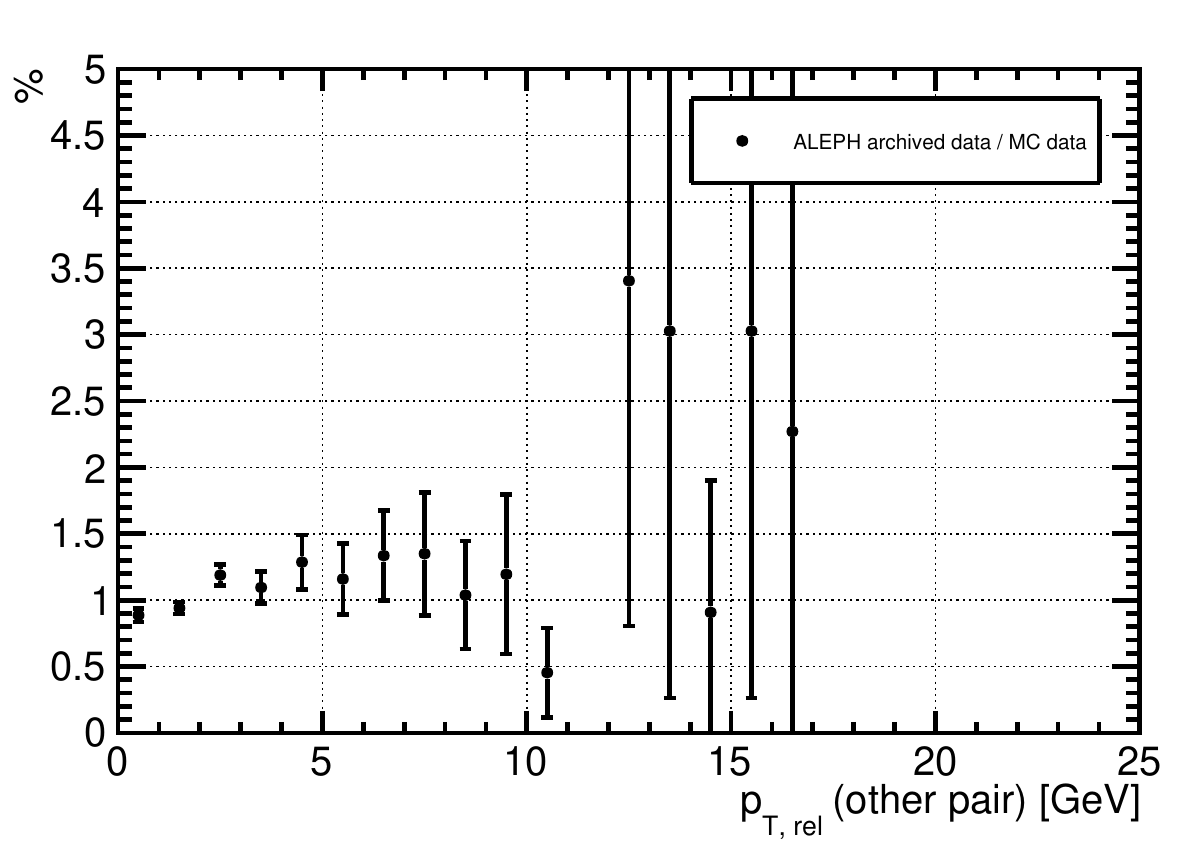}}
      \caption{The relative transverse momentum distribution $p_{\rm T, rel}$ of the closest muon-jet pair in comparison to the prediction from the full ALEPH Monte-Carlo simulation (left) and for the other muon-jet pair combination (right).}
      \centering
      \vspace{1 mm}
      \subcaptionbox{\label{fig:ptrel_vs_mass_data}}
      {\includegraphics[width=0.48\textwidth]{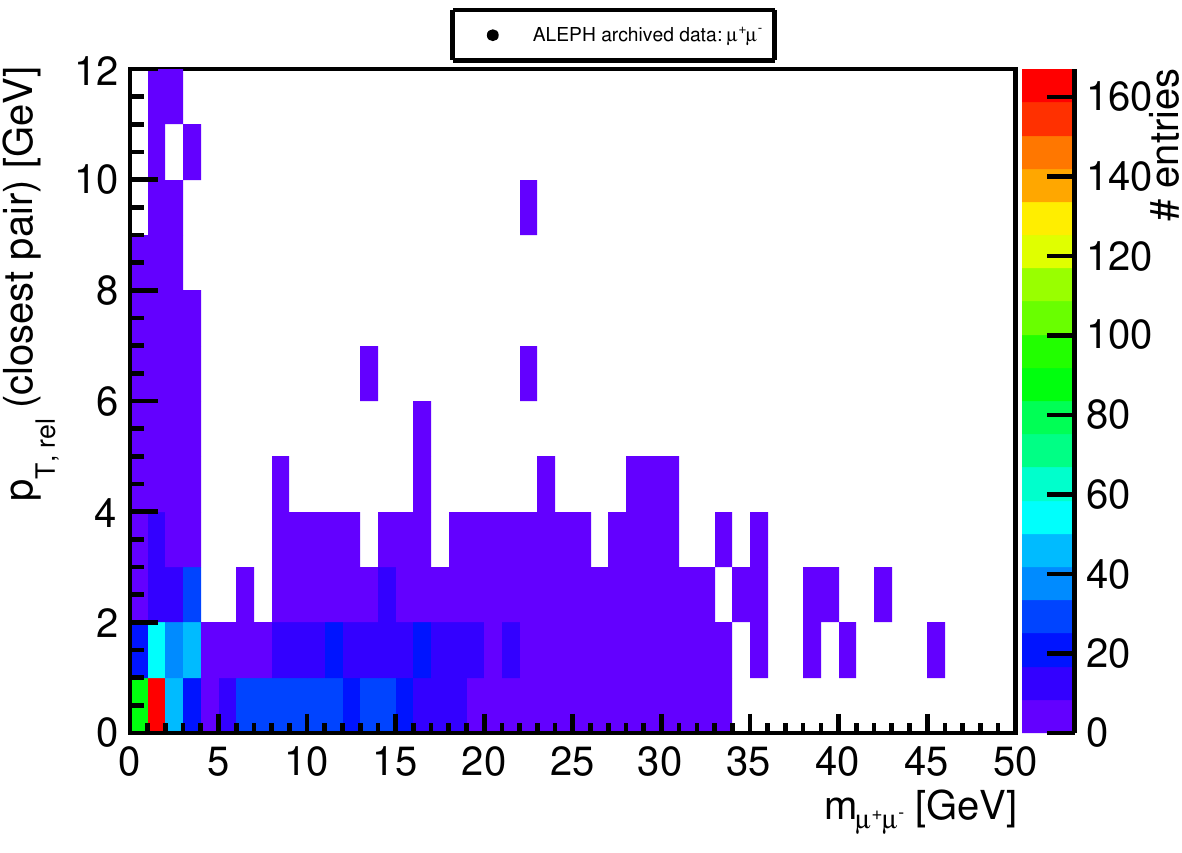}}
      \subcaptionbox{\label{fig:ptrel_vs_mass_mcdata}}
      {\includegraphics[width=0.48\textwidth]{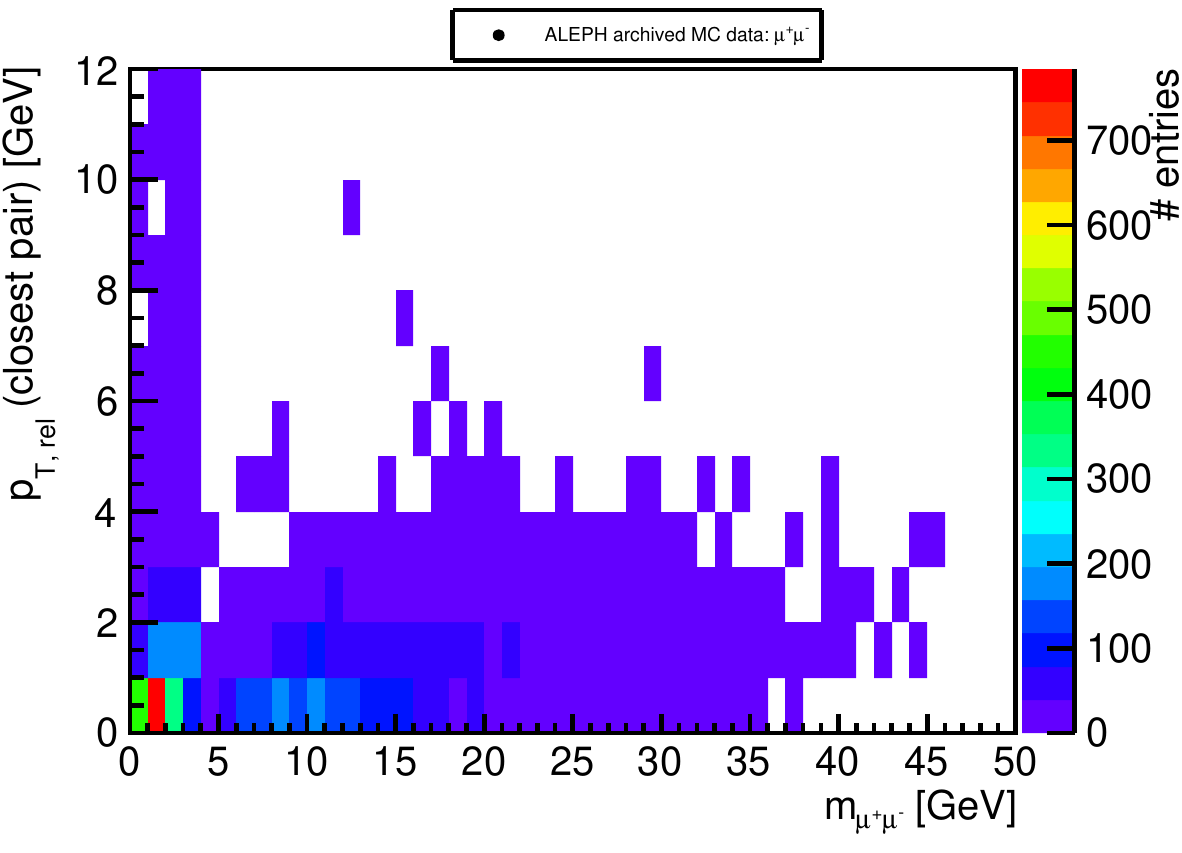}}
      \caption{ The relative transverse momentum distribution $p_{\rm T, rel}$ of the closest muon-jet pair versus the opposite sign di-muon mass spectrum in ALEPH data (left) and in full simulation (right).}
\end{figure*}

Electrons or muons from semi-leptonic b-decays will be contained inside the hadronic jets in almost all cases. As described before, we made several checks for events in the signal region around 30~GeV to find out how close the two selected muons with an invariant mass of about 30~GeV are to the hadronic jets. We find that each of the two muons which form the excess around 30~GeV are also close to one of the leading jets. However, if the structure around 30~GeV were due to semi-leptonic b-decays, then it should also be seen in the opposite sign electron-muon mass spectrum, which is definitely not the case. We use this mass spectrum as background model as described earlier in Sec.~\ref{sec:background model}.

\subsection{Relative transverse momenta of muons with respect to the closest jet}
The relative transverse momentum distribution $p_{\rm T, rel}$ of the closest muon-jet pair in comparison to the prediction of the full ALEPH Monte-Carlo simulation is shown in Figs.~\ref{fig:ptrelbest} and~\ref{fig:ptrelbestdiv}. The comparison for the remaining other muon-jet combination is shown in  Figs.~\ref{fig:ptrelother} and~~\ref{fig:ptrelotherdiv}. Taking into account the small fraction of signal like events in the sample, we do not expect and do not find any large deviation of the data from the simulation.

Figs.~\ref{fig:ptrel_vs_mass_data} and~\ref{fig:ptrel_vs_mass_mcdata} show the relative momentum distribution $p_{\rm T, rel}$ of the closest muon-jet pair versus the opposite sign di-muon mass spectrum in ALEPH data and in full simulation. No striking difference is visible in those distributions as well.

\subsection[Expected number of reconstructed muons fulfilling the event hypothesis: ${\rm Z}^{0} \to {\rm b}\overline{\rm b} \,  \mu^{+}\mu^{-}$]{\boldmath Expected number of reconstructed muons fulfilling the event hypothesis: ${\rm Z}^{0} \to {\rm b}\overline{\rm b} \,  \mu^{+}\mu^{-}$}
Assuming the "signal" like final state:  ${\rm Z}^{0} \to {\rm b}\overline{\rm b} \,  \mu^{+}\mu^{-}$ as described in Sec.\ref{sec:introduction}, we can expect with a probability of about 20\% to have at least three muons per selected event. Each b-quark can decay semi-leptonically into a muon with a probability of about 10\%~\cite{Agashe:2014kda}. Taking into account  the ALEPH detector muon identification efficiency of 86\% (see Sec.~\ref{sec:alephdetector}), we should expect on average about 17\% of our selected events to contain at least three muons.

In hadronic ${\rm Z}^{0} \to {\rm b}\overline{\rm b} + {\rm X}$ decays, there are many possibilities resulting in three or more final state muons, e.g.~a ${\rm Z}^{0} \to {\rm b}\overline{\rm b}$ decay where only one b-quark decays semi- leptonically into a muon and in addition  production of a J/$\psi$ during hadronization, which then decays into two additional muons, thus giving three final state muons. For a ${\rm Z}^{0} \to {\rm b}\overline{\rm b}$ where both b-quarks decay semi-leptonically into one muon each, we expect the average number of at least two muons with opposite charge to be close to 100\%.

\begin{figure*}[!h]
      \centering
      \subcaptionbox{\label{fig:nom_vs_mass_data}}
      {\includegraphics[width=0.44\textwidth]{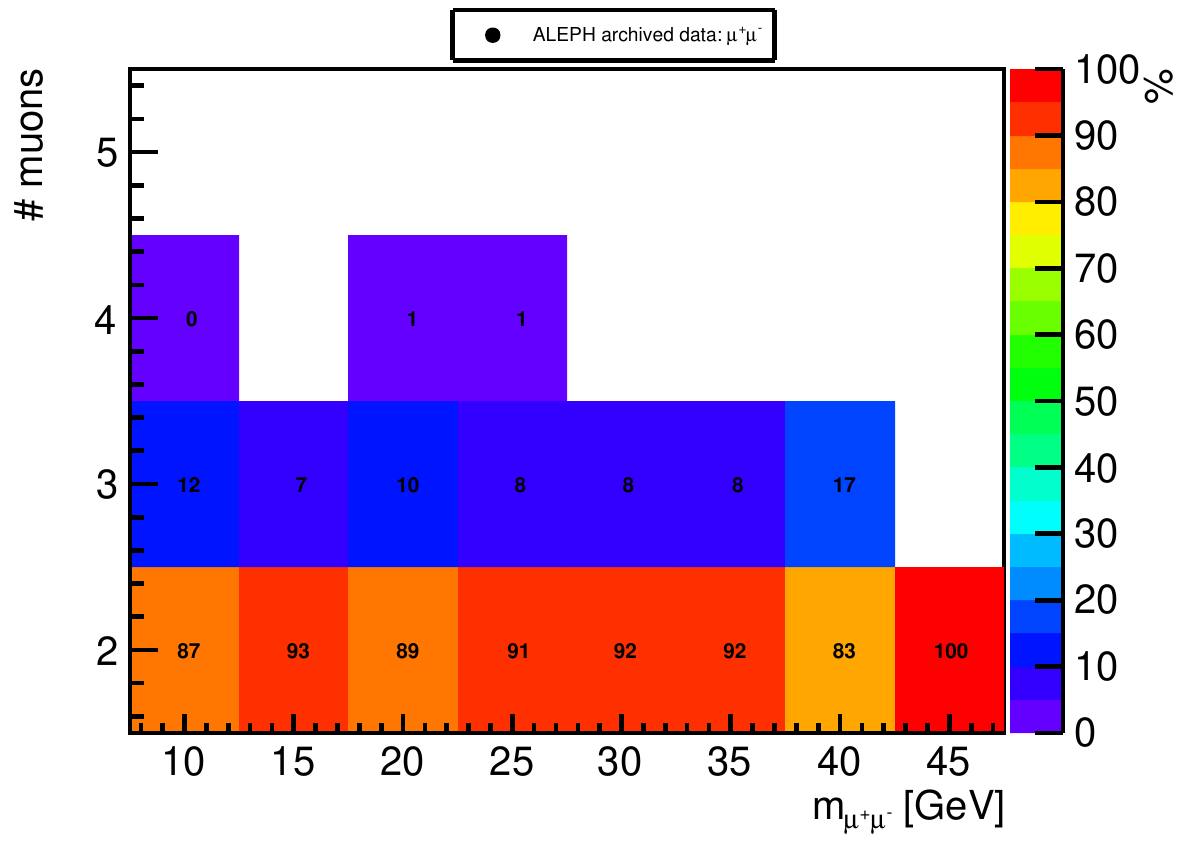}}
      \subcaptionbox{\label{fig:nom_vs_mass_mcdata}}
      {\includegraphics[width=0.44\textwidth]{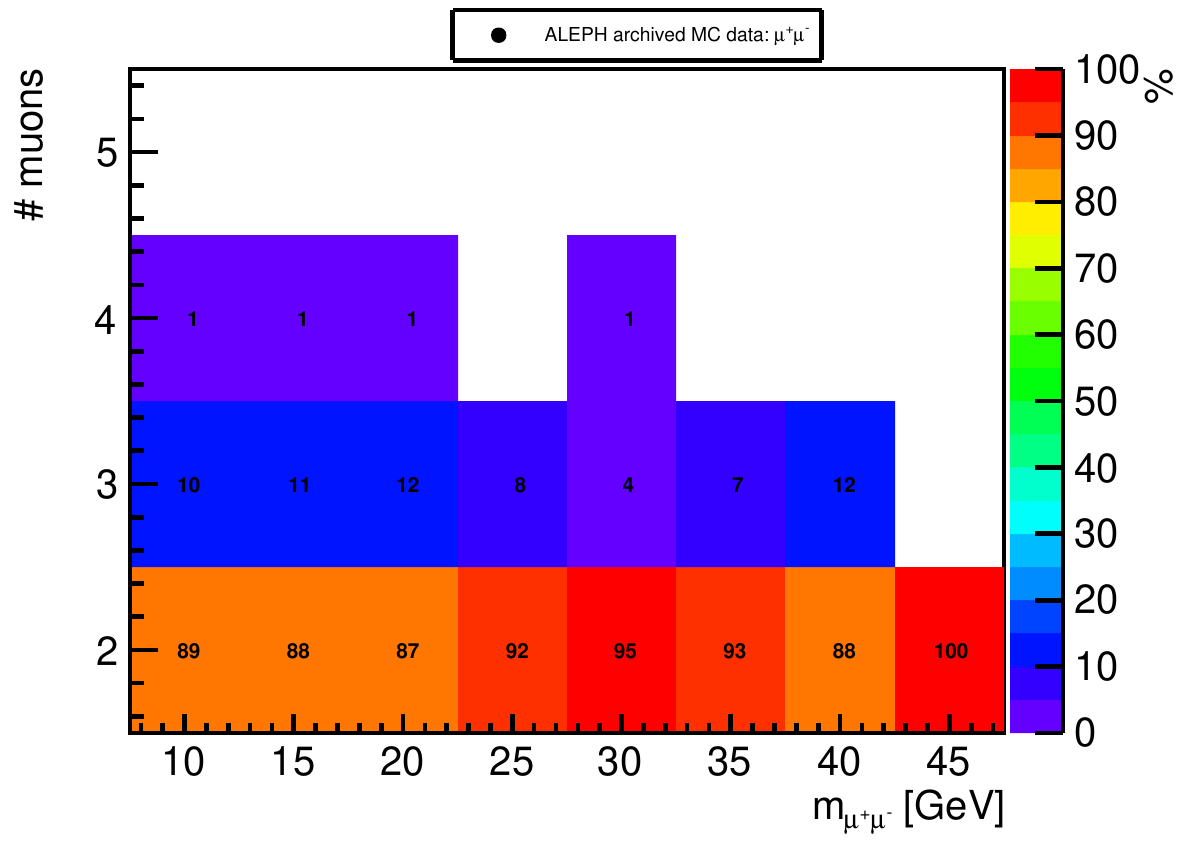}}
      \caption{ The percentage of reconstructed number of muons versus the opposite di-muon mass spectrum for ALEPH data (left) and ALEPH MC data (right).}
      \subcaptionbox{\label{fig:nom_vs_mass_data_numb}}
      {\includegraphics[width=0.44\textwidth]{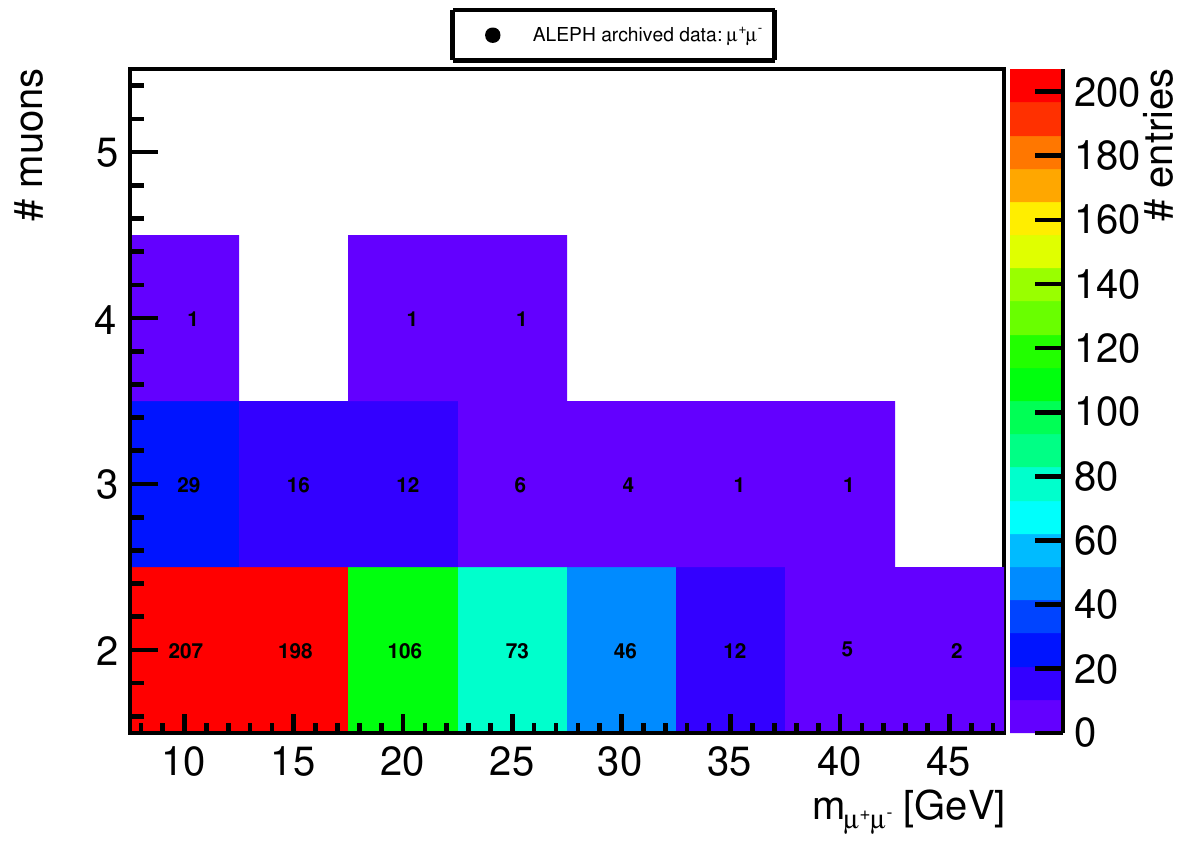}}
      \subcaptionbox{\label{fig:nom_vs_mass_mcdata_numb}}
      {\includegraphics[width=0.44\textwidth]{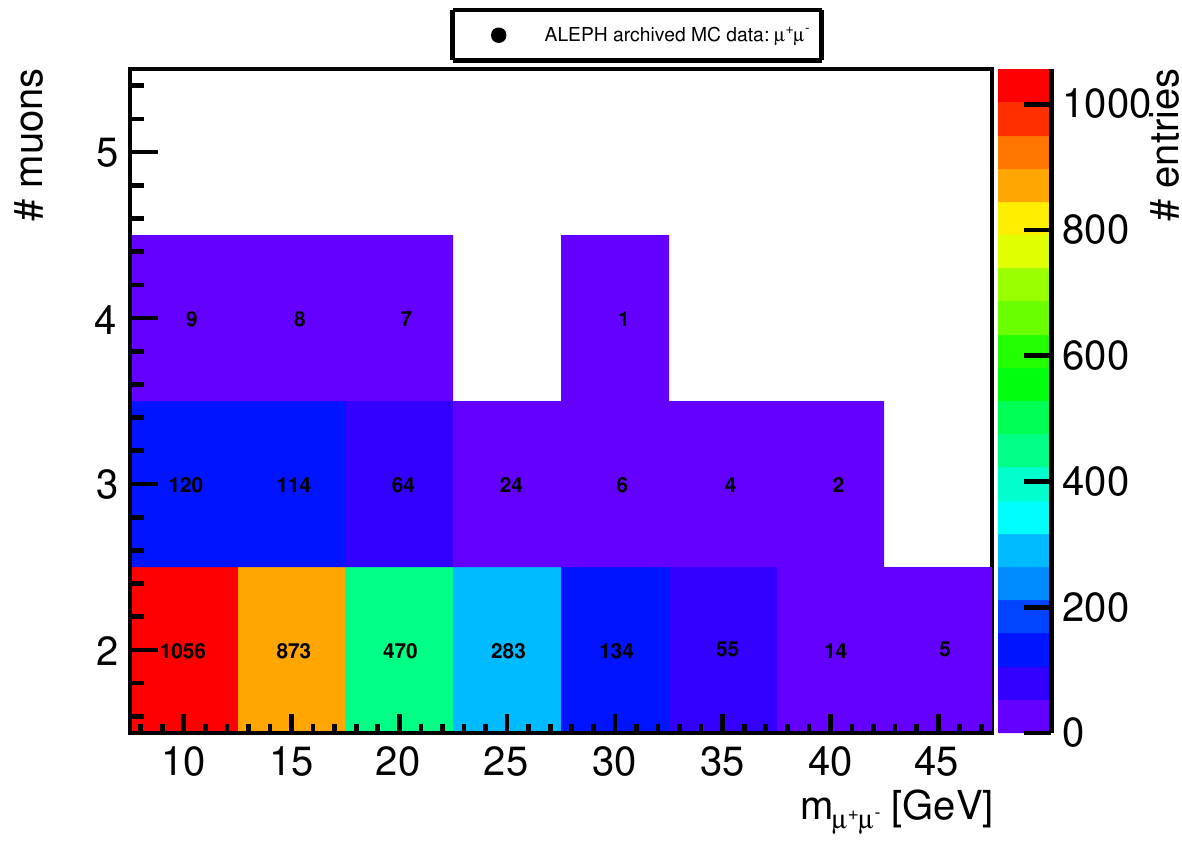}}
      \caption{ The number of reconstructed muons versus the opposite di-muon mass spectrum for ALEPH data (left) and ALEPH MC data (right).}
      \subcaptionbox{\label{fig:di-muon_signal_data_mc}}
       {\includegraphics[width=0.44\textwidth]{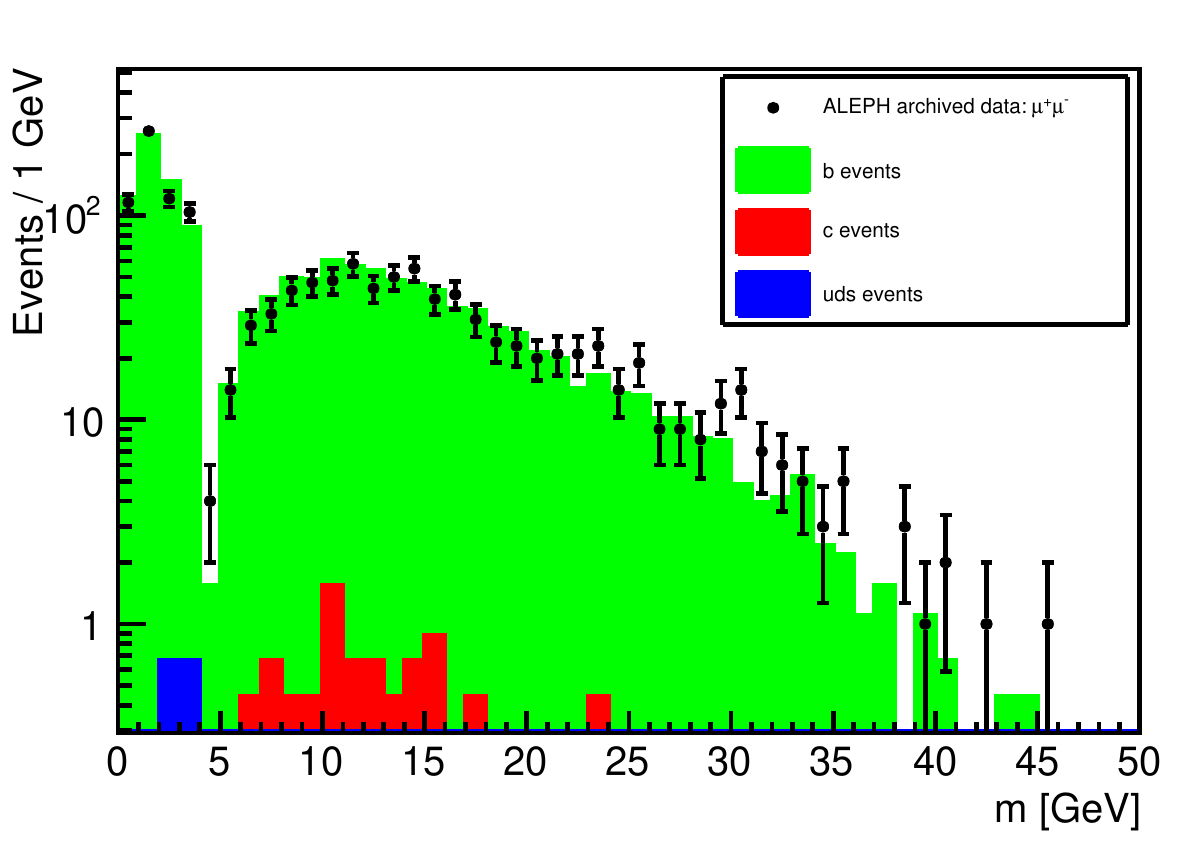}}
       \subcaptionbox{\label{fig:di-muon_signal_data_ss}}
      {\includegraphics[width=0.44\textwidth]{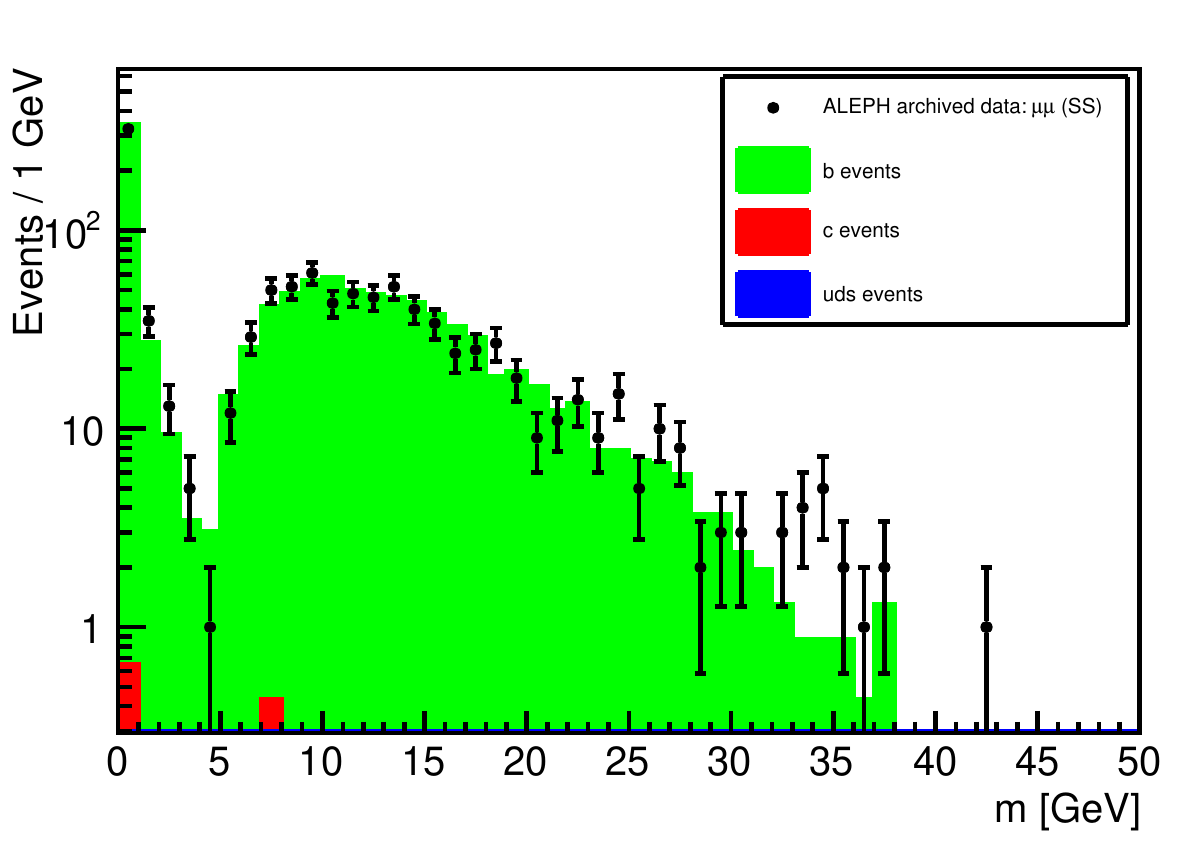}}
\caption{The data/MC data description of the opposite sign di-muon mass spectrum (left) and the same sign di-muon mass spectra obtained from ALEPH data compared to the ALEPH MC data (right). Contributions from different hadronic ${\rm Z}^{0}$ decays are identified by different colors.}
\end{figure*}

The percentage of reconstructed muons versus the opposite di-muon mass spectrum is shown for ALEPH data in Fig.~\ref{fig:nom_vs_mass_data} and for the ALEPH full simulation in Fig.~\ref{fig:nom_vs_mass_mcdata}. We do find that about 90\% of the events contain exactly two muons in the final state both in data and simulation. Near the mass region of the excess around 30~GeV, about 8\% of the real data events contain at least three muons in the final state, while simulation predicts that we should find three or muons in our selection in 5 to 8\% of the cases. This discrepancy between data and simulation could also be explained by statistical fluctuations (see Figs.~\ref{fig:nom_vs_mass_data_numb} and~\ref{fig:nom_vs_mass_mcdata_numb}).

The fractions obtained from the ALEPH data, i.e.~that in 90\% of the selected events we find exactly two muons and in 8\% three or more, suggest that we mostly have final states from semi-leptonic b-decays in our selection.

\begin{figure*}[b]
      \centering
      {\includegraphics[width=0.50\textwidth]{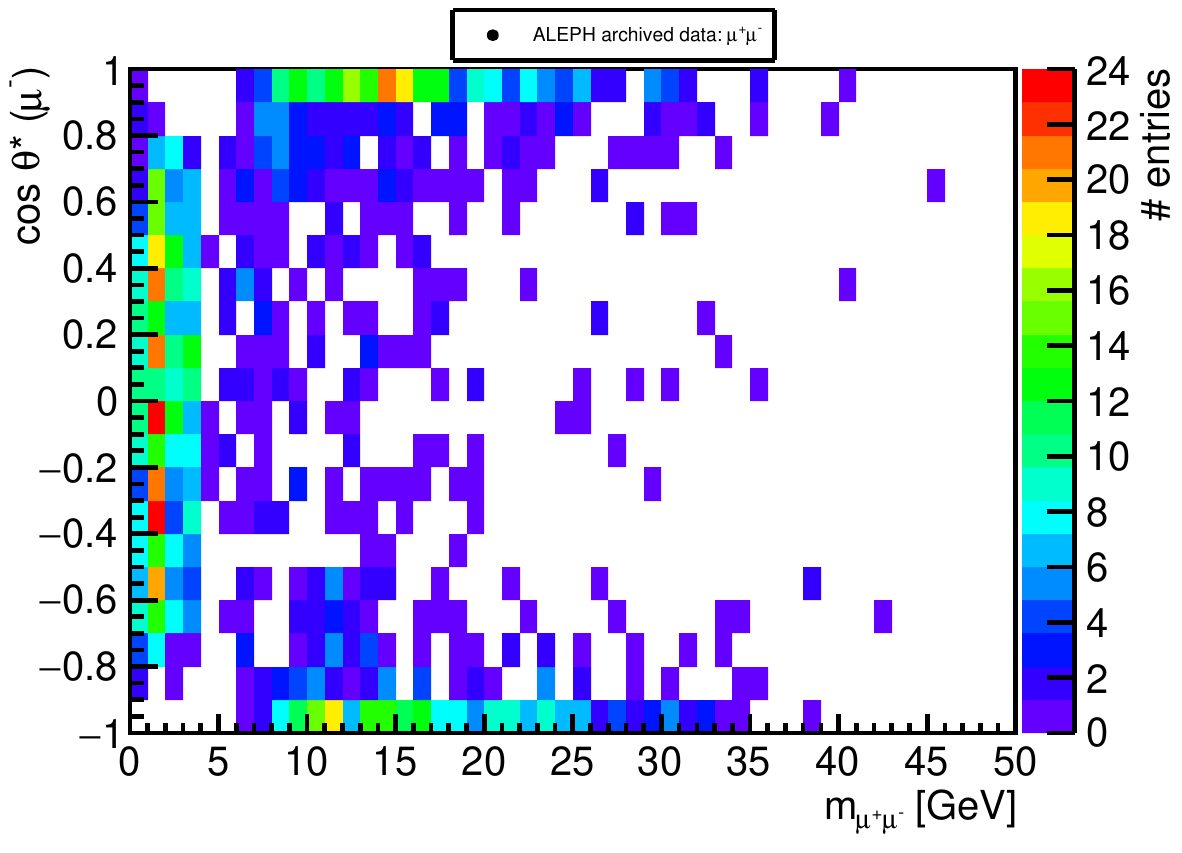}}
      \caption{The decay angle ${\rm cos}\,\theta^{*}$ for muons ($\mu^{-}$) in the di-muon rest frame with respect to the boost axis versus the opposite sign di-muon mass ${\rm m}_{\mu^{+}\mu^{-}}$ spectrum. \label{fig:spin_vs_mass}}
      \subcaptionbox{\label{fig:spin_amw2}}
      {\includegraphics[width=0.45\linewidth]{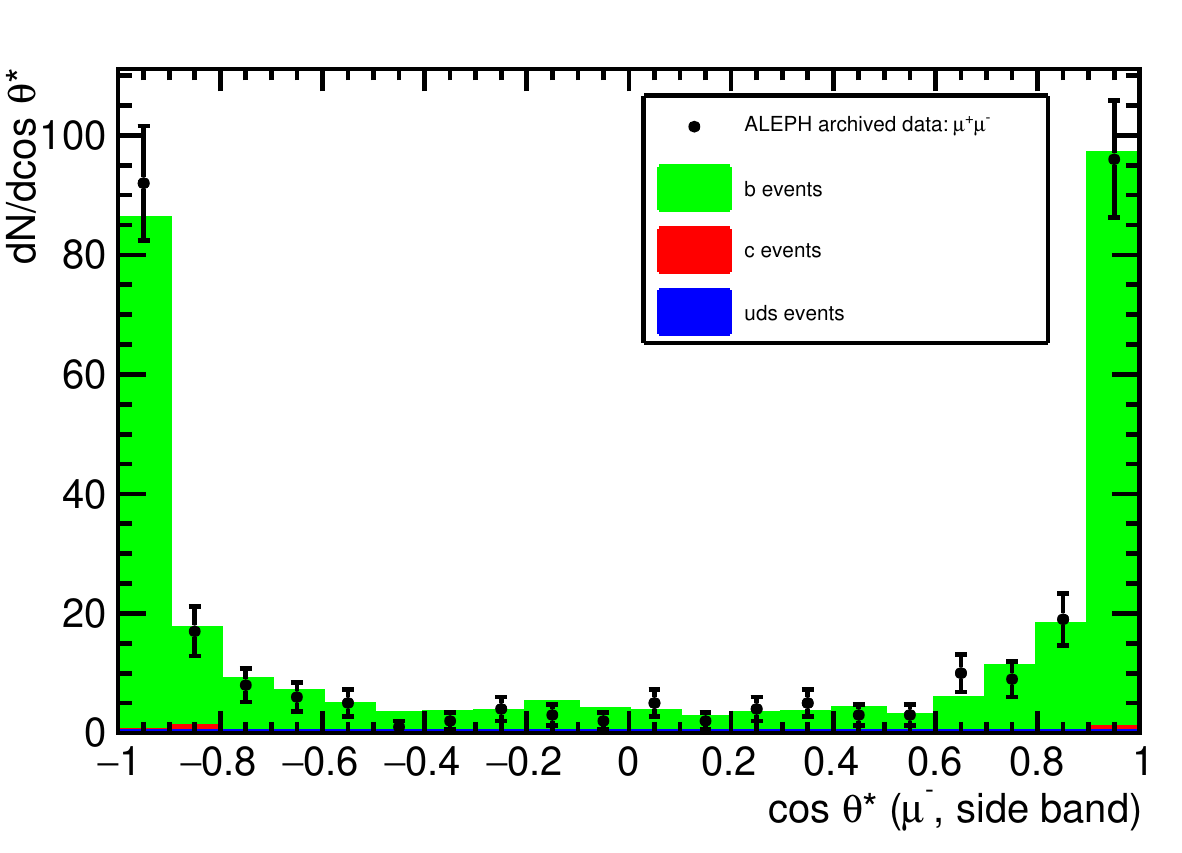}}
      \subcaptionbox{\label{fig:spin_masswindow}}
      {\includegraphics[width=0.45\linewidth]{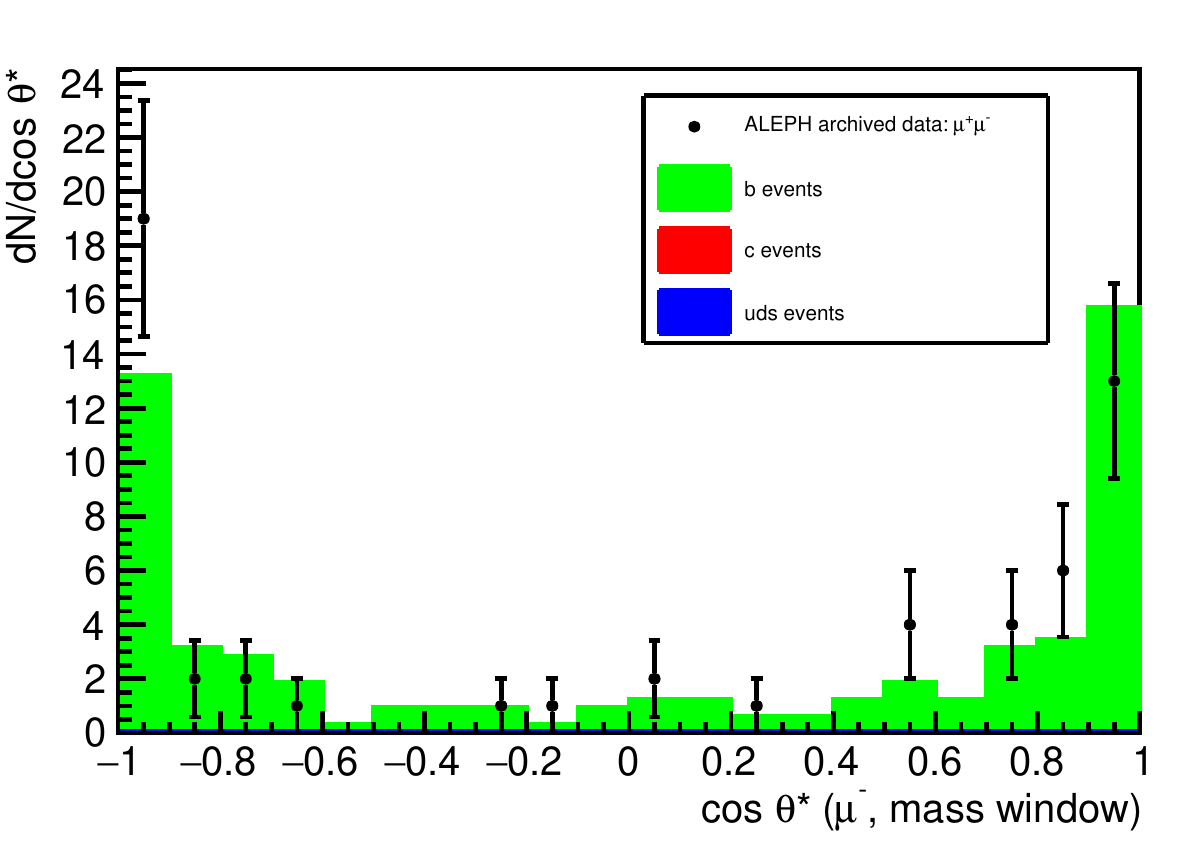}}
      \caption{The decay angle ${\rm cos}\,\theta^{*}$ for muons ($\mu^{-}$) in the di-muon rest frame with respect to the boost axis for di-muon side band events, i.e.~$15 < {\rm m}_{\mu^{+}\mu^{-}} < 50$~[GeV] excluding a mass window $\pm 2\,\sigma$ around 30~GeV (left) and for events inside this window (right). Both distributions include a comparison to the prediction from the ALEPH MC simulation.}
\end{figure*}

\subsection{Opposite sign di-muon mass spectrum: MC description of the data}
\label{sec:opposite_sign}
Fig.~\ref{fig:di-muon_signal_data_mc} shows the opposite sign di-muon mass spectrum obtained from ALEPH data compared with the MC description. Contributions from different hadronic ${\rm Z}^{0}$ decays are identified by different colors. For opposite sign di-muon masses $m > 25\,{\rm\; GeV}$, only semi-leptonic decays of ${\rm b}\overline{\rm b}$ final states contribute according to  the MC simulation. The overall description of the data by the MC data is good. In addition to the J/$\psi$ peak around 3~GeV, which is well described by the MC, the excess around 30~GeV is clearly visible.

\subsection{Study if the excess is visible in the same sign di-muon mass spectrum}
\label{sec:same_sign}
We do not observe any excess in the same sign (SS) di-muon mass spectrum, which is shown in Fig.~\ref{fig:di-muon_signal_data_ss}. The amount of same sign di-muon candidates is comparable with the opposite sign di-muon candidates, e.g.~see Fig.~\ref{fig:di-muon_signal_data_mc}. We obtain a similar result using the Monte-Carlo event generator SHERPA 2.2.0~\cite{Gleisberg:2008kq,Hoche:2014kca} (see Appendix~\ref{App:Sherpa}).

\subsection{Angular distributions}
\label{sec:angular_distributions}
The decay angle ${\rm cos}\,\theta^{*}$ for muons ($\mu^{-}$) in the di-muon rest frame with respect to the boost axis is shown in Fig.~\ref{fig:spin_vs_mass} as a function of the opposite sign di-muon mass ${\rm m}_{\mu^{+}\mu^{-}}$  . For low di-muon masses ${\rm m}_{\mu^{+}\mu^{-}} < 20\,{\rm\; GeV}$ no preferred direction of the $\mu^{-}$ is visible. For higher masses, around ${\rm m}_{\mu^{+}\mu^{-}} \simeq 30\,{\rm\; GeV}$, the forward/backward direction is preferred.

Figs.~\ref{fig:spin_amw2} and \ref{fig:spin_masswindow} show the decay angle ${\rm cos}\,\theta^{*}$ for side band events and for events in a mass window $\pm 2\,\sigma$ around 30~GeV. The di-muon side band mass region is defined as $15 < {\rm m}_{\mu^{+}\mu^{-}} < 50$~[GeV] excluding a mass window $\pm 2\,\sigma$ around 30~GeV (see Tab.~\ref{table:dimuon_fit}). The preference for the forward/backward direction mentioned above, which is a kinematic effect due to the closeness of the muons forming the di-muon resonance to the jets (see Sec.~\ref{sec:isolation}), is described by the ALEPH MC simulation. We obtain compatible results with the Monte-Carlo event generator SHERPA 2.2.0 (see Appendix~\ref{App:Sherpa}).

\subsection[Study if the excess is associated with ${\rm b}\overline{\rm b}$ final states only]{\boldmath Study if the excess is associated with ${\rm b}\overline{\rm b}$ final states only}
\begin{figure}[h]
      \centering
       \includegraphics[width=0.49\textwidth]{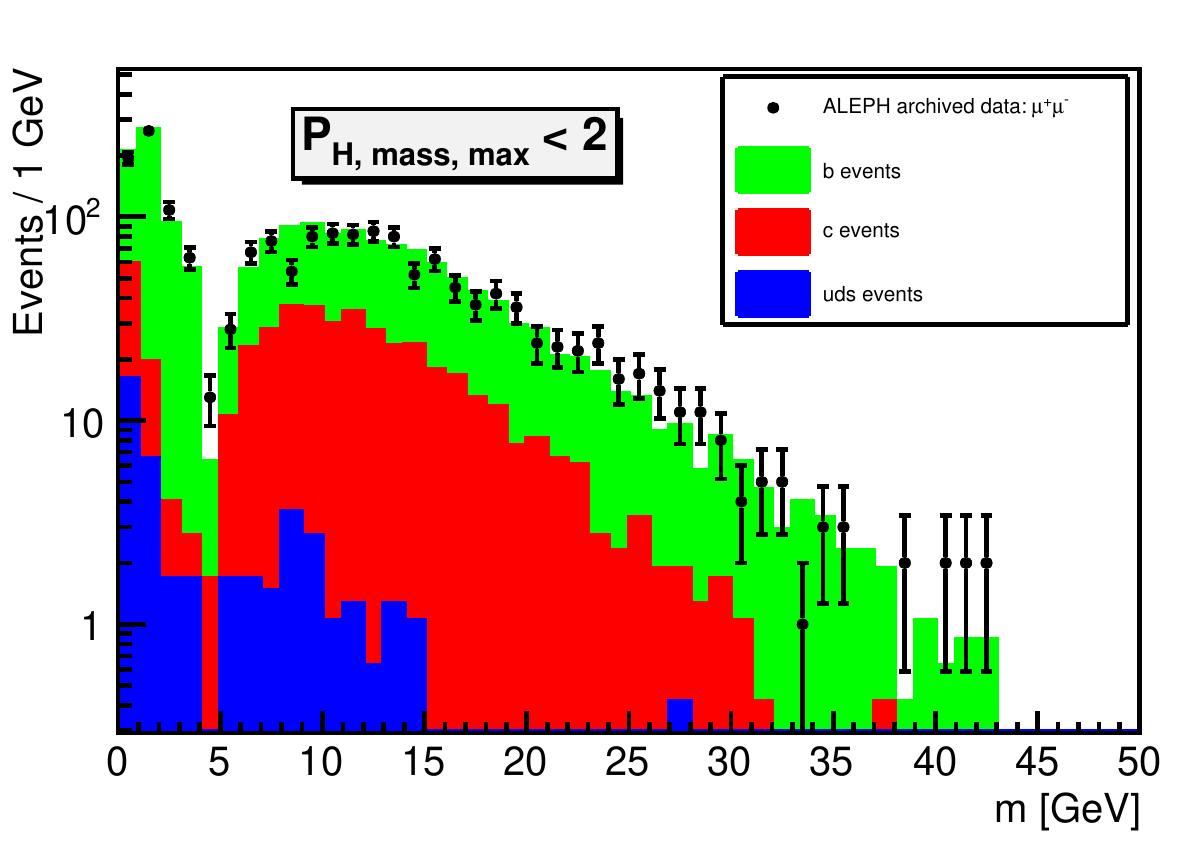}
      \caption{\label{fig:di-muon_signal_data_mc_btag_inverted}The opposite sign di-muon mass spectrum for events where the b-tag was inverted, i.e. $P_{\rm H,\,mass, max} < 2$. Contributions from different hadronic ${\rm Z}^{0}$ decays are identified by different colors.}
\end{figure}

The observed excess is indeed associated with ${\rm b}\overline{\rm b}$ final states. If we invert the  ${\rm b}\overline{\rm b}$ final state identification described in Sec. \ref{sec:analysis_btag}, i.e. we require $P_{\rm H,\,mass, max} < 2$, no excess is visible anymore (Fig.~\ref{fig:di-muon_signal_data_mc_btag_inverted}).

\begin{figure*}[!h]
      \centering
       \includegraphics[width=8.5cm]{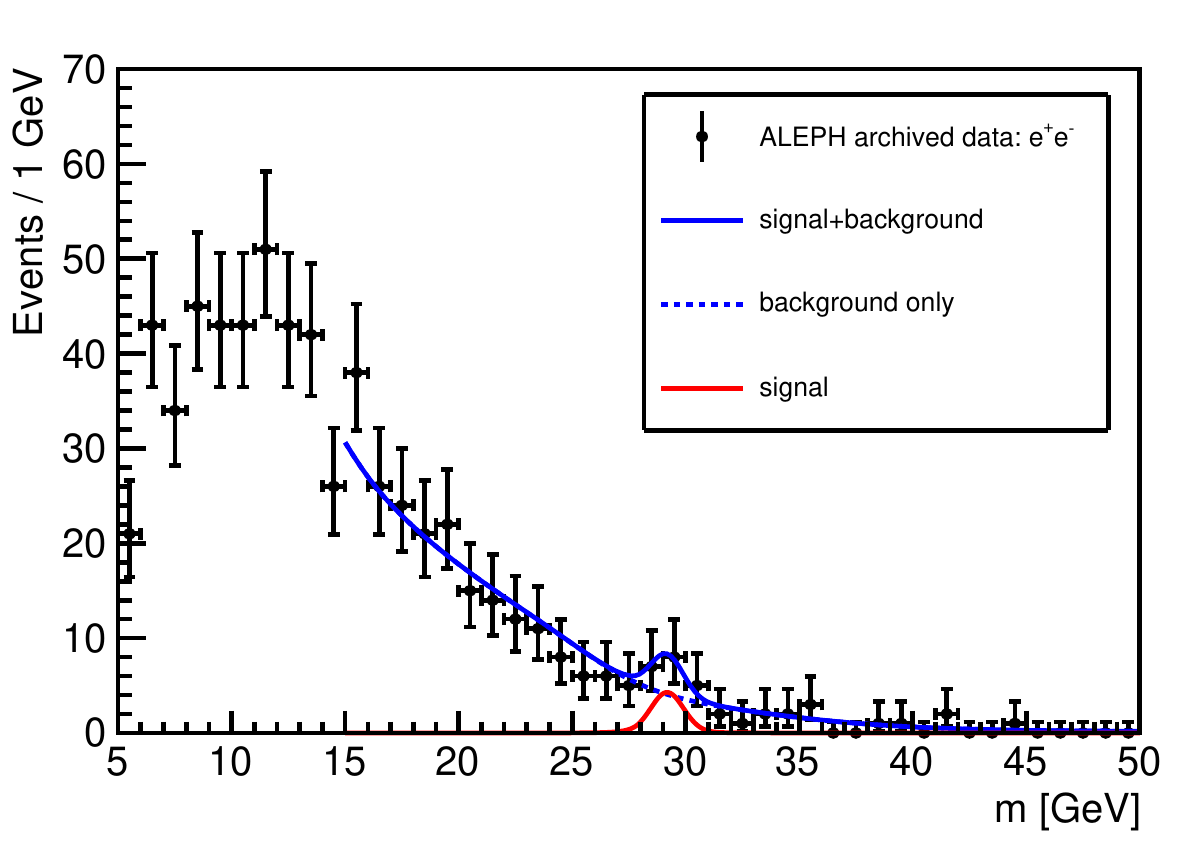}
      \caption{\label{fig:di-electron_signal}The result of the extended maximum likelihood fit of the signal + background model to the unbinned opposite sign di-electron mass spectrum.}
      \subcaptionbox{\label{fig:di-electron_signal_binned}}
        {\includegraphics[width=0.44\textwidth]{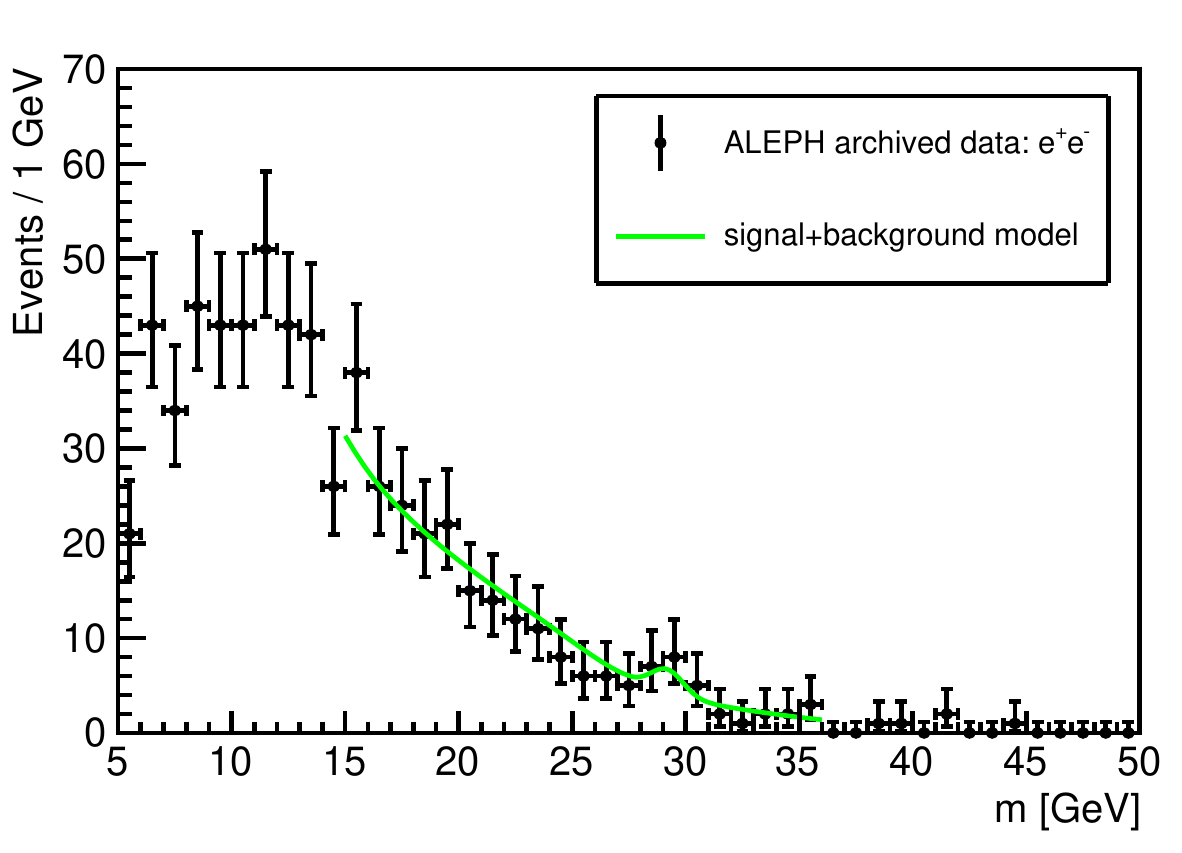}}
      \subcaptionbox{\label{fig:di-electron_signal_pull}}
        {\includegraphics[width=0.44\textwidth]{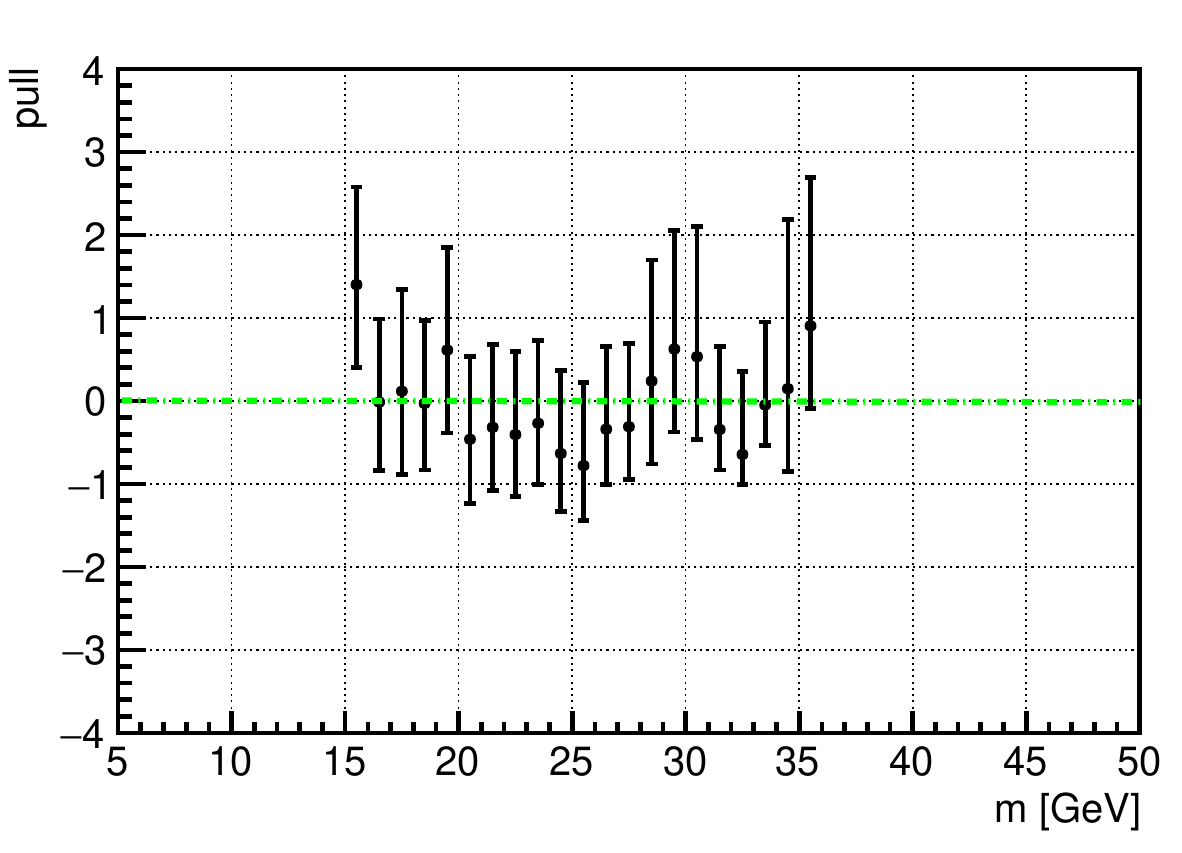}}
      \caption{The result of the $\chi^{2}$-fit of the signal + background model to the binned version of the opposite sign di-electron spectrum (left) and the corresponding distribution of pulls (right). $\chi^{2}/{\rm ndof} = 0.39$. To avoid bins with zero entries the fit range for this test is restricted to [15, 36]~GeV.}
       \vspace{4 mm}
\begin{subtable}[c]{0.49\textwidth}
 \centering
 \scalebox{0.7}{ 
 \begin{tabular}{lrl}    
        \hline\noalign{\smallskip}     
        {\bf Parameter}& {\bf Value} & {\bf Error} \\
        \noalign{\smallskip}\hline\noalign{\smallskip}
        \# signal events & 8.00  & $\pm$ 4.53  \\ 
	\# background events (overall) & 1036.91 & $\pm$ 69.45 \\
        \noalign{\smallskip}\hline
	mass [GeV] & 29.18 & $\pm$ 0.47 \\ \hline
	width (Crystal Ball) [GeV] & 0.10 & $\pm$ 1.82 \\
	alpha (Crystal Ball) [GeV] & 1.67 & $\pm$ 1.94 \\ 
	n (Crystal Ball) [GeV] & 8.79 & $\pm$ 9.38 \\
        \noalign{\smallskip}\hline
	width (Gaussian) [GeV] &	0.70 & $\pm$ 0.10 \\
        \noalign{\smallskip}\hline
  \end{tabular} 
  } 
  \subcaption{\label{table:dielectron_fit}}
\end{subtable}
   \begin{subtable}[]{0.49\textwidth}
   \centering
   \scalebox{0.8}{
   \begin{tabular}{ll}    
        \hline\noalign{\smallskip}     
        {\bf Observable}& {\bf Value} \\
        \noalign{\smallskip}\hline\noalign{\smallskip}
 	$Z_{\rm Bi}$ & $1.15\,\sigma$ \\
	\noalign{\smallskip}\hline
        $Z_{\rm asym}$ & $1.53\,\sigma$  \\
        p-value & 0.062995 \\
        \noalign{\smallskip}\hline
  \end{tabular}  }
   \subcaption{\label{table:dielectron_sig}}
\end{subtable}
 \caption{Parameter values of the extended maximum likelihood fit to the opposite sign di-electron mass spectrum obtained from ALEPH data (left) and significance of the excess (right).\label{table:dielectron}}
 \end{figure*}

\subsection{Study if the excess is visible in opposite sign di-electron final states}
\label{sec:result_dielectron}
A slight excess is visible in di-electron final states using the same selection criteria as in the di-muon case (Sec.~\ref{sec:analysis}). Fig.~\ref{fig:di-electron_signal} shows a MLE fit of a signal + background model to the unbinned di-electron mass spectrum. The background model described in Sec.~\ref{sec:background model} is utilized again. To take into account the different signal shape due to Bremsstrahlung of the electrons forming the di-electron resonance, a convolution of a Gaussian with a Crystal Ball distribution~\cite{Skwarnicki:1986xj} is used. We expect a shift of the obtained mean mass value due to Bremsstrahlung also (see Sec.~\ref{sec:jpsi}). Again as described in Sec.~\ref{sec:results} the width of the Gaussian used is constrained by a penalty function, due to the correlation of the Gaussian width with that of the Crystal Ball function.

The obtained probability of equality for the KS test is 96\% for  the mass range 20 to 40~GeV. The $\chi^{2}$-fit to a binned version of the di-electron mass spectrum (Figs.~\ref{fig:di-electron_signal_binned} and \ref{fig:di-electron_signal_pull}) results in $\chi^{2}/{\rm ndof} = 0.39$. Again, to avoid bins with zero entries the fit range for this test is restricted to [15, 36]~GeV.

Tab.~\ref{table:dielectron_fit} and \ref{table:dielectron_sig} summarize the result of the fit as well as the signal significance. As expected the obtained mass value of 29.18~GeV is slightly lower compared to the di-muon final state (Tab.~\ref{table:dimuon_fit}). The detector mass resolution for di-electrons of about 0.70~GeV is reasonable. The natural width of the excess is masked by its large error.

\begin{figure*}[t]
      \centering
      \subcaptionbox{\label{fig:comb_dimuon}}
        {\includegraphics[width=0.49\textwidth]{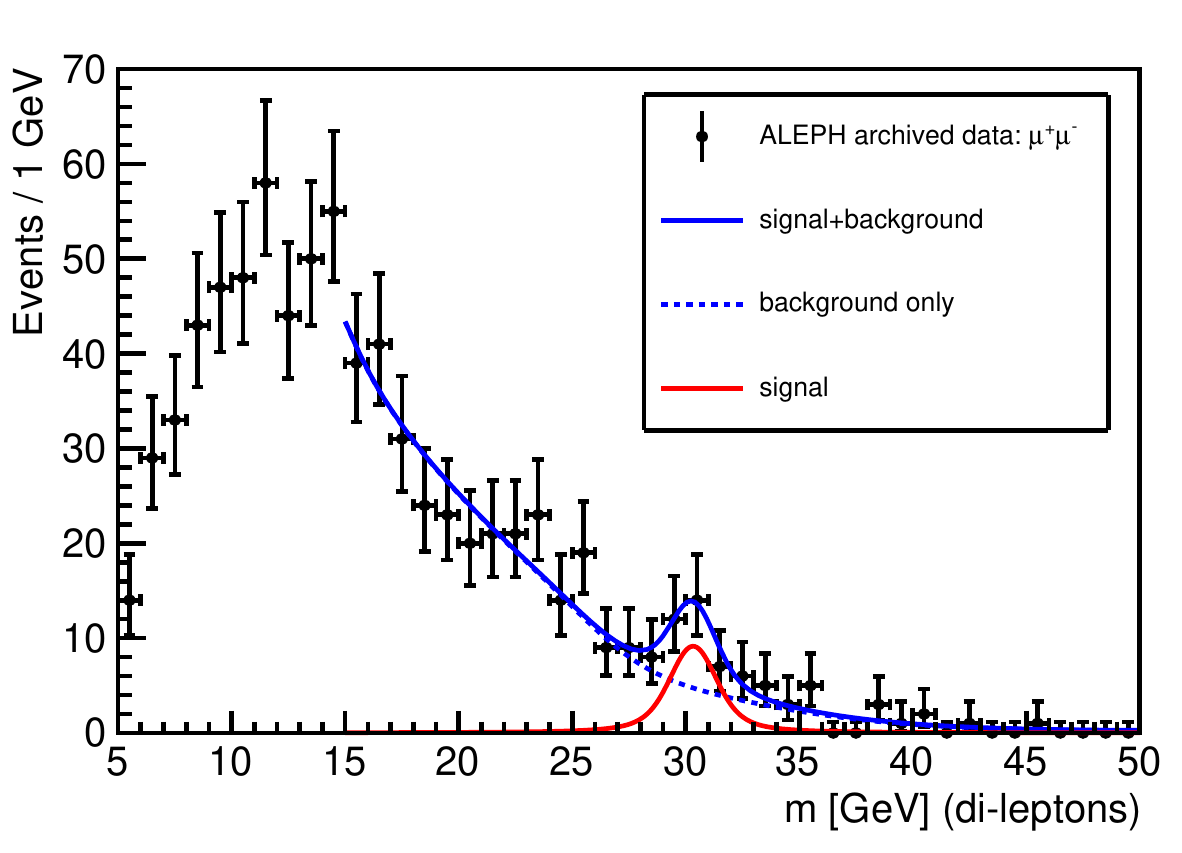}}
      \subcaptionbox{\label{fig:comb_dielectron}}
        {\includegraphics[width=0.49\textwidth]{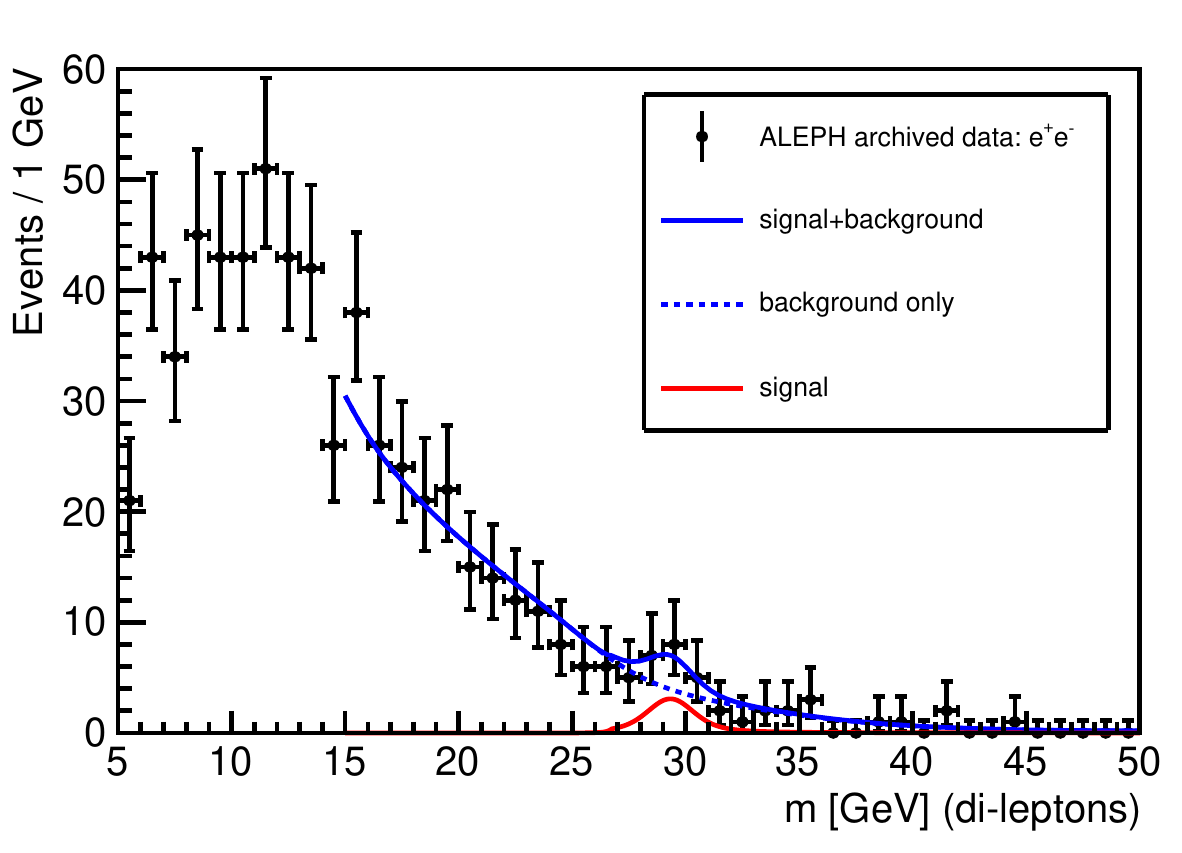}}
      \caption{The result of the simultaneous extended maximum likelihood fit of the corresponding signal + background models to the unbinned opposite sign di-muon (left) and di-electron (right) mass spectrum.}
 \centering
 \begin{subtable}[c]{0.5\linewidth}
 \centering 
\scalebox{0.86}{
  \begin{tabular}{lrl}    
        \hline\noalign{\smallskip}     
        {\bf Parameter}& {\bf Value} & {\bf Error} \\
        \noalign{\smallskip}\hline\noalign{\smallskip}
        \# signal events & 29.53  & $\pm$ 10.39  \\ 
	\# background events (overall) & 1468.2 & $\pm$ 89.3 \\ 
        \noalign{\smallskip}\hline
	mass [GeV] & 30.33 & $\pm$ 0.37 \\
        \noalign{\smallskip}\hline
	width (Breit-Wigner) [GeV] & 1.26 & $\pm$ 0.99 \\
        \noalign{\smallskip}\hline
	width (Gaussian) [GeV] &	0.75 & $\pm$ 0.10 \\
        \noalign{\smallskip}\hline
  \end{tabular}
}
  \subcaption{\label{table:comb_dimuon_fit}}
\end{subtable}
   \begin{subtable}[h]{0.45\linewidth}
   \centering
\scalebox{0.86}{
  \begin{tabular}{lrl}    
        \hline\noalign{\smallskip}     
        {\bf Parameter}& {\bf Value} & {\bf Error} \\
        \noalign{\smallskip}\hline\noalign{\smallskip}
       \# signal events & 9.19  & $\pm$ 5.47 \\ 
	\pbox{20cm}{\# background events \\ (overall)} & 1031.8 & $\pm$ 70.5 \\
        \noalign{\smallskip}\hline
	mass [GeV] & 29.33 & $\pm$ 0.37 \\ 
	mass shift [GeV] & -1 & const. \\
        \noalign{\smallskip}\hline
	width (Crystal Ball) [GeV] & 1.26 & $\pm$ 0.99 \\
	alpha (Crystal Ball) [GeV] & 3.53 & $\pm$ 0.19 \\ 
	n (Crystal Ball) [GeV] & 5.46 & $\pm$ 7.12 \\
        \noalign{\smallskip}\hline
	width (Gaussian) [GeV] &	0.73 & $\pm$ 0.10 \\
        \noalign{\smallskip}\hline
   \end{tabular}   
}
   \subcaption{\label{table:comb_dielectron_fit}}
\end{subtable}
 \caption{Parameter values of the simultaneous extended maximum likelihood fit to the opposite sign di-muon (Tab.~\ref{table:comb_dimuon_fit}) and di-electron (Tab.~\ref{table:comb_dielectron_fit}) mass spectrum obtained from ALEPH data. The obtained result is compatible with the single fits to the opposite sign di-muon and di-electron mass spectra (Tab.~\ref{table:dimuon} and \ref{table:dielectron}). The expected mass shift of the obtained mean for di-electron final states is clearly visible.}
 \end{figure*}
 
\subsection{Check if the two excesses in the di-muon and di-electron mass spectrum are compatible}
\label{sec:combinedresult}
The mean and natural width of the excess observed in the di-muon and di-electron invariant mass spectra are indeed statistically compatible. Fig~\ref{fig:comb_dimuon} and \ref{fig:comb_dielectron} show the result obtained from a simultaneous extended maximum likelihood fit of the corresponding signal + background models to the unbinned opposite sign di-muon and di-electron mass spectrum. For both spectra, the background model described in Sec.~\ref{sec:background model} is used. For the di-muon mass spectrum a Breit-Wigner distribution convoluted with a Gaussian is used again (Sec.~\ref{sec:results}). A convolution of a Crystal Ball distribution with a Gaussian is used for the di-electron mass spectrum due to the effects of Bremsstrahlung as outlined in Sec.~\ref{sec:result_dielectron}.
For the simultaneous fit the RooFit package of ROOT is again utilized~\cite{ROOT,Antcheva:2009zz,Moneta:2010pm}. The common parameters in the fit are: 
\begin{enumerate}
  \item The mean mass value,
  \item The natural width of the resonance, i.e. the width of the Breit-Wigner and Crystal Ball functions. 
\end{enumerate}
Free parameters are:
\begin{enumerate}
  \item The width of the two Gaussians, which reflect the different reconstruction performance of the ALEPH detector for electrons and muons,
  \item The parameters of the Crystal-Ball function, namely alpha and n.
\end{enumerate}
As described earlier in Sec.~\ref{sec:results} and \ref{sec:result_dielectron} the widths of the two Gaussians are again constrained by a penalty function. Since the mean mass of the di-electron mass spectrum is shifted to lower values due to Bremsstrahlung (Sec.~\ref{sec:jpsi} and \ref{sec:result_dielectron}), the mean value of the fitted di-electron excess is allowed to have a (constant) shift to a lower mass. In Tabs.~\ref{table:comb_dimuon_fit} and \ref{table:comb_dielectron_fit} the obtained fit parameters are listed.

To compute a combined significance of the excess around 30~GeV in the opposite di-muon and di-electron mass spectra a {\it signal strength} needs to be defined. The signal strength depends on the production and decay of a specific final state, i.e.~on a specific theoretical model predicting and describing the observations. Thus the signal strength depends on the assumptions made in a specific model. At present we did not find a model which can be applied to our observation, therefore no combined significance is given. Besides the common excess around 30~GeV, Fig~\ref{fig:comb_dimuon} and \ref{fig:comb_dielectron} show very few coincidences of common up- or downward fluctuations in the mass spectra. Hence the calculation of a look elsewhere effect calculated for the combined significance using those distributions will probably be small.  

\section{Summary}
\label{sec:summary}
The re-analysis of events recorded at the ${\rm Z}^{0}$ resonance by the ALEPH experiment during 1992 to 1995 shows an excess in the opposite sign di-muon invariant mass spectrum for ${\rm Z}^{0} \to {\rm b}\overline{\rm b} + {\rm X}$ events. A smaller excess is also visible in the opposite sign di-electron  invariant mass spectrum. We did not find any excess in the opposite sign electron-muon invariant mass spectrum nor in any same sign di-muon or di-electron mass spectrum. 

The obtained mass from a MLE fit to the unbinned data of the excess is 30.40~$\pm$~0.46~GeV. Its natural width is 1.78 $\pm$~1.14~GeV.

The di-muon excess has a significance of at least $2.6\,\sigma$ ($Z_{\rm Bi}$) using the calculations presented.

The local significance of the excess is around $5\,\sigma$ ($Z_{\rm asym}$), depending on the background model used (see Sec~\ref{sec:results}). The significances for background models based on a kernel density approximation stay close to $3\,\sigma$ ($Z_{\rm freq,\,lee}$) when including a look elsewhere effect. As discussed in Sec.~\ref{sec:combinedresult} a combined significance for di-muons and di-electrons can be calculated if the source of the excess is understood.

Several experiments have data samples that include the di-lepton mass region discussed here. The excess described in this paper may be present in data of other experiments at LEP, the Tevatron and the LHC\footnote{A basic set of observables used by this analysis is implemented in the Rivet toolkit, which is a system for validation of Monte Carlo event generators~\cite{Buckley:2010ar}. It can be accessed by using the reference number: ALEPH\_2016\_I1492968.
}.

\ack
I wish to thank the ALEPH collaboration for access to the archived data since the closure of the collaboration~\cite{ALEPH-archived-data}. I also wish to thank the CERN accelerator divisions for the successful operation of LEP, and I am indebted to the researchers, engineers and technicians for their contribution to the excellent performance of ALEPH.

I would like to thank especially Duccio Abbaneo, Werner Bernreuther, Marcello Maggi, Lorenzo Moneta, Andrei Ostaptchouk, Bernhard Spaan and G\"unter Quast for their technical help, fruitful discussions, advice and support during the development of this analysis. A big thank you to Demetris Pandoulas for carefully reading the paper. Many thanks also to Steven Murray and Holger Schulz.

\clearpage

\bibliographystyle{spphys} 

\bibliography{ALEPH_di_lepton.bib}

\clearpage

\appendix

\section{SHERPA generator study of opposite and same sign di-muon pairs}
\label{App:Sherpa}
The Monte-Carlo event generator SHERPA 2.2.0~\cite{Gleisberg:2008kq,Hoche:2014kca} was used to generate 1 Million
\begin{eqnarray*}
{\rm e}^{+}{\rm e}^{-} \to {\rm Z}^{0} \to {\rm q}\overline{\rm q}
\end{eqnarray*}
events. The generated events had to pass the following selection criteria:
\begin{enumerate}
  \item \# b-jets >= 2,
  \item ${\rm p}_{\,T, \,{\rm muon}} > 2.5\,{\rm\; GeV}$
  \item $\cos \theta_{\,{\rm muon}} < 0.7$
\end{enumerate}
The momentum and angular selection criteria for the muons are designed to mimic the muon particle identification (see Sec.~\ref{sec:analysis_overview}) and the detector acceptance.

\begin{figure*}[h]
      \centering
      \subcaptionbox{The opposite sign di-muon mass spectra ${\rm m}_{\mu^{+}\mu^{-}}$ OS of identified ${\rm Z}^{0} \to {\rm b}\overline{\rm b} + {\rm X}$ decays.\label{fig:sherpa_di-muon_OS}}
      {\includegraphics[width=0.49\textwidth]{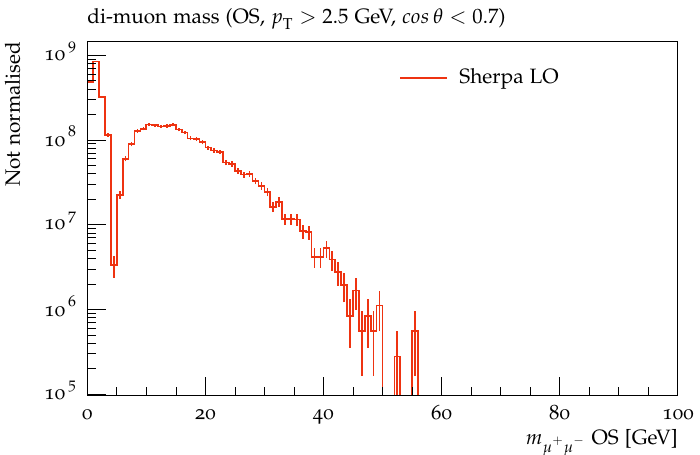}}
      \subcaptionbox{The same sign di-muon mass spectra ${\rm m}_{\mu^{+}\mu^{-}}$ SS of identified ${\rm Z}^{0} \to {\rm b}\overline{\rm b} + {\rm X}$ decays.\label{fig:sherpa_di-muon_SS}}
      {\includegraphics[width=0.49\textwidth]{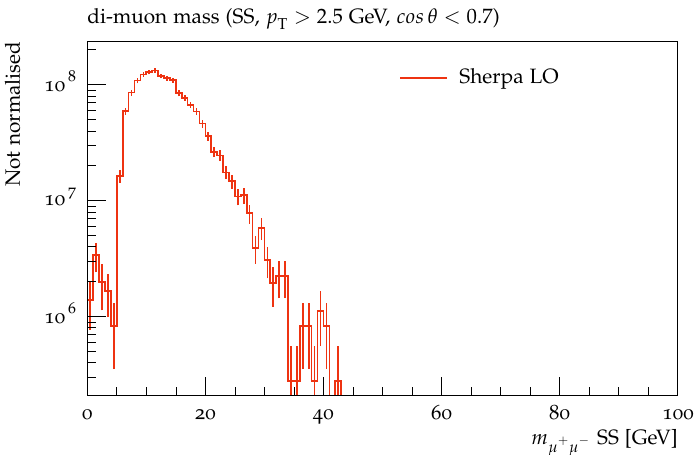}}
      \centering
       \subcaptionbox{The decay angle ${\rm cos}\,\theta^{*}$ for muons ($\mu^{-}$) in the di-muon rest frame with respect to the boost axis for identified ${\rm Z}^{0} \to {\rm b}\overline{\rm b} + {\rm X}$ decays.\label{fig:sherpa_di-muon_spin}}
      {\includegraphics[width=0.8\textwidth]{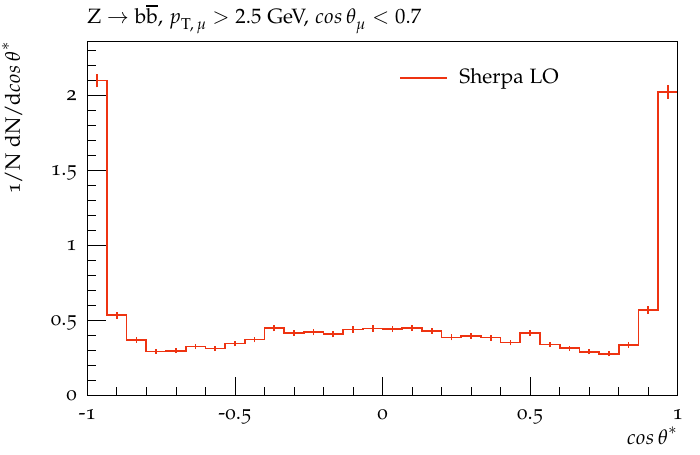}}
      \caption{Selected observables of the SHERPA generator study.}
\end{figure*}

Figs.~\ref{fig:sherpa_di-muon_OS} and \ref{fig:sherpa_di-muon_SS} show the mass spectra for opposite sign and same sign di-muon pairs. Always, the di-muon pair with the highest mass in the event is chosen. No excess is visible in both figures. The amount of di-muon pairs in the mass range from 10 to 30~GeV is similar (see Secs.~\ref{sec:opposite_sign} and \ref{sec:same_sign}) The decay angle ${\rm cos}\,\theta^{*}$ for muons ($\mu^{-}$) in the di-muon rest frame with respect to the boost axis for identified ${\rm Z}^{0} \to {\rm b}\overline{\rm b} + {\rm X}$ decays is shown in Fig.~\ref{fig:sherpa_di-muon_spin} (see Sec.~\ref{sec:angular_distributions}). The presented observables are implemented in the Rivet toolkit~\cite{Buckley:2010ar} using the reference number: ALEPH\_2016\_I1492968.

\clearpage

\section{3-momentum distribution of oppositely charged di-muon pairs}
\label{App:3Mom}
\begin{figure*}[!h]
      \centering
      \subcaptionbox{Low side band events: $20 < m_{\mu^{+}\mu^{-}} \lesssim 26.55$~[GeV].\label{fig:dimuon_P_low}}
      {\includegraphics[width=0.55\textwidth]{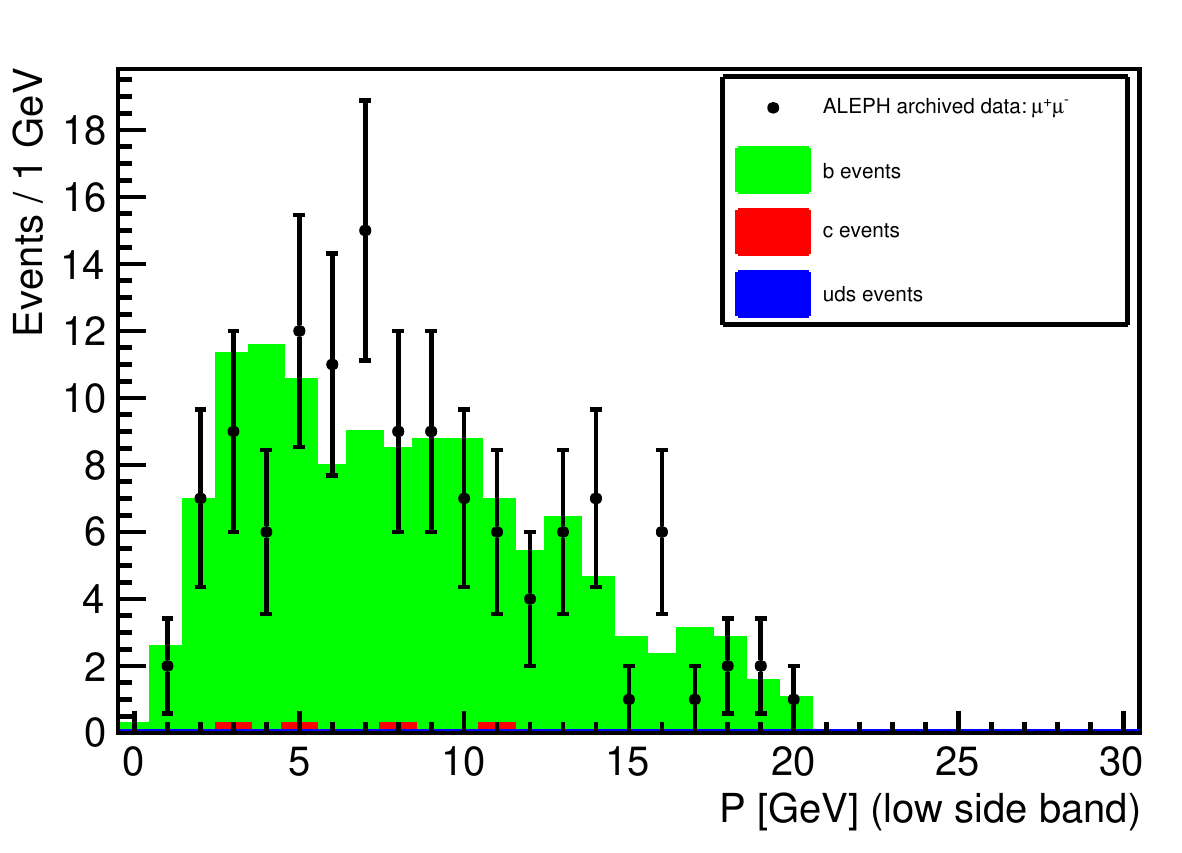}}\\
      \subcaptionbox{2$\sigma$ mass window events: $26.55 \lesssim m_{\mu^{+}\mu^{-}} \lesssim 34.25$~[GeV].\label{fig:dimuon_P_mw}}
       {\includegraphics[width=0.55\textwidth]{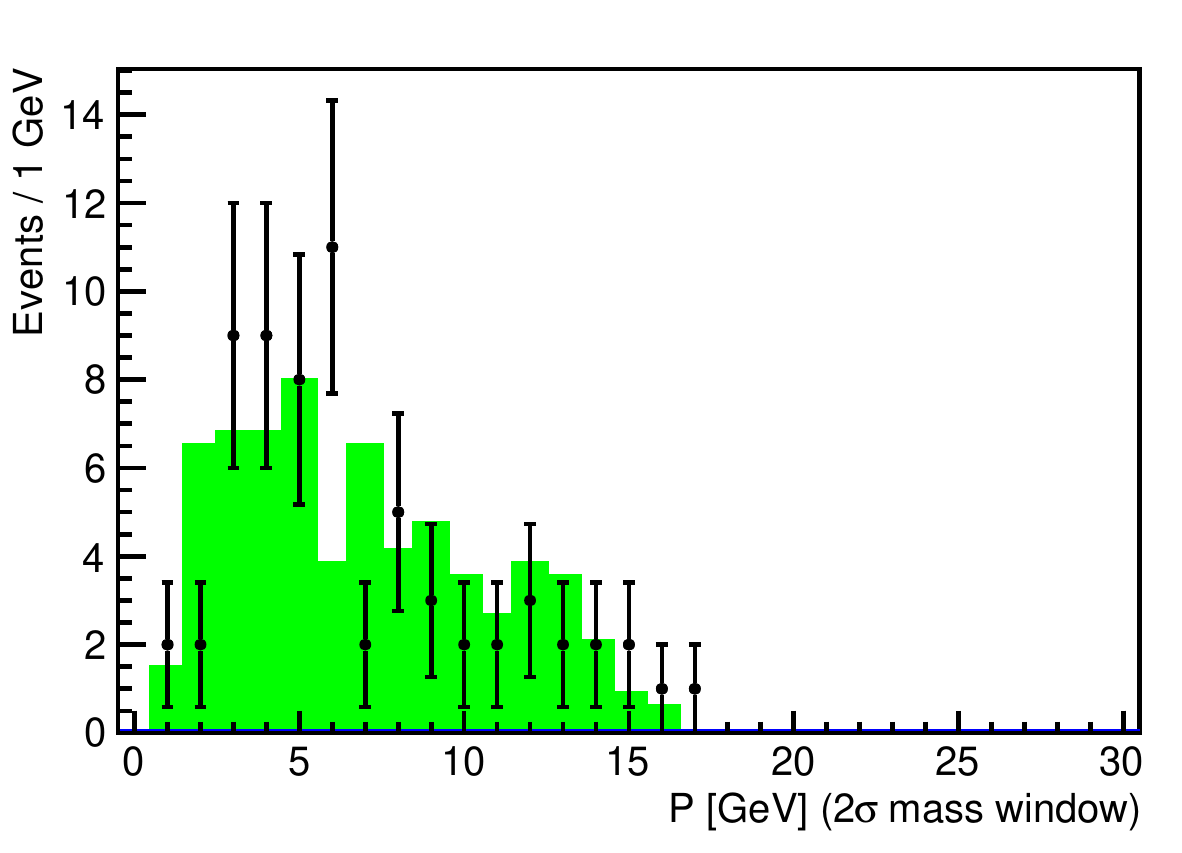}}\\
       \subcaptionbox{High side band events: $m_{\mu^{+}\mu^{-}} \gtrsim 34.23$~GeV.\label{fig:dimuon_P_high}}
       {\includegraphics[width=0.55\textwidth]{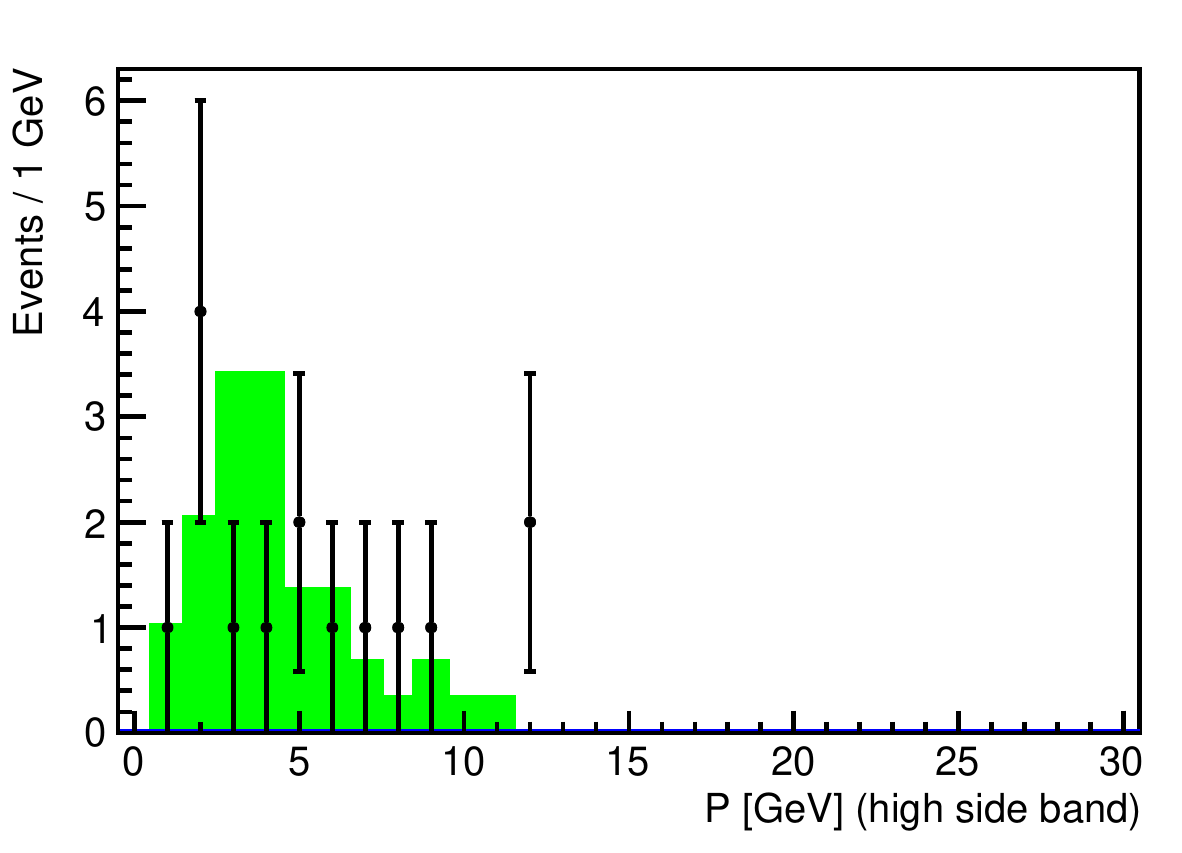}}
      \caption{The 3-momentum P of oppositely charged di-muon pairs for three different di-muon mass $m_{\mu^{+}\mu^{-}}$ regions.\label{fig:dimuon_P}}
\end{figure*}

The distributions of the 3-momentum P of oppositely charged di-muon pairs for three different di-muon mass $m_{\mu^{+}\mu^{-}}$ regions in comparison with the ALEPH MC simulation are shown in Figs.~\ref{fig:dimuon_P_low}, \ref{fig:dimuon_P_mw} and \ref{fig:dimuon_P_high}. The three different regions are defined by a 2$\sigma$ di-muon mass window:
\begin{enumerate}[(a)]
\item The {\it low side band} is defined by  $20 < m_{\mu^{+}\mu^{-}} \lesssim 26.55$~[GeV].
\item Events in the {\it 2$\sigma$ mass window} have a mass of $26.55 \lesssim m_{\mu^{+}\mu^{-}} \lesssim 34.25$~[GeV].
\item The {\it high side band} contains events, which have a di-muon mass $m_{\mu^{+}\mu^{-}} \gtrsim 34.23$~GeV.
\end{enumerate}

\clearpage

\section{Event displays of events from the signal region around 30~GeV using the ALEPH offline event display DALI}
\label{App:event_displays}

\begin{figure*}[!h]
      \centering
 \begin{overpic}[scale=0.72]{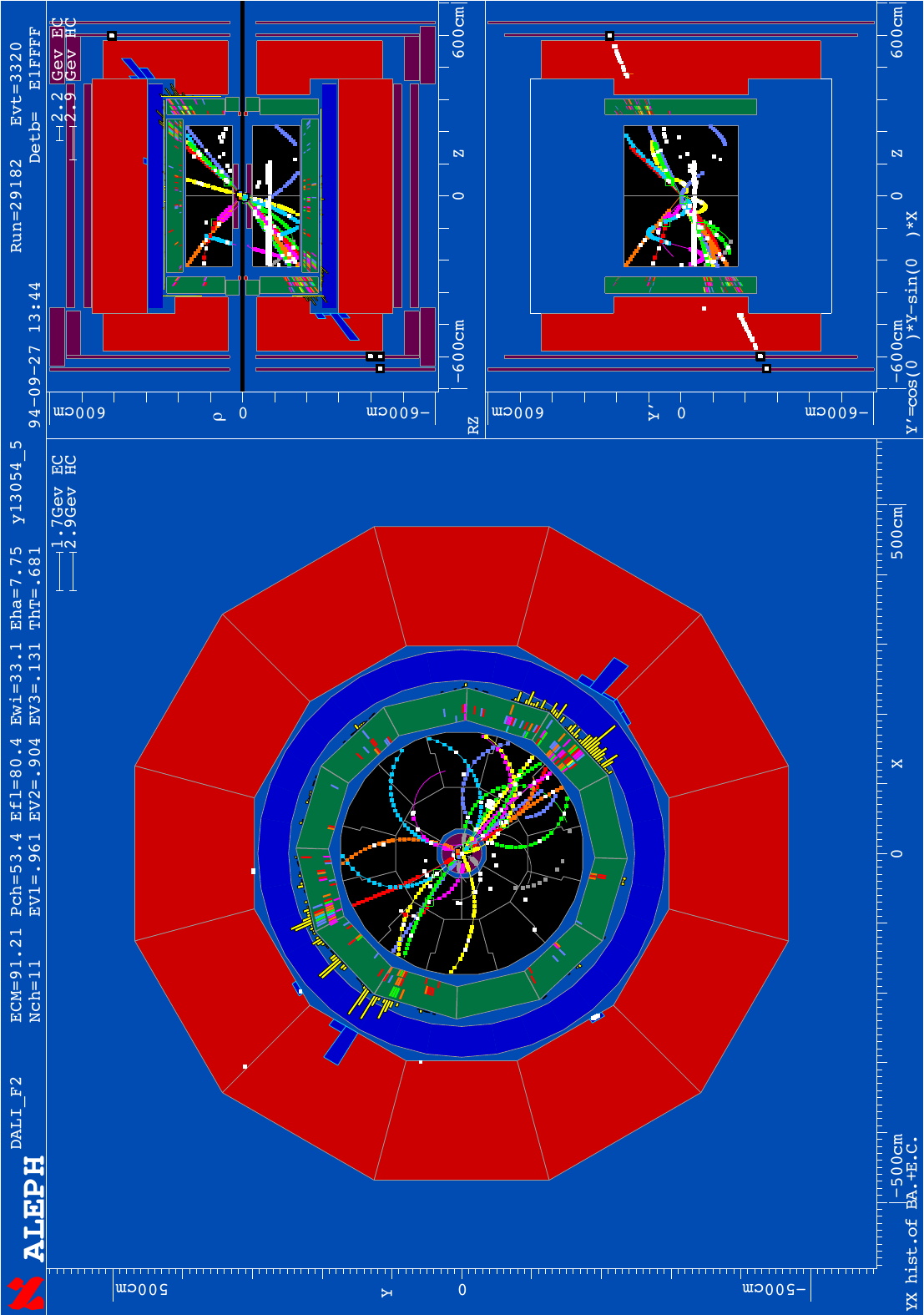}
     \put(4,4){\includegraphics[width=0.165\linewidth]{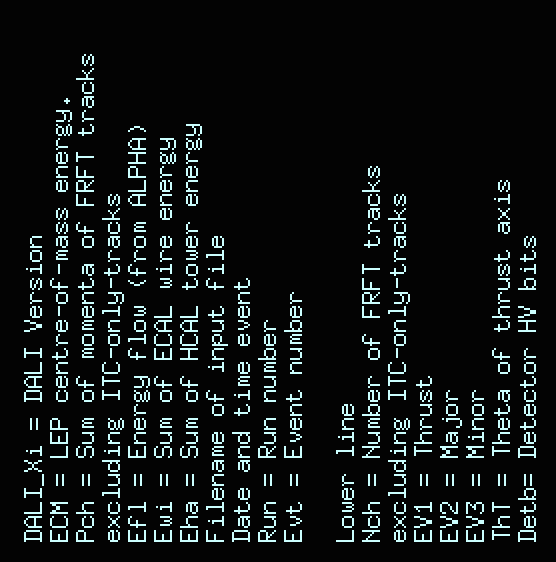}}  
  \end{overpic}
      \caption{This event display contains a legend describing the numbers given in the head lines.\label{fig:ed1}}
\end{figure*}

\begin{figure*}[!ht]
      \centering
       \includegraphics[width=0.9\linewidth]{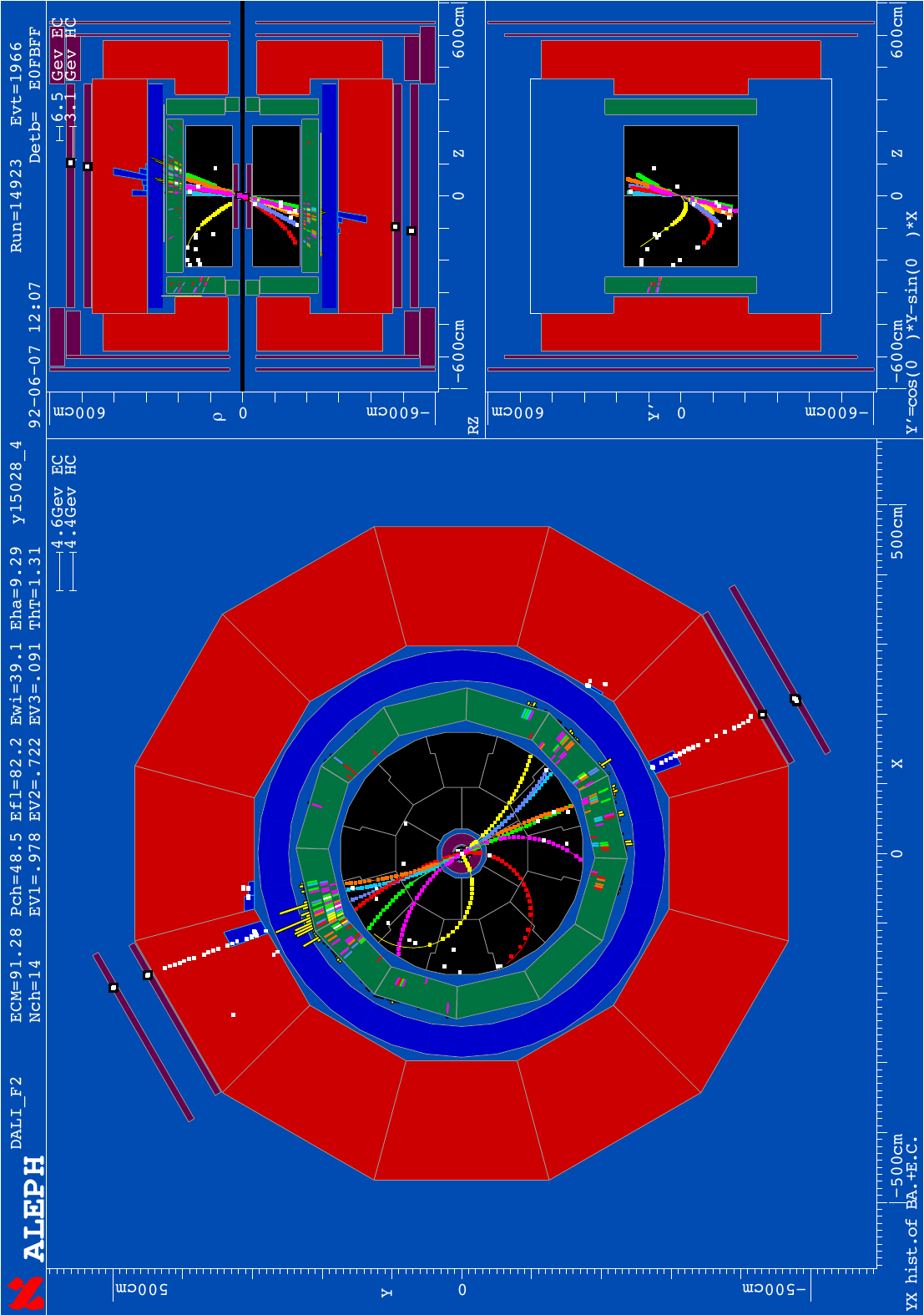}
      \caption{\label{fig:ed2}}
\end{figure*}

\begin{figure*}[!ht]
      \centering
       \includegraphics[width=0.9\linewidth]{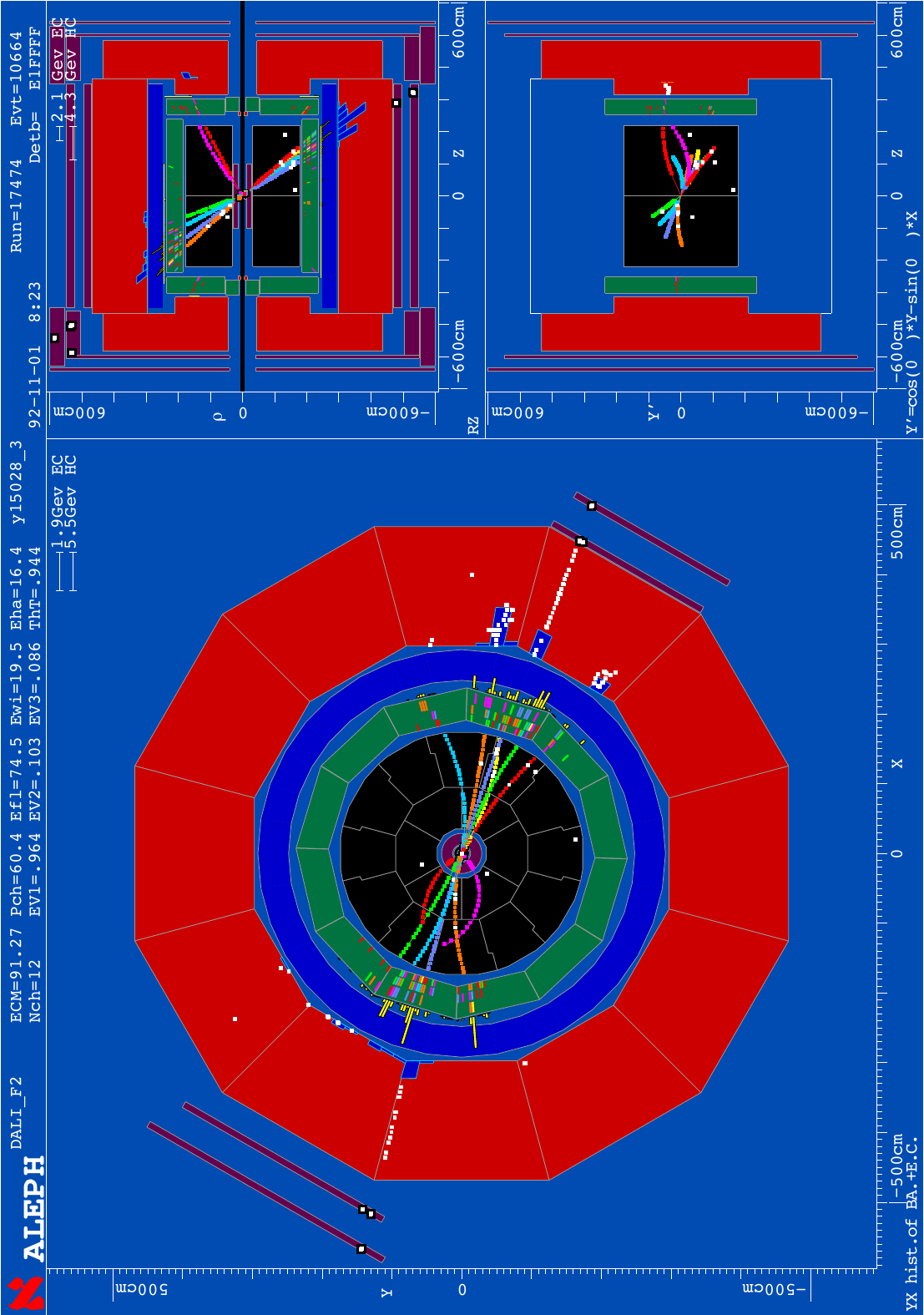}
      \caption{\label{fig:ed3}}
\end{figure*}

\begin{figure*}[!ht]
      \centering
       \includegraphics[width=0.9\linewidth]{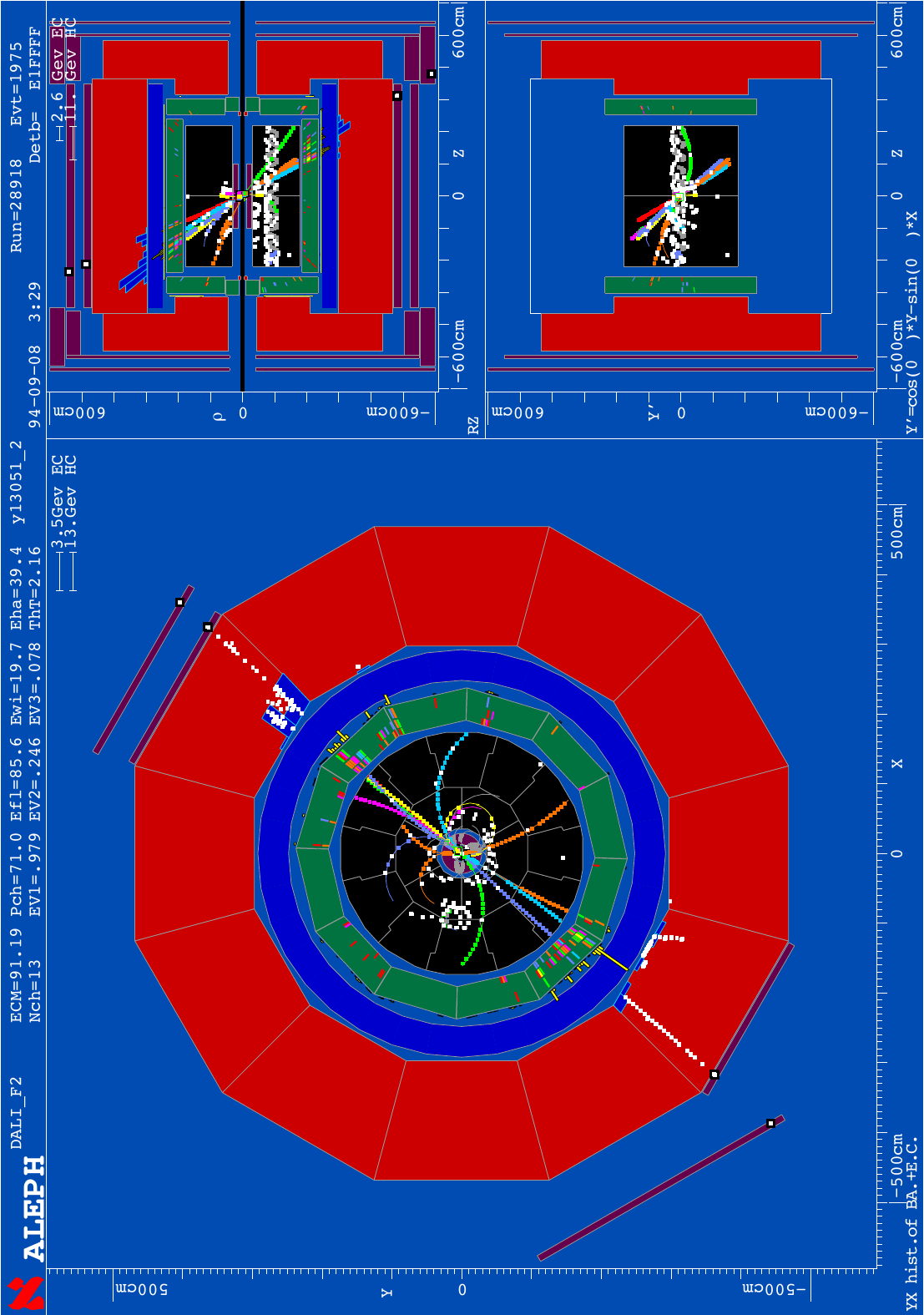}
      \caption{\label{fig:ed4}}
\end{figure*}

\begin{figure*}[!ht]
      \centering
       \includegraphics[width=0.9\linewidth]{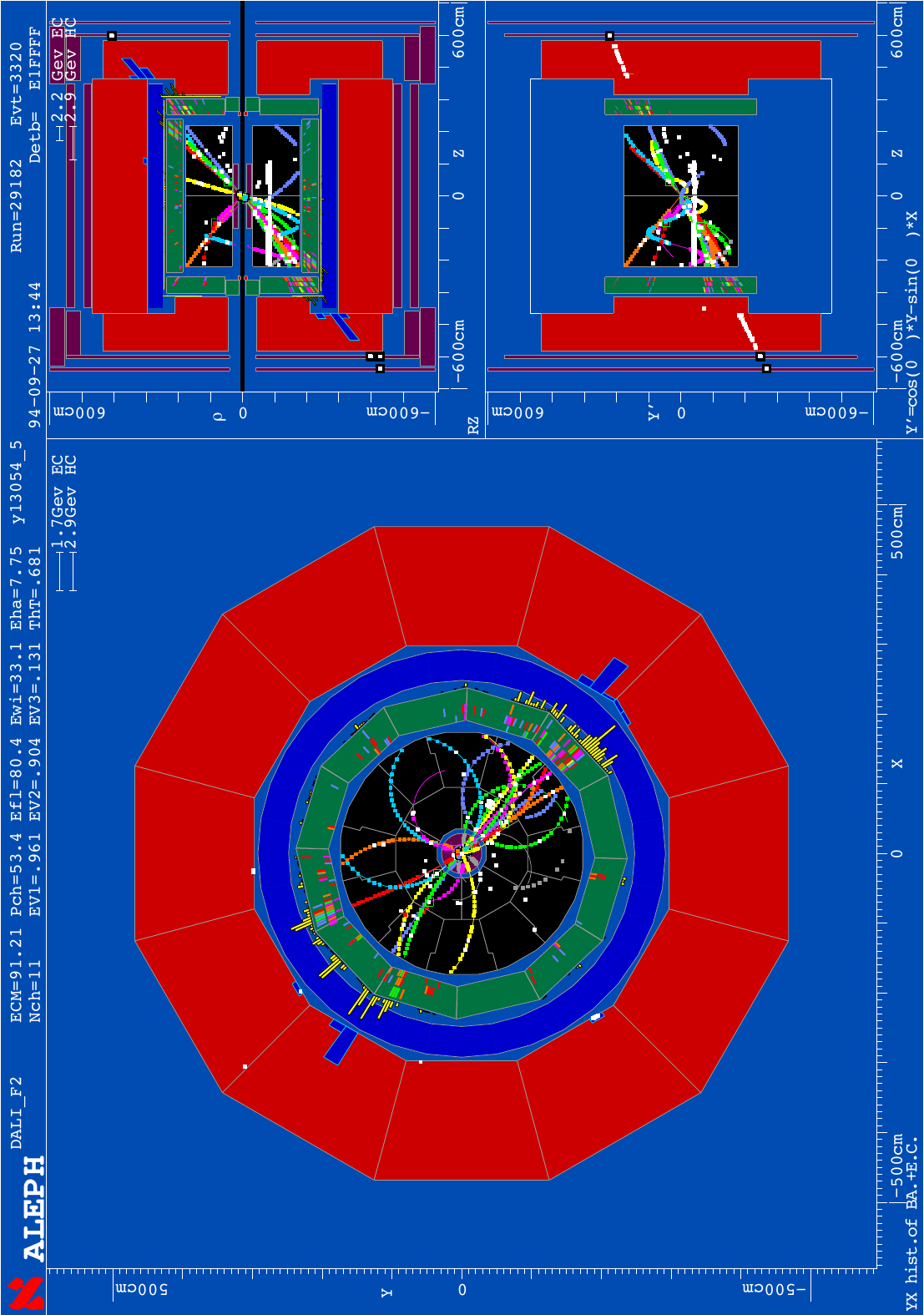}
      \caption{\label{fig:ed5}}
\end{figure*}

\begin{figure*}[!ht]
      \centering
       \includegraphics[width=0.9\linewidth]{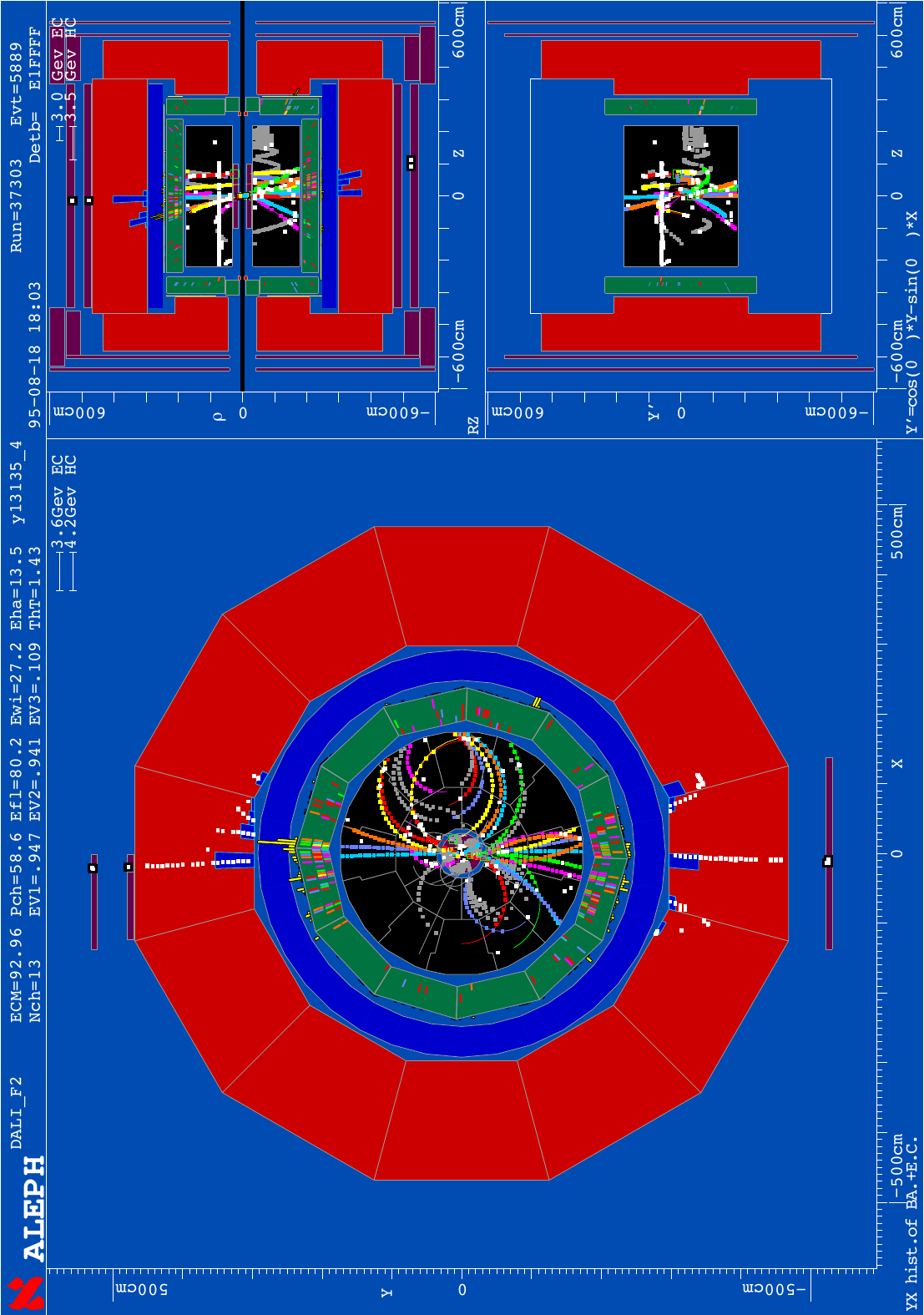}
      \caption{\label{fig:ed6}}
\end{figure*}

\end{document}